\pdfoutput=1   

\documentclass[pra,aps,onecolumn,superscriptaddress,showpacs,showkeys]{revtex4-1}

\usepackage{amssymb}
\usepackage{amsmath}
\usepackage{graphicx}
\usepackage[fleqn]{mathtools}
\usepackage[separate-uncertainty = true,multi-part-units=single]{siunitx}
\usepackage{xcolor}
\usepackage{hyperref}

\renewcommand{\vec}[1]{\mathbf{{#1}}}                         
\newcommand{\pvec}[1]{\mathbf{{#1}}_\parallel }
\newcommand{\vecUnit}[1]{\mathbf{\hat{#1}}}                         
\newcommand{\pvecUnit}[1]{\mathbf{\hat{#1}}_\parallel }
\newcommand{\nn}{\nonumber}
\newcommand{\imu}{\mathrm{i}}  
\newcommand{\dint}[2][]{\!\mathrm{d}^{#1}#2\,}              
\newcommand{\dfint}[3][]{\!\frac{\mathrm{d}^{#1}#2}{#3}\,}  

\newcommand{\etal}{{\it et al.}}

\renewcommand{\Re}{\mathrm{Re}\,}
\renewcommand{\Im}{\mathrm{Im}\,}

\DeclareMathOperator{\sgn}{\,sgn}

\newcommand{\red}[1]{#1}               

\newcommand{\sgnQK}{
  \red{
    \sgn ( \vec{\hat{q}}_\parallel\cdot \vec{\hat{k}}_\parallel )
  }
 }

\newcommand{\bqe}{\begin{eqnarray}}
\newcommand{\eqe}{\end{eqnarray}}
\newcommand{\bseq}{\begin{subequations}}
\newcommand{\eseq}{\end{subequations}}

\newcommand{\zxp}{\zeta({\bf x}_{\|})}
\newcommand{\zxpp}{\zeta({\bf x}\, '\!\!_{\|} )}
\newcommand{\bkp}{{\bf k}_{\|}}
\newcommand{\bqp}{{\bf q}_{\|}}

\newcommand{\bal}{\begin{align}}
\newcommand{\eal}{\end{align}}
\newcommand{\qp}{q_{\|}}
\newcommand{\pp}{p_{\|}}

\newcommand{\bQp}{{\bf Q}_{\|}}
\newcommand{\bQpp}{{\bf Q}\, '\!\!_{\|}}
\newcommand{\bxp}{{\bf x}_{\|}}

\newcommand{\bxpp}{{\bf x}\, '\!\!_{\|}}

\newcommand{\p}{\partial }

\newcommand{\kp}{k_{\|}}

\newcommand{\zhq}{\hat{\zeta}({\bf Q}_{\|})}
\newcommand{\zhqp}{\hat{\zeta}({\bf Q}\, '\!\!_{\|})}

\newcommand{\w}{\omega }

\newcommand{\la}{\langle}
\newcommand{\ra}{\rangle}

\newcommand{\xp}{x_{\|}}

\graphicspath{{Figures/},.}         

\begin{document} 

\title{Inversion of simulated and experimental light scattering data for characterization of  two-dimensional randomly rough metal surfaces}

\author{I. Simonsen}
\email{ingve.simonsen@ntnu.no}
\affiliation{Surface du Verre et Interfaces, UMR 125 CNRS/Saint-Gobain, F-93303 Aubervilliers, France}
\affiliation{Department of Physics, NTNU -- Norwegian University of Science and Technology, NO-7491 Trondheim, Norway}

\author{J.B. Kryvi} 
\affiliation{Department of Physics, NTNU -- Norwegian University of Science and Technology, NO-7491 Trondheim, Norway}

\author{A.A. Maradudin}
\affiliation{Department of Physics and Astronomy, University of California, Irvine CA 92697, U.S.A.}

\date{\today}

\begin{abstract}
  An approach is presented for the inversion of simulated and experimental in-plane, co-polarized light scattering data in p and s polarization to obtain the normalized surface-height autocorrelation function and the rms-roughness of a two-dimensional randomly rough metal surface. The approach is based on an expression, obtained on the basis of second-order phase perturbation theory, for the contribution to the mean differential reflection coefficient from the light scattered diffusely by the rough surface. The inversion scheme is validated by using several sets of computer generated scattering data for rough silver surfaces defined by Gaussian surface height correlation functions.  The reconstructions obtained by this approach are found to be rather accurate for weakly rough surfaces illuminated by p- and s-polarized incident light; this is also true in cases where the contributions to the input data from multiple scattering of surface plasmon polaritons is not insignificant. Finally, the inversion scheme is applied to experimental scattering data obtained for characterized two-dimensional randomly rough gold surfaces, and the results obtained in this way, compare favorably to what is obtained by directly analyzing the surface morphology. Such results testify to the attractiveness of the computationally efficient inversion scheme that we propose.
\end{abstract}


\maketitle

\section{Introduction}

Randomly rough surfaces are abundant in natural and man-made systems alike~\cite{Ogilvy1987}. There is therefore a strong desire to characterize such surfaces and, in particular, to determine their statistical properties. Statistical information on a randomly rough surface is contained in its distribution of heights and in its normalized surface-height autocorrelation function~\cite{Ogilvy1987,Book:Stout2000,Simonsen2004-3}. These two functions are often parametrized by the root-mean-square~(rms) height and the transverse correlation length of the surface roughness. In principle, the most direct and straightforward approach to the statistical properties of a randomly rough surface, is first to obtain a map of its surface morphology over a grid of points in the mean plane and then to calculate the statistical properties of interest on the basis of these data. For instance, a map of the surface height can be obtained by atomic force microscopy and/or by contact profilometry~\cite{Book:Kaupp2006,Book:Thomas1999} two of many possible scanning probe microscopy techniques. What is measured by such techniques is the surface morphology convoluted by the tip of the probe. Moreover, such techniques are typically contact methods which  make them time-consuming to perform and typically put constraints on the size of the spatial region in the mean plane of the surface that can be measured with sufficient spatial resolution. In addition, the morphology data obtained in this way can, at times, be non-trivial to collect and  analyze~\cite{Simonsen2018-07}.

Indirect methods represent alternative approaches to the determination of the statistical properties of a randomly rough surface for which maps of the surface morphologies are not required. Such methods are typically non-contact methods. A wave approach of this kind is based on \textit{inverse scattering theory}. Here one first measures the angular dependence of the intensity of the light that has been scattered diffusely into the far field by the rough surface when it is illuminated by a plane incident wave and next use these measured data  to determine the statistical properties of the rough surface. There are several advantages of this approach. The technique is non-invasive, fast, relatively cheap to implement, and typically can cover a large region of the mean plane of the surface which is important when aiming to determine the statistical properties of the rough surface. For two-dimensional randomly rough surfaces, this problem was initially studied by Chandley~\cite{Chandley1976}, and later by Marx and Vorburger~\cite{Marx1990}. Chandley assumed that the intensity of the light scattered by the randomly rough surface could be modeled by scalar diffraction theory in combination with the use of a thin phase screen model. In this way and due to the form of the expressions for the scattered intensity, the angular dependence of the mean scattered intensity could be readily inverted for the statistical properties of the surface by the use of a two-dimensional Fourier transform. Furthermore, Chandley also made the additional assumption that the autocorrelation function of the wave front is the same as the surface-heigh autocorrelation of the surface. This is only a good approximation for sufficiently small polar angles of incidence and scattering and for surfaces which are not too rough. An additional disadvantage of this approach is that one needs to measure the full angular distribution of the scattered light, something that is both time-consuming and requires rather specialized optical equipment (photo goniometers) to perform.

Almost one and a half decades after the initial work of Chandley~\cite{Chandley1976}, Marx and Vorburger~\cite{Marx1990} assumed the Kirchhoff approximation and used it to calculate the mean scattered intensity resulting from a plane incident scalar wave that is scattered from a perfectly conducting surface. By assuming a particular form of the normalized surface-height autocorrelation function, such a model was used in a least-square procedure to determine the transverse correlation length and the rms-roughness of the surface from experimental scattering data.  This approach had the advantage over the approach of Chandley that the full angular intensity distribution of the scattered light was not required; for instance, it was sufficient to only consider the intensity distribution in the plane of incidence (in-plane scattering).

%
Two decades after the initial publication by Chandley~\cite{Chandley1976}, a generalization of part of his work was conducted by Zhao and colleagues~\cite{Zhao1996}. Like  Chandley, these authors used a Fourier technique for the inversion of the scattering data, while the scattering model that they used was based on the Kirchhoff approximation, unlike Chandley, but similar to what was assumed by Marx and Vorburger~\cite{Marx1990}. The advantage of the approach of Zhao~\etal is that it is computationally efficient, due to the use of the Fourier transform, and that it does not require the full angular intensity distribution of the scattered light as an input. Moreover, this approach is not limited to small polar angles of incidence and scattering and it is not restricted to surfaces for which the  rms-roughness over the wavelength of the incident light is much less than unity (as was the case for the approach of Chandley).

Assuming scalar wave theory and the Kirchhoff approximation, Zamani~\etal\cite{Zamani2016} recently derived, within certain additional approximations, an analytic relation between the surface-height autocorrelation function and the intensity of the scalar wave that has been scattered diffusely by the rough surface; also see Refs.~\onlinecite{Zamani2012,Zamani2012a}. This expression these authors were able to invert analytically for the surface-height autocorrelation function of the rough surface. It should be remarked that previously a similar analytic inversion approach had been introduced by Zhao~\etal\cite{Zhao1996}. However, the expressions that the analytic inversions were based on in Refs.~\onlinecite{Zhao1996} and \onlinecite{Zamani2016} are different, and so are the results for the correlation functions obtained by inversion.  These differences are due to the two groups of authors using different assumptions and mathematical approximations in deriving these expressions.

\textit{All} the inversion schemes reported in Refs.~\onlinecite{Chandley1976,Marx1990,Zhao1996,Zamani2016} assume that scalar wave theory is adequate to model the scattering of light from two-dimensional randomly rough dielectric or metal surfaces. It is well known that light is vector waves, and that the polarization states of the incident and scattered light have to be taken into account in order to obtain a quantitatively accurate description of the intensity of the light that is scattered from a randomly rough surface~\cite{Simonsen2004-3}. When the transverse correlation length of the rough surface is significantly larger than the wavelength of the incident light, one typically finds little difference between the in-plane angular intensity distributions of the p-to-p and the s-to-s scattered light [for instance, see Figs.~\ref{Fig:Ex1:ppol}(a) and \ref{Fig:Ex1:spol}(a)]; in such cases a scalar wave theory may (or, may not) be sufficient. However, when the lateral correlation length is smaller than the wavelength of the incident light, the angular intensity distributions for in-plane p- and s-polarized light are typically rather different [for an example, see Figs.~\ref{Fig:Ex4:ppol}(a) and \ref{Fig:Ex4:spol}(a)]; in these cases, scalar wave theory is inadequate and will produce incorrect results. We also note that when the lateral correlation length of the surface is smaller than the wavelength of the incident light, the Kirchhoff approximation is not expected to be valid; therefore, a reconstruction approach based on it is not expected to produce reliable results.

Recently a vector theory was used successfully for the reconstruction of the statistical properties of a randomly rough \textit{dielectric} surface performed on the basis of the in-plane angular dependence of s-to-s scattering data obtained for the surface~\cite{Simonsen2014-05}. The vector theory used here to model the light that is scattered from the rough surface is second order phase perturbation theory~\cite{Shen1980,NavarreteAlcala2009}. The reconstructions that Simonsen~\etal\cite{Simonsen2014-05} performed using this vector theory were otherwise preformed in a manner that resembles how they were performed by Marx and Vorburger~\cite{Marx1990}. By assuming a particular form of the normalized surface-height autocorrelation function, a model for the scattering of light derived within phase perturbation theory, and the input scattering data, a least-square minimization procedure was used to determine the transverse correlation length and the rms-roughness of the surface. In this way, successful and accurate reconstructions were demonstrated for dielectric surfaces in s polarization for a set of different input and trial correlation functions. It ought to be remarked that a similar approach based on phase perturbation theory and applied to the reconstruction of surface parameters based on p-to-p scattering data for dielectric surfaces could in principle have been performed. In practice, however, such an approach turned out not to be very accurate due to the Brewster angle that exists in the scattering from a planar dielectric surface in p-polarization~\cite{Simonsen2016-06}. Since a Brewster angle does not exist for the scattering of s-polarized light from a planar dielectric surface, a similar problem does not exist for this polarization, something that is also shown explicitly by the results reported in Ref.~\onlinecite{Simonsen2014-05}.

Another electromagnetic inversion approach has also quite recently been developed and it is based on the vectorial Kirchhoff approximation~\cite{Simonsen2019-04}. By invoking the stationary phase method and a few approximations, the expressions for the scattered intensity can be inverted analytically, up to the evaluation of an integral, for the surface height correlation function and rms-roughness without any adjustable parameters. The main attractiveness of this approach is that no parametrization of the correlation function is needed; from only the in-plane angular dependence of the co-polarized scattered intensity, the correlation function of the surface can be calculated by the evaluation of an integral. This approach has been applied successfully for the reconstruction of the roughness parameters of both dielectric and metallic surfaces.

%
Only recently have inversion schemes been applied to experimental scattering data. For instance, Zamani~\etal applied their approach to experimental scattering data obtained when light is scattered from rough silicon surfaces~\cite{Zamani2016}. They compared the results for the surface-height autocorrelation functions obtained by inversion to what was obtained directly from the surface morphologies measured by microscopy. For the surfaces that they considered, good agreement is reported between the two classes of results. In their study, Zamani~\etal argue that their approach is more accurate than previous approaches, and this in particular applies to the approach of Zhao~\etal\cite{Zhao1996}. Also in Ref.~\onlinecite{Simonsen2019-04} the introduced electromagnetic inversion approach is applied to experimental data. For instance, here two experimental data sets that correspond to the scattering of light from two rough gold surfaces are reconstructed for their roughness parameters. The results that are obtained in this way agree well with what were obtained by analyzing the measured surface morphology of the surfaces.

The electromagnetic inverse scattering problem, as we discuss it here, is intimately related to the corresponding direct (or forward) scattering problem. Over the years, numerous perturbative and approximate approaches towards the solution of the direct scattering problem have been developed~\cite{Ogilvy1987,Elfouhaily2004,Simonsen2004-3,Warnick2001} and some of these are compared to experimental data in Ref.~\onlinecite{NavarreteAlcala2009}. However, there are rather few \emph{rigorous} approaches to the scattering of electromagnetic waves from two-dimensional randomly rough surfaces~\cite{Tran1994,Pak1997,Tran1994a,Wagner1997,Simonsen2009-1,Simonsen2009-9,Simonsen2011-05,Simonsen2012-05} and those that exit, are computationally expensive to apply. As the size of the randomly rough surface grows, the numerical cost of applying such methods very quickly becomes excessive so that they cannot readily be used in practice. For the study of the scattering of light from two-dimensional randomly rough surfaces, several integral equation methods are the predominant approaches used for numerical simulations. One such method is based on the Stratton-Chu integral equation where the unknowns are surface currents and the kernels of the integral equations are related to the Green's functions~\cite{Book:Kong2005,Simonsen2009-1,Simonsen2009-9}. Another approach that can be used for the simulation of scattering of light from rough surfaces is the reduced Rayleigh equation approach~\cite{Brown1984,Simonsen2012-05,Simonsen2011-02} which is a spectral method~\footnote{When the Rayleigh hypothesis is satisfied this approach is also rigorous.}. Here one solves a set of coupled inhomogeneous integral equations where the unknowns are either the reflection or transmission amplitudes of the reflected or transmitted light, respectively. The advantage of the Rayleigh approach over the Stratton-Chu  approach is that for the same size and discretization  of the surface, the former is significantly more efficient. It is the Rayleigh approach that we use in this work to produce simulation results used for evaluating our inversion approach.

The purpose of this paper, is to use second order phase perturbation theory ---  a vector theory for light ---  to develop the needed formalism and to use it to perform reconstructions of the statistical properties of randomly rough \emph{metal} surfaces based on the polarization dependent in-plane angular intensity distributions. This will be possible for both p- and s-polarized light, since for a metal surface a Brewster angle does not exist for which the scattered intensity vanishes. First, the formalism that we propose will be applied to the reconstruction based on computer generated scattering data for both linear polarizations and for a set of roughness parameters. Next, the reconstruction will be performed on the basis of experimental scattering data. The intensity of the light scattered from weakly and moderately rough metal surfaces, unlike light scattered from rough dielectric surfaces, can obtain, depending on the roughness parameters of the surface, significant contributions from the multiple scattering of surface plasmon polaritons~\cite{Book:Maier2007,Simonsen2004-3}.   
An important issue that we address in this work is how multiple scattering processes, contributing to the input scattering data, influence the quality of the height autocorrelation functions and the  surface parameters that are reconstructed by the approach that we propose. It is demonstrated that even when multiple scattering contributions are non-neglectable, reconstruction based on the proposed approach can produce fairly accurate results.

\smallskip
The remaining part of the paper is organized in the following way: Section~\ref{Sec:geomtry} presents the scattering system that we will be concerned with, followed  by some useful elements of scattering theory on which the subsequent discussion relies~[Sec.~\ref{Sec:theory}]. The inversion scheme that we will use for the reconstruction of the surface roughness parameters as well as the surface-height autocorrelation function is presented in Sec.~\ref{Sec:Inversion}. Section~\ref{sec:Results} presents and discusses the results that can be obtained by applying the proposed reconstruction procedure to a set of different scattering geometries and computer generated and experimental scattering data. The conclusions and outlook that can be drawn from this study is presented in Sec.~\ref{Sec:Conclusions}. The paper ends with an Appendix presenting a detailed derivation of the expressions, central to the present work, for the first few moments of the scattering matrix for p-to-p scattering obtained on the basis of phase perturbation theory.

\section{The System Studied} 
\label{Sec:geomtry}

%
\begin{figure}
  \includegraphics{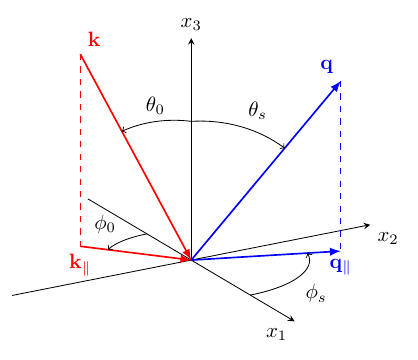}
  \caption{Schematics of the scattering geometry considered in this work.}
  \label{fig:geometry}
\end{figure}

The scattering system that we consider in this work is depicted in Fig.~\ref{fig:geometry}. It consists of vacuum in the region $x_3 > \zeta(\pvec{x})$, and a metal, characterized by a frequency dependent dielectric function $\varepsilon(\omega)$ in the region $x_3 < \zeta(\pvec{x})$. At the angular frequency $\omega$ of the incident light, this dielectric function has a negative real part while its imaginary part is positive or zero. The vector  $\pvec{x} = (x_1, x_2, 0)$ is a position vector in the plane $x_3  =0$. The surface profile function $\zeta(\pvec{x})$ is assumed to be a single-valued function of $\pvec{x}$ that is differentiable with respect to $x_1$ and $x_2$. It is also assumed to constitute a stationary, zero-mean, isotropic, Gaussian random process defined by 
\begin{subequations}
  \label{eq:1}
  \begin{align}
    \label{eq:1b}
    \langle \zeta({\bf x}_{\|}) \rangle &= 0 \\
    \label{eq:1a}
    \langle \zxp \zxpp \rangle &= \delta^2 W(|\bxp - \bxpp|), 
  \end{align}
\end{subequations}
where the angle brackets denote an average over the ensemble of realizations of $\zeta(\pvec{x})$, $\delta= \left<\zeta^2({\bf x}_{\|})\right>^{1/2}$ is the root-mean-square~(rms) height of the surface, and $W(|\bxp|)$ denotes  the {\it normalized surface-height autocorrelation function}, with the property that $W(0) = 1$. 

To aid in the later discussion, it will be convenient to introduce the following Fourier integral representation of the surface profile function
\begin{align}
  \label{eq:2}
  \zxp &= \int \dfint[2]{Q_{\parallel}}{(2 \pi)^2} \zhq \exp(\imu \bQp \cdot \bxp ),
\end{align}
where $\bQp = (Q_1, Q_2, 0)$ is a two-dimensional wave vector, so that
\begin{subequations}
\label{eq:3a}
\begin{align}
  \zhq &= \int \dint[2]{x_{\parallel}} \zxp \exp (-\imu \bQp \cdot \bxp ).
\end{align}
In addition we also introduce the notation
\label{eq:3b}
\begin{align}
  \hat{\zeta}^{(n)}(\pvec{Q}) &= \int \dint[2]{x_{\parallel}} \zeta^n({\bf x}_{\|}) \exp(-\imu \bQp \cdot \bxp ),
\end{align}
\end{subequations}
where for simplicity when $n=1$ we simply will write $\hat{\zeta}(\pvec{Q})\equiv\hat{\zeta}^{(1)}(\pvec{Q})$.

The Fourier coefficient $\zhq$ is also a zero-mean Gaussian random process defined by 
\begin{align}
  \label{eq:4}
  \left< \zhq \zhqp \right> &=(2 \pi)^2 \delta(\bQp + \bQpp) \, \delta^2 g(|\bQp|),
\end{align}
where $g(|\bQp|)$, the {\it power spectrum} of the surface roughness, is defined by
\begin{align}
  \label{eq:5}
  g(|\bQp|) &= \int \dint[2]{x_{\parallel}} W(|\bxp |)  \exp(- \imu \bQp \cdot \bxp ).
\end{align}

From Eqs.~\eqref{eq:1} and \eqref{eq:5}, it follows that $g(|\bQp|)$ is normalized to unity,
\begin{align}
\label{eq:6}
\int \frac{\dint[2]{Q_{\parallel}}}{(2 \pi)^2} g(|\bQp|) &= 1.
\end{align}

\section{Scattering Theory}
\label{Sec:theory}

The metallic surface $x_3 = \zeta(\pvec{x})$ is illuminated from the vacuum by a plane-wave electromagnetic field of angular frequency $\omega$. The  electric field component of this (incident) field is given by 
\begin{subequations}
  \label{eq:7}
  \begin{align}
    \label{eq:7a}
    \vec{E}^{(i)}(\vec{x};t) & = \Big\{
                                   -\frac{c}{\omega} \left[ \pvecUnit{k} \alpha_0(\kp) + \vecUnit{x}_3 \kp \right] B_p(\bkp)   
                                   + ( \vecUnit{x}_3 \times \pvecUnit{k} )B_s(\bkp) 
                            \Big\}\,
                            \exp\left\{ \imu [\bkp - \vecUnit{x}_3\alpha_0(\kp)] \cdot \vec{x} - \imu \omega t\right\},
  \end{align}
where $B_p(\bkp)$ and $B_s(\bkp)$ are known amplitudes. The \emph{total} electric field in the vacuum region above the surface consists of the sum of this incident field and a scattered field, ${\bf E}({\bf x}; t) = {\bf E}^{(i)}({\bf x};t) 
+ {\bf E}^{(s)}({\bf x};t)$, where the scattered electric field has the form 
  \begin{align}
    \label{eq:7b}
    \vec{E}^{(s)}(\vec{x};t) & = \int \dfint[2]{\qp}{(2 \pi)^2}
        \Big\{ 
              \frac{c}{\omega} \left[ \pvecUnit{q} \alpha_0(\qp) - \vecUnit{x}_3 \qp \right] A_p(\bqp)   
              + (\vecUnit{x}_3 \times \pvecUnit{q} )A_s(\bqp) 
        \Big\} \,
        \exp \left\{ \imu [ \bqp + \vecUnit{x}_3\alpha_0(\qp) ] \cdot \vec{x} - \imu\omega t\right\},
  \end{align}
\end{subequations}
where the amplitudes $A_p(\bqp)$ and $A_s(\bqp)$ have to be determined. In Eq.~\eqref{eq:7} the subscripts p and s denote the p-polarized~(TM) and s-polarized~(TE) components of each of these fields, respectively, and a caret over a vector indicates that it is a unit vector. The function $\alpha_0(\qp)$ that appears in  Eq.~\eqref{eq:7} is 
defined as
\begin{align}
  \label{eq:def-alpha0}
  \alpha_0(\qp) &= \left[ \frac{\w^2}{c^2} - \qp^2 \right]^{1/2}, 
     \qquad \quad \Re\alpha_0(\qp) > 0, \,\, \Im\alpha_0(\qp) > 0. 
\end{align}
Due to the linearity of the Maxwell's equations, a linear relationship exists between the amplitudes $A_\alpha(\bqp)$ and $B_\beta(\bkp)$ of Eq.~\eqref{eq:7}. We write this relation in the form $(\alpha = p, s, \beta = p, s)$
\begin{align} 
  \label{eq:8}
  A_\alpha(\bqp) & = \sum\limits_\beta R_{\alpha \beta}(\bqp|\bkp) B_\beta(\bkp),
\end{align}
where the quantities $\{R_{\alpha \beta}(\bqp|\bkp)\}$ denote the reflection amplitudes which play a significant role in the present theory since the mean differential reflection coefficient~(DRC) is defined in terms of them. It can be shown that the contribution to the mean~DRC  from the light scattered incoherently by the rough surface is given by~\cite{Simonsen2014-05,Simonsen2009-1,Simonsen2004-3} 
\begin{align}
  \label{eq:10}
  \left \langle \frac{\partial R_{\alpha \beta}(\bqp|\bkp)}{\partial \Omega_s} \right\rangle_{\textrm{incoh}} 
  &= 
  \frac{1}{{\mathcal S}}\left(\frac{\omega}{2 \pi c}\right)^2 \frac{\cos^2 \theta_s}{\cos \theta_0}
  \left[ \left< \left|R_{\alpha \beta}(\bqp|\bkp)\right|^2 \right> - \left| \Big< R_{\alpha \beta}(\bqp|\bkp)\Big> \right|^2 \right].
\end{align}
Here ${\mathcal S}$ denotes the area of the plane $x_3 = 0$ covered by the rough  surface, $(\theta_0, \phi_0)$  and $(\theta_s, \phi_s)$ are the polar and azimuthal angles of incidence and scattering, respectively, as defined in Fig.~\ref{fig:geometry}, and finally 
\begin{subequations}
  \label{eq:9-1}
  \begin{align}
    \pvec{k} &= \frac{\omega}{c} \sin \theta_0(\cos \phi_0, \sin \phi_0, 0)
    \label{eq:9-1a}
  \end{align}
  and
  \begin{align}
    \pvec{q} &= \frac{\omega}{c} \sin \theta_s(\cos \phi_s, \sin \phi_s, 0).
    \label{eq:9-1b} 
  \end{align}
\end{subequations}

Next we defined the scattering matrix ${\bf S}(\bqp|\bkp)$, whose elements $\{S_{\alpha \beta}(\bqp|\bkp)\}$  are expressed in terms of the elements of the matrix of reflection amplitudes ${\bf R}(\bqp|\bkp)$ by the relation~\cite{Welford1977}
\begin{align}
  \label{eq:11}
  S_{\alpha \beta}(\bqp|\bkp) & = \frac{\alpha_0^{1/2}(\qp)}{\alpha_0^{1/2}(\kp)} R_{\alpha \beta}(\bqp|\bkp),
\end{align}
and satisfy the reciprocity relations~\cite{Saxon1955,Book:Jones1964}
\begin{subequations}
\label{eq:12}
\begin{align}
  S_{pp}(\bqp|\bkp) & = S_{pp}(-\bkp|-\bqp) \label{eq:12a} \\
  S_{ss}(\bqp|\bkp) & = S_{ss}(-\bkp|-\bqp) \label{eq:12b} \\
  S_{ps}(\bqp|\bkp) & = -S_{sp}(-\bkp|-\bqp) \label{eq:12c}. 
\end{align}
\end{subequations}
The expression for the mean~DRC in Eq.~\eqref{eq:10} can alternatively be given in terms of the  elements of the scattering matrix as
\begin{align}
  \label{eq:13}
 \left< \frac{\partial R_{\alpha \beta}(\bqp|\bkp)}{\partial \Omega_s} \right>_{\textrm{incoh}} 
  &  = 
  \frac{1}{{\mathcal S}}\left(\frac{\omega}{2 \pi c}\right)^2 \cos \theta_s
  \left[ \left< \left| S_{\alpha \beta}(\bqp|\bkp) \right|^2 \right> - \left| \Big< S_{\alpha \beta}(\bqp|\bkp) \Big> \right|^2 \right].
\end{align}
It is this form we will work with here for the expression for the contribution to the mean DRC from the light scattered incoherently by the rough surface. The advantage of using the form in Eq.~\eqref{eq:13}, instead of the form in Eq.~\eqref{eq:10}, is that the the reciprocity relations~\eqref{eq:13} can be used to check the correctness of the expressions for the mean DRC that we obtain.

%
\medskip
Here we will be concerned with co-polarized scattering, that is, with the $\alpha\alpha$ element of Eq.~\eqref{eq:13}. To obtain them, we need $S_{\alpha \alpha}(\bqp|\bkp)$ which we will obtain on the basis of second-order phase perturbation theory. In the Appendix we detail that, within second-order phase perturbation theory, the $\alpha\alpha$ element of Eq.~\eqref{eq:13} can be expressed as
\begin{subequations}
  \label{eq:MDRC_PPT_total}
  \begin{align}
  \label{eq:MDRC_PPT}
  \left< \frac{\partial R_{\alpha\alpha}(\bqp|\bkp)}{\partial \Omega_s} \right>_{\textrm{incoh}}
  &=
    \frac{ \red{ \left| \varepsilon - 1 \right|^2 }   }{ 2\pi}
    \left(\frac{\omega}{c}\right)^2 \cos \theta_s
  \frac{|f_\alpha( q_\parallel )f_\alpha( k_\parallel )|}{|d_\alpha( q_\parallel )d_\alpha( k_\parallel )|^2} \exp [-2M_\alpha(\pvec{q} |\pvec{k} ) ] 
    \nonumber \\
  & \quad                 
   \times
   \int\limits^{\infty}_{0} \dint{u_\parallel} 
           u_\parallel J_0(|\pvec{q} - \pvec{k} | u_\parallel)
  \left\{ 
    \exp \left[  4\delta^2 
      \left| 
        \frac{\alpha_0( q_\parallel )\alpha_0( k_\parallel )}{f_{\alpha}( q_\parallel )f_{\alpha}( k_\parallel )}
      \right| 
    \left|H_\alpha(\pvec{q} |\pvec{k} )\right|^2 
    W(u_\parallel)
  \right]
  - 1 
\right\},
\end{align}
where $J_0(\cdot)$ denotes the Bessel function of the first kind and order zero, and 
\begin{align}
  M_{\alpha}(\pvec{q} |\pvec{k} ) 
  &= 
  -2\delta^2 \Re \left[ \frac{\alpha_0( q_\parallel )\alpha_0( k_\parallel )}{f_\alpha( q_\parallel )f_\alpha( k_\parallel )}\right]^\frac{1}{2}
  \int \dfint[2]{p_\parallel}{(2\pi )^2}   
  F_\alpha(\pvec{q} |\pvec{p} |\pvec{k} ) \,
  g(|\pvec{p} - \pvec{k} |)    . 
  \label{eq:M_alpha-definition}
\end{align}
\end{subequations}
In writing this expression it has been assumed that the surface is isotropic and we have introduced the functions
\begin{subequations}
  \label{eq:A.4}
  \begin{align}
    \label{eq:A.4a}
    d_p( q_\parallel )
    &=
      \varepsilon\alpha_0( q_\parallel ) + \alpha ( q_\parallel )
    \\
    \label{eq:4.b}
    d_s( q_\parallel )
    &=
      \alpha_0( q_\parallel ) + \alpha ( q_\parallel )             
  \end{align}
\end{subequations}
where $\alpha_0(q_\parallel)$ is defined in Eq.~\eqref{eq:def-alpha0}  and 
\begin{align}
  \label{eq:def-alpha}
  \alpha(\qp) &= \left[\varepsilon \left(\frac{\w}{c}\right)^2 - \qp^2 \right]^{1/2} 
     \qquad \quad \Re\alpha(\qp) > 0, \,\, \Im\alpha(\qp) > 0. 
\end{align}
Furthermore, we have also defined the functions
%
\begin{subequations}
  \label{eq:f_defintion}
\begin{align}
  \label{eq:f_p}
  f_p( q_\parallel )
  &=
    \varepsilon \left( \frac{\omega}{c} \right)^2 - (\varepsilon + 1)  q_\parallel^2 
  \\
  \label{eq:f_s}
  f_s( q_\parallel )
  &= \left( \frac{\omega}{c} \right)^2         
\end{align}
\end{subequations}
and
\begin{subequations}
  \label{eq:H_defintion}
  \begin{align}
    \label{eq:H_p}
    H_p(\pvec{q} |\pvec{k} ) 
    &=
      \sgnQK
      \left[
      \varepsilon q_\parallel k_\parallel 
      - \alpha ( q_\parallel ) 
      \pvecUnit{q} \cdot \pvecUnit{k}   
      \alpha ( k_\parallel )
      \right]
    \\ 
    \label{eq:H_s}
    H_s( \pvec{q} | \pvec{k} )
    &=
      - \sgnQK \left( \frac{ \omega }{ c } \right)^2 \pvecUnit{q} \cdot \pvecUnit{k}.
 \end{align}
\end{subequations}
Finally the functions $F_\alpha(\pvec{q} |\pvec{p} |\pvec{k} )$ are defined as (see the Appendix)
\begin{subequations}
  \label{eq:F_defintion}
  \begin{align}
    \label{eq:F_p}
      F_p(\pvec{q} |\pvec{p} |\pvec{k} ) 
  =
  \sgnQK
  &
  \Bigg[  
  \frac{1}{2} 
  \left[\alpha ( q_\parallel )+\alpha ( k_\parallel ) \right]
  \left[ 
    q_\parallel k_\parallel -\alpha ( q_\parallel )
    \pvecUnit{q} \cdot \pvecUnit{k} 
    \alpha ( k_\parallel ) 
   \right]
  \nonumber\\ & \quad 
  + \left( \frac{\varepsilon - 1}{\varepsilon} \right) 
  \bigg\{ 
  \alpha( q_\parallel ) 
  \pvecUnit{q} \cdot \pvecUnit{p}  
  \alpha ( p_\parallel )
  \pvecUnit{p} \cdot \pvecUnit{k}  
  \alpha ( k_\parallel )
  \nonumber\\ & \quad \quad
  - \frac{
      \left[\varepsilon q_\parallel p_\parallel - \alpha ( q_\parallel )
        \pvecUnit{q} \cdot \pvecUnit{p}  
        \alpha ( p_\parallel ) 
      \right]
      \big[\varepsilon p_\parallel k_\parallel -\alpha ( p_\parallel ) 
        \pvecUnit{p} \cdot \pvecUnit{k}  
        \alpha ( k_\parallel ) 
      \big]
    }{
      d_p( p_\parallel) 
    }
  \nonumber\\& \quad \quad 
  - \varepsilon \left( \frac{\omega}{c}\right)^2 
  \frac{
    \alpha ( q_\parallel )
    \left[ \pvecUnit{q} \times \pvecUnit{p} \right]_3   
    \big[ \pvecUnit{p} \times \pvecUnit{k} \big]_3   
    \alpha ( k_\parallel )
  }{
    d_s ( p_\parallel )
  }
   \bigg\}
   \Bigg],
  \end{align}
  and (see Eq.~\eqref{eq:FF_s})
\begin{align}
  \label{eq:F_s}
  F_s(\pvec{q} |\pvec{p} |\pvec{k} )
  =&
     - 
     \sgn(\pvecUnit{q} \cdot \pvecUnit{k})
     \left( \frac{\omega}{c}\right)^2
     \bigg\{ 
     \frac{1}{2}[\alpha(\qp) + \alpha(\kp)] (\pvecUnit{q} \cdot \pvecUnit{k})  
  + (\varepsilon - 1)   
    [\pvecUnit{q} \times \pvecUnit{p}]_3          
   \frac{\alpha_0(\pp) \alpha(\pp)}{d_p(\pp)} 
   [\pvecUnit{p} \times \pvecUnit{k}]_3   \nn \\  
  & \qquad \qquad  \qquad \qquad \qquad 
  -(\varepsilon - 1)\left(\frac{\w}{c} \right)^2
  \frac{ (\pvecUnit{q} \cdot \pvecUnit{p}) (\pvecUnit{p} \cdot \pvecUnit{k}) }{d_s(\pp)} 
   \bigg\}.
\end{align}
\end{subequations}
Here the notation $[\cdot]_3$ denotes the third component of the vector argument. It ought to be remarked that the expressions for the mean DRC in Eqs.~\eqref{eq:MDRC_PPT_total}--\eqref{eq:F_defintion} are obtained using one, of several possible, formulations of second-order phase perturbation theory. One alternative formulation of phase perturbation theory is presented in  Ref.~\onlinecite{NavarreteAlcala2009} where it is used to derive the corresponding expressions for the mean DRC. It is still not known which of the two formulations are the most accurate, or if one formulation is more accurate for metallic scattering systems, say, and the other for dielectric scattering systems. It might also be that the formulation to prefer depends on the wavelength regime of interest. There are questions that further research will have to answer.

\smallskip
We now turn to the calculation of the function $M_{\alpha}(\pvec{q} |\pvec{k} )$ that is defined by Eqs.~\eqref{eq:M_alpha-definition} and \eqref{eq:F_defintion}. This calculation is facilitated by expanding the power spectrum $g(|\pvec{p} - \pvec{k} |)$, defined by Eq.~\eqref{eq:5}, in the following way 
\begin{align}
  \label{eq:g-expansion}
  g(|\pvec{p} - \pvec{k} |)
  &=
    2\pi
    \sum_{n=-\infty}^\infty \exp\left[\imu n (\phi_p - \phi_k ) \right]
    \int_0^\infty \dint{u_\parallel} u_\parallel W(u_\parallel)\, J_n( p_\parallel u_\parallel)\, J_n\big( k_\parallel u_\parallel \big),
\end{align}
where $J_n(\cdot)$ denotes the Bessel function of the first kind and order $n$. The expansion~\eqref{eq:g-expansion} is established by first introducing polar representations of the vectors $\pvec{p}$, $\pvec{k}$,  and $\pvec{u}$, that is $\pvec{p}=p_\parallel(\cos\phi_p,\sin\phi_p,0)$  with $\phi_p$ the azimuthal angle of this vector with the positive $x_1$ axis (see Fig.~\ref{fig:geometry}) etc., then using the  Jacobi-Anger identity~\cite[Ch.~9]{Book:Abramowitz1964}, $\exp(\imu mz \cos\phi) =\sum_{m=-\infty}^\infty J_m(z)\exp(\imu \phi)$, and the property of the Bessel function $J_{-m}(z)=(-1)^mJ_{m}(z)$. When the expansion~\eqref{eq:g-expansion} is introduced into Eq.~\eqref{eq:M_alpha-definition} a lengthy,  but in principle straight forward calculation, shows that the angular integration (over $\phi_p$) can be performed analytically. For p-polarized light [$\alpha=p$], the result of such a calculation is 
\begin{subequations}
\begin{align}
  M_{p}(\pvec{q} |\pvec{k} )
  &=
    - \delta ^2
    \red{ \mathrm{sgn} \left( \pvecUnit{q} \cdot \pvecUnit{k} \right) }
    \Re 
    \sqrt{
    \frac{
    \alpha_0(\qp) \alpha_0(\kp)
    }{
    f_p(\qp) f_p(\kp)}
    }
    m_{p}(\pvec{q} |\pvec{k} )
\end{align}
where
\begin{align}
  m_{p}(\pvec{q} |\pvec{k} )
  &=
         q_\parallel  [\alpha( q_\parallel) + \alpha( k_\parallel)]  k_\parallel
    -
    \red{ ( \pvecUnit{q} \cdot \pvecUnit{k} ) }
    \alpha( q_\parallel)
    \left[\alpha( q_\parallel) + \alpha( k_\parallel)\right] \alpha( k_\parallel)
    \nn 
    \\ 
  & \quad {}
    +
    \frac{\varepsilon -1}{\varepsilon}\,
    \int_0^\infty \dint{ p_\parallel}  p_\parallel \bigg\lbrace
    -
    \frac{2 \varepsilon ^2 q_\parallel  p_\parallel ^2  k_\parallel}{d_p ( p_\parallel)}
    +
    \red{ ( \pvecUnit{q} \cdot \pvecUnit{k} ) }
    \alpha( q_\parallel)
    \left[ 
    \alpha( p_\parallel)
    -
    \frac{ \alpha ^2 ( p_\parallel) }{d_p ( p_\parallel)}
    +
    \frac{ \varepsilon (\omega/c )^2   }{d_s( p_\parallel)}
    \right]
    \alpha( k_\parallel)
    \bigg\rbrace
    \nn \\
  & \qquad \qquad \qquad \qquad  \quad
    \times
    \int_0^\infty \dint{ x_\parallel}  x_\parallel W( x_\parallel){J_0( p_\parallel  x_\parallel)J_0( k_\parallel  x_\parallel)}
    \nn \\
  & \quad {}
    +
    2
    (\varepsilon - 1)
    \int_0^\infty \dint{ p_\parallel}  p_\parallel
    \frac{ p_\parallel \alpha ( p_\parallel ) }{d_p ( p_\parallel)}
    \left[
    q_\parallel  
    \alpha ( k_\parallel)
    +
    \red{ ( \pvecUnit{q} \cdot \pvecUnit{k} ) }
    \alpha( q_\parallel) 
    k_\parallel
    \right]
    %
    \int_0^\infty \dint{ x_\parallel}  x_\parallel W( x_\parallel)
    {J_1( p_\parallel  x_\parallel)J_1( k_\parallel  x_\parallel)}
    \nn \\
  & \quad {}
    +
    \red{ ( \pvecUnit{q} \cdot \pvecUnit{k} ) }
    \frac{\varepsilon - 1}{\varepsilon}
    \int_0^\infty \dint{ p_\parallel}  p_\parallel
    \alpha ( q_\parallel )
    \left\lbrace
    \alpha ( p_\parallel ) 
    -
    \frac{ \alpha ^2 (  p_\parallel )  }{d_p ( p_\parallel )}
    -
    \frac{ \varepsilon (\omega/c )^2 }{d_s ( p_\parallel )}
    \right\rbrace
    \alpha (  k_\parallel )
    \int_0^\infty \dint{ x_\parallel}  x_\parallel W( x_\parallel)
    {J_2( p_\parallel  x_\parallel)J_2( k_\parallel  x_\parallel)}.
\end{align}

A similar calculation for s-polarized light was recently performed in  Ref.~\onlinecite{Simonsen2014-05} with the result that (see Eqs.~(22) and (A9) in Ref.~\onlinecite{Simonsen2014-05}) 
\begin{align}     
  M_{s}(\pvec{q} |\pvec{k} )
  &=
    - \delta^2
    \left| \pvecUnit{q} \cdot \pvecUnit{k} \right|
    \alpha_0^{1/2}(\qp) \alpha_0^{1/2}(\kp)
    \Re
    \bigg\{
    -\alpha(\qp) - \alpha(\kp)   \nn \\
  & \qquad \quad
    + 
    (\varepsilon - 1)
     \int_0^\infty \dint{\pp} \pp 
     \left( 
       \frac{\alpha_0(\pp) \alpha(\pp)}{d_p(\pp)} 
       +
       \frac{(\w/c)^2}{d_s(\pp)}
     \right)  
     \int_0^\infty \dint{\xp} \xp W(\xp) J_0(\pp \xp)J_0(\kp \xp) 
  \nn \\
  & \qquad \quad
    + 
    (\varepsilon - 1)
   \int_0^\infty \dint{\pp} \pp 
     \left( 
       -\frac{\alpha_0(\pp) \alpha(\pp)}{d_p(\pp)} 
       +
       \frac{(\w/c)^2}{d_s(\pp)}
     \right)  
     \int_0^\infty \dint{\xp} \xp W(\xp) J_2(\pp \xp)J_2(\kp \xp)
    \bigg\}. 
\end{align}
\end{subequations}

Note that for normal incidence, according to Eq.~\eqref{eq:9-1a},  one has $\pvecUnit{k} = (\cos\phi_0 \sin\phi_0, 0)$ even if $\pvec{k}/k_\parallel$ is not well defined in this case. Below we will be concerned with in-plane scattering in which case  $| \pvecUnit{q} \cdot \pvecUnit{k
} | =1$.

\section{The Inverse Problem}
\label{Sec:Inversion}

The aim of this work is to determine the rms-roughness $\delta$ and the height autocorrelation function $W(x_\parallel)$ of the rough metal surface from the knowledge of the in-plane angular dependence of the co-polarized light scattered incoherently from it. This will be done by minimizing the cost function
 \begin{align}
   \chi^2({\cal P}) = \int\limits^{\frac{\pi}{2}}_{-\frac{\pi}{2}} d\theta_s
   \sum_{\alpha}
   \bigg[\bigg\la \frac{\p R_{\alpha\alpha}(\theta_s)}{\p\Omega_s}\bigg\ra_{\textrm{incoh,input}}
  - \left\la \frac{\p R_{\alpha\alpha}(\theta_s)}{\p\Omega_s}\right\ra_{\textrm{incoh,calc}}\bigg]^2 , 
\label{eq:27}
 \end{align}
 where we to simplify the notation have introduced
 \begin{align}
  \left< \frac{ \p R_{\alpha\alpha}(\theta_s)}{\p\Omega_s}\right>_{\textrm{incoh}} 
  \equiv \left \langle \frac{\partial R_{\alpha\alpha}(\bqp|\bkp)}{\partial \Omega_s} \right\rangle_{\textrm{incoh}}\Bigg|_{|\pvecUnit{q} \cdot \pvecUnit{k}|=1},
\end{align}
and where ${\cal P}$ denotes the set of variational parameters used to characterize $\la \p R_{\alpha\alpha}(\theta_s)/\p\Omega_s\ra_{\textrm{incoh,calc}}$.  The $\alpha$-summation is present in Eq.~\eqref{eq:27} to facilitate the reconstruction of surface parameters based on the simultaneous use of scattering data in both p and s polarization. If data for only one polarization is used in the reconstruction, the summation is redundant. The minimization of $\chi^2({\cal P})$ with respect to the elements of ${\cal P}$ was carried out by the use of the routine ``\textrm{lmdif1}'' contained in the Fortran package MINPACK which is part of  the general purpose mathematical library SLATEC~\cite{SLATEC}. The routine \textrm{lmdif1} implements a modified version of the Levenberg-Marquardt algorithm~\cite{Levenberg1944,Marquardt1963}, and it calculates an approximation to the Jacobian by a forward-difference approach.

The functions $\la \p R_{\alpha\alpha}(\theta_s)/\p \Omega_s\ra_{\textrm{incoh,input}}$ with $\alpha={p,s}$ were either obtained from rigorous, nonperturbative, purely numerical solutions~\cite{Simonsen2011-03,Simonsen2011-05} of the reduced Rayleigh equation for the scattering of polarized light from a two-dimensional randomly rough metal surface~\cite{Brown1984} or from measurements~\cite{NavarreteAlcala2009}. The calculations were carried out for an ensemble of random surfaces generated~\cite{Simonsen2011-05} on the basis of expressions for $W(|\bxp |)$ of the Gaussian form
  \begin{align}
    W(|\pvec{x} |) = \exp \left\{- \left( \frac{x_\parallel}{a} \right)^2 \right\}. 
    \label{eq:28_Gaussian}
  \end{align}
In Eq.~\eqref{eq:28_Gaussian}, $a$ denotes the transverse correlation length of the surface roughness.

On the other hand, the functions $\la \p R_{\alpha\alpha}(\theta_s)/\p \Omega_s\ra_{\textrm{incoh,calc}}$ for $\alpha=p,s$ were obtained by evaluating the expression for it obtained using phase perturbation theory~[Eq.~\eqref{eq:MDRC_PPT_total}] for the trial function assumed to represent $W(|\pvec{x} |)$.  Several forms for this trial function were used in our calculations; either we assumed the Gaussian trial function
\begin{align}
    W(|\pvec{x} |) = \exp \left\{- \left( \frac{x_\parallel}{a_\star} \right)^2 \right\}, 
    \label{eq:29_Gaussian}
  \end{align}
or a stretched exponential trial function of the form 
\begin{align}
  W(|\bxp |) = \exp \left\{-\left(\frac{\xp}{a_\star}\right)^{\gamma_\star} \right\}.
  \label{eq:30}
\end{align}
When the form~\eqref{eq:29_Gaussian} is used for the trial function the variational parameters of the reconstruction are $\delta_\star$ and $a_\star$.  On the other hand, when the trial function~\eqref{eq:30} is used, the variational parameters of the reconstruction are $\delta_\star$, $a_\star$, and $\gamma_\star$. Note that stretched exponential trial function reduces to the Gaussian forms~\eqref{eq:29_Gaussian} when $\gamma_\star=2$.

\section{Results}
\label{sec:Results}

When visible light is scattered from a weakly rough metal surface, like silver, one of the prominent features that can be observed for some roughness parameters in the angular distribution of the scattered light is the enhanced backscattering phenomenon~\cite{Maradudin1990,Zayats2005,Simonsen2004-3} which expresses itself as a well defined peak in the retroreflection direction.
This phenomenon is also known under the alternative name of coherent backscattering phenomenon~\cite{Corey1995}. For weakly rough surfaces, it is known to result from the excitation and multiple scattering of surface waves known as surface plasmon polaritons~(SPPs)~\cite{Book:Maier2007}. Hence, when the enhanced backscattering peaks are present in the angular distribution of the scattered light, it is a clear signature of the contribution from multiple scattering processes to the scattered intensity. In the case when the contribution from single scattering processes can be neglected, the intensity at the position of the enhanced backscattering peak is expected to be twice of the intensity of its background~\cite{Maradudin1990}.

\subsection{Inversion of computer generated scattering data}

The second-order phase perturbation theoretical approach that we base the reconstruction on is not a complete multiple scattering approach. Even if second-order phase perturbation theory includes some (but not all) multiple scattering processes, it is unable to predict the existence of the enhanced backscattering phenomenon. Therefore, it is of interest to study how well the reconstruction approach based on it is able to handle different levels of contribution from multiple scattering to the scattered intensity. To this end, we will illustrate the inversion (or reconstruction) procedure developed here by applying it to the reconstruction of the rms-roughness $\delta$  and $W(\pvec{x}|)$ based on computer generated scattering data that receive various levels of contribution from multiple scattering processes (via the excitation and multiple scattering of SPPs). In the preceding subsection, we will apply our reconstruction approach to experimental scattering data.

%
\subsubsection{Scattering system no.~1}

\begin{figure}[tbp] 
  \centering
  \centering
  \includegraphics*[height=0.33 \columnwidth]{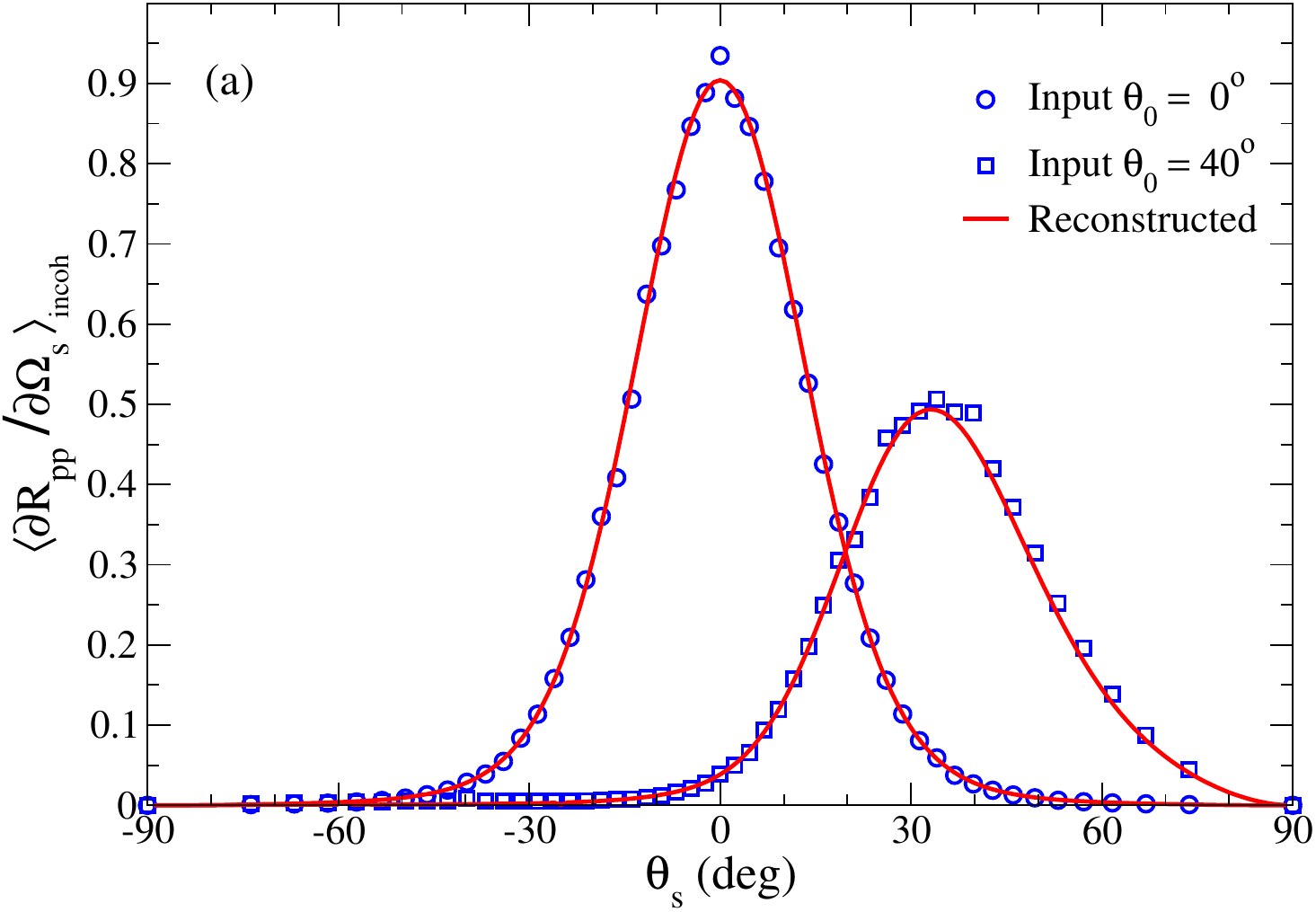}
  \\ 
  \includegraphics*[height=0.33 \columnwidth]{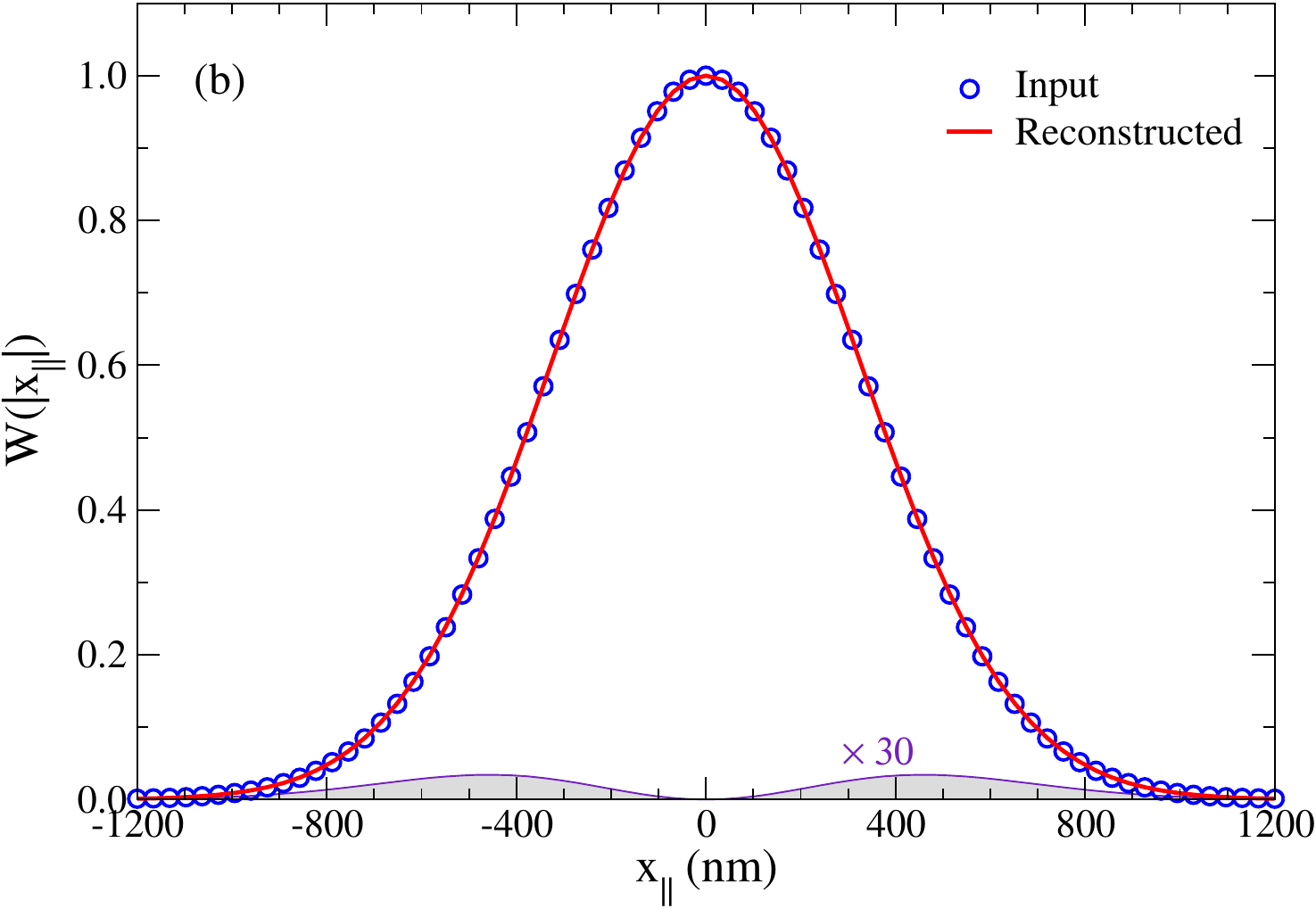}
  \\ 
  \includegraphics*[height=0.33 \columnwidth]{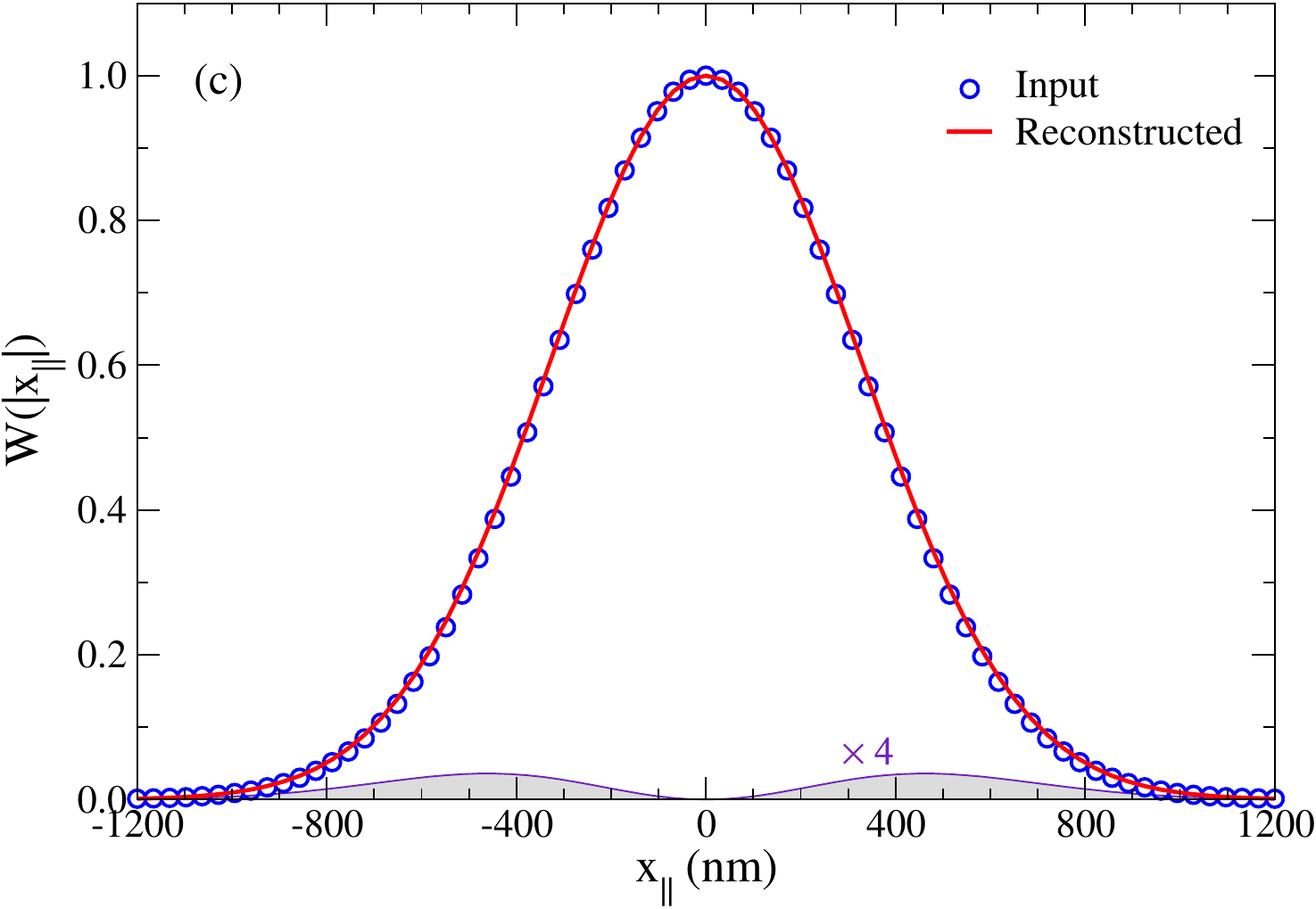}
  \caption{(Scattering system no.~1) Reconstruction of the rms-roughness $\delta_\star$ and the transverse correlation length $a_\star$ from p-polarized in-plane scattering data obtained for a Gaussian correlated rough silver surface. The wavelength (in vacuum) of the incident light is $\lambda = \SI{457.90}{\nano\meter}$ for which the dielectric function of silver is  $\varepsilon(\omega) = -7.50+0.24\imu$. The surface roughness parameters assumed in the computer simulations have the values $\delta =\lambda/20=\SI{22.90}{\nano\meter}$ and $a = \lambda=\SI{457.90}{\nano\meter}$.
    (a) The incoherent components of the in-plane, co-polarized (p-to-p) mean differential reflection coefficient $\left< \partial R_{pp}/\partial\Omega_s\right>_{\textrm{incoh}}$ as a function of the polar angle of scattering $\theta_s$ obtained from computer simulations for the polar angles of incidence $\theta_0=\ang{0}$~(open circles) and $\theta_0=\ang{40}$~(open squares), and from second-order phase perturbation theory for these polar angles of incidence with the use of the reconstructed surface roughness parameters (solid lines), for a two-dimensional randomly rough silver surface whose correlation function is defined by Eq.~\protect\eqref{eq:28_Gaussian}. The mean DRC results (input data) produced by computer simulations were obtained on the basis of \num{2500} surface realizations. 
    For normal incidence~[$\theta_0=\ang{0}$] the reconstructed values of the surface roughness parameters are $\delta_\star = \SI{22.72(0.20)}{\nano\metre}$ and $a_\star = \SI{458.6(3.6)}{\nano\metre}$, while for $\theta_0=\ang{40}$ one finds $\delta_\star = \SI{22.96(0.23)}{\nano\metre}$ and $a_\star =\SI{463.5(4.7)}{\nano\metre}$.
    In performing these reconstructions the form~\eqref{eq:29_Gaussian} for the autocorrelation function was assumed, and the minimization of the cost function~\eqref{eq:27}
was started from the initial values $\delta_\star=\SI{8.00}{nm}$ and $a_\star=\SI{150.00}{nm}$. For both angles of incidence, the computer simulation results for the in-plane angular dependence of  $\left< \partial R_{sp}/\partial\Omega_s\right>_{\textrm{incoh}}$ (p-to-s cross-polarization) were essentially indistinguishable from zero and therefore such results were not included.
    (b) For normal incidence~[$\theta_0=\ang{0}$], the input (open circles) and reconstructed (solid line) surface-height autocorrelation function $W(|\pvec{x}|)$ for the random surface. The shaded gray region represents the absolute difference between the input and reconstructed surface-height autocorrelation functions. The notation ``$\times 30$'' means that this difference has been multiplied by a factor \num{30} for better visibility. (c) The same as Fig.~\protect\ref{Fig:Ex1:ppol}(b) but now for the polar angle of incidence $\theta_0=\ang{40}$. 
  }
  \label{Fig:Ex1:ppol}
\end{figure}

\begin{figure}[tbp]  
  \centering
  \includegraphics*[height=0.33 \columnwidth]{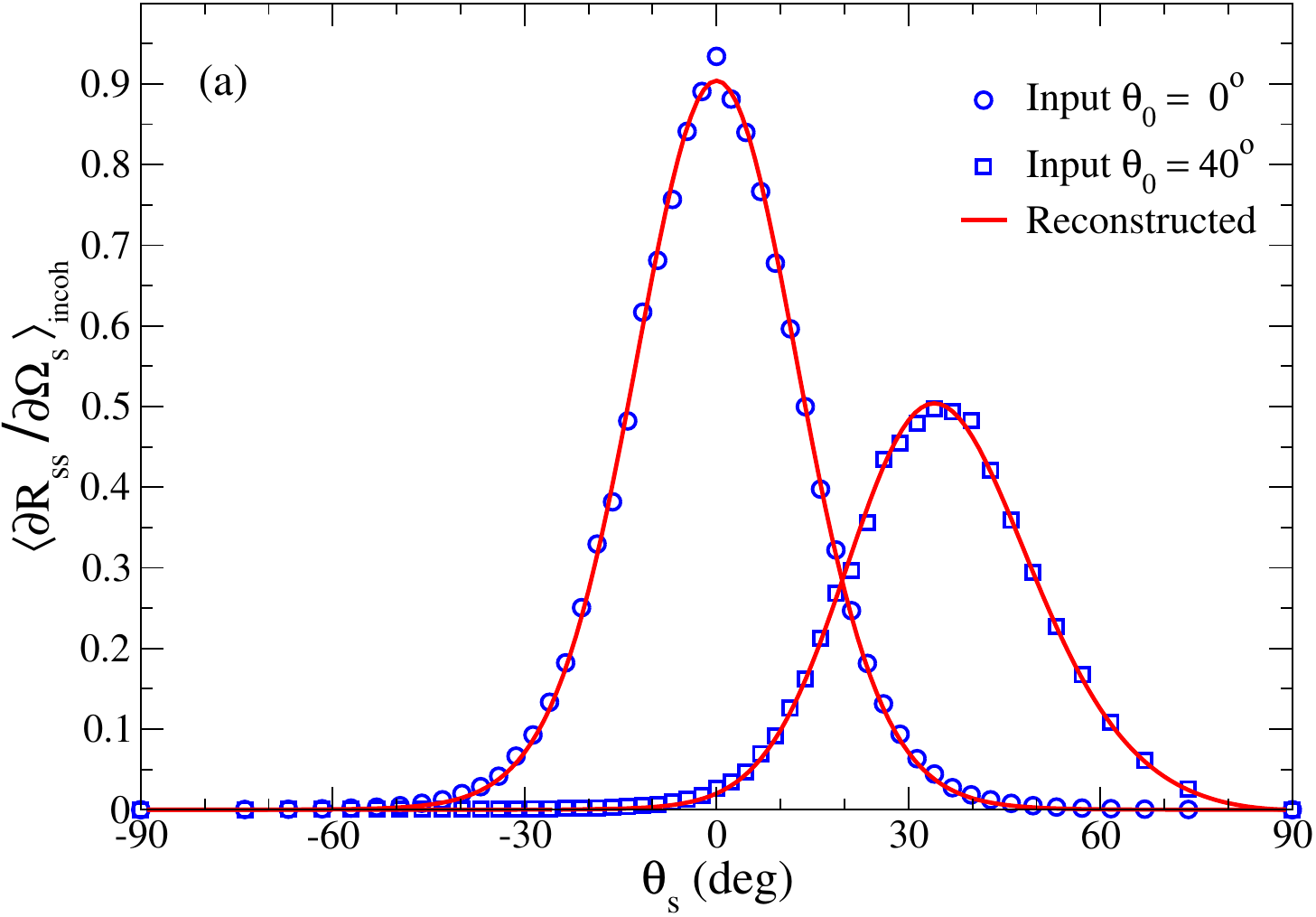}
  \\ 
  \includegraphics*[height=0.33 \columnwidth]{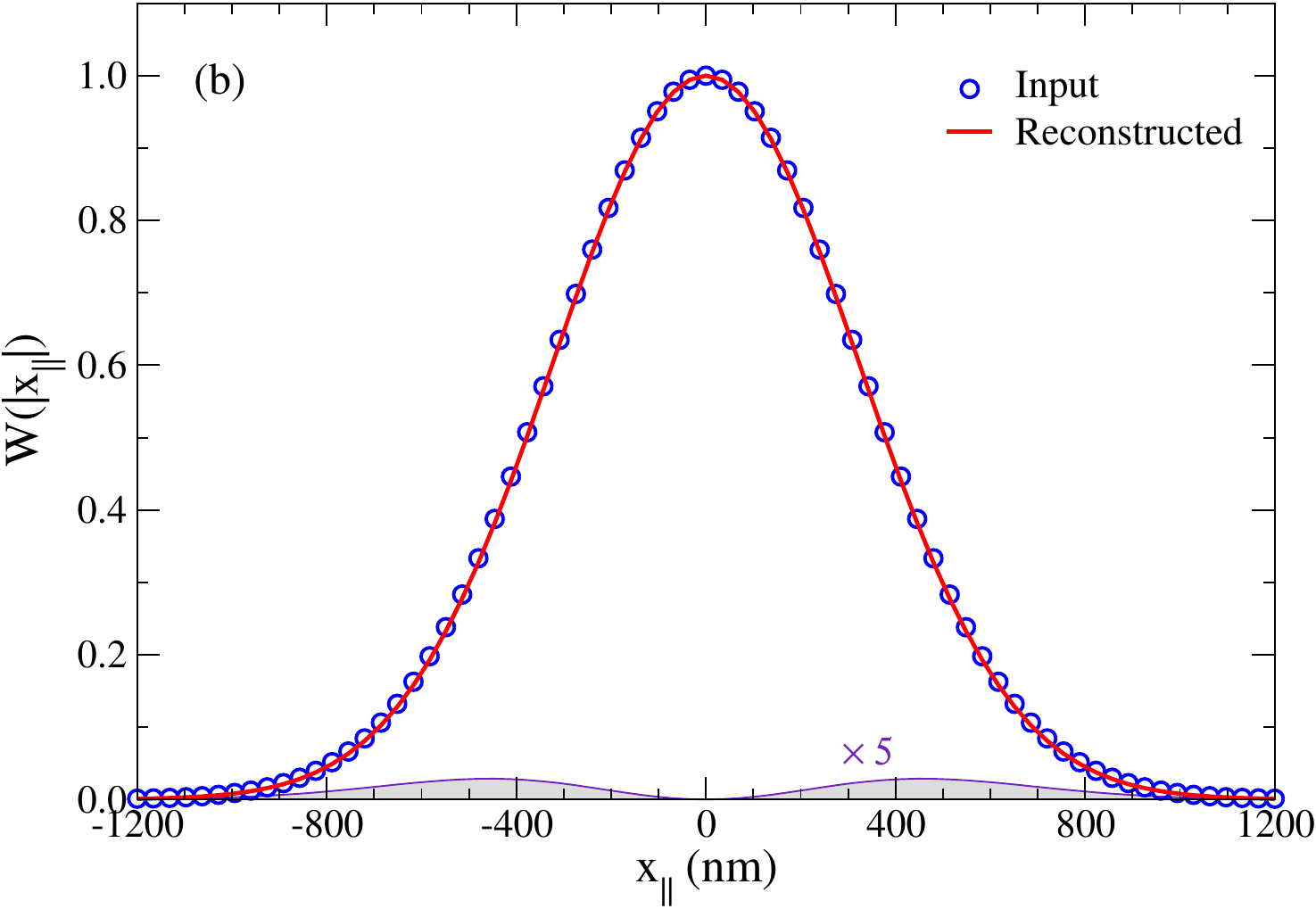}
  \\ 
  \includegraphics*[height=0.33 \columnwidth]{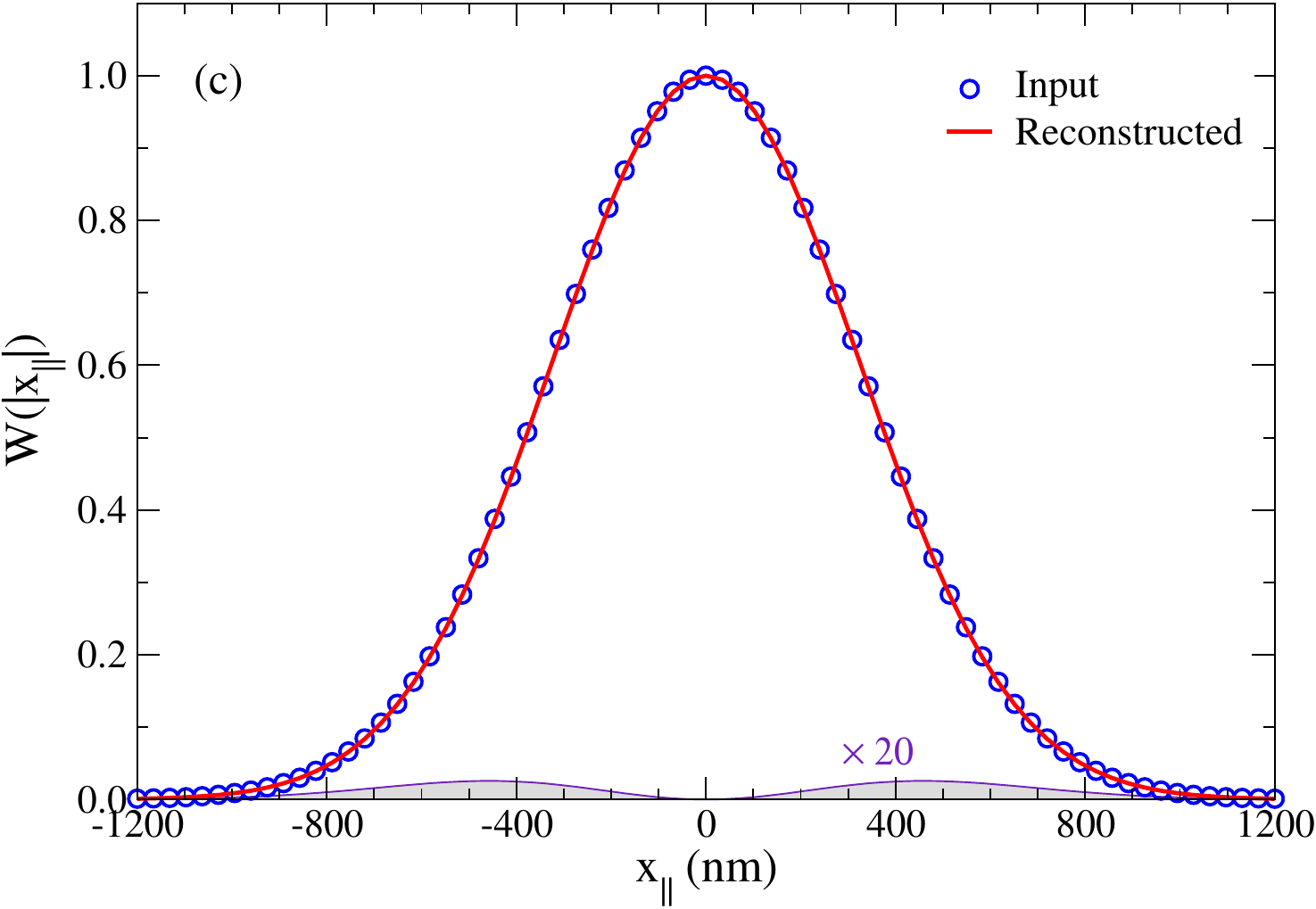}
  \caption{(Scattering system no.~1) Reconstruction of the rms-roughness $\delta_\star$ and transverse correlation length $a_\star$ from in-plane s-to-s scattering data obtained for a Gaussian correlated silver surface. The same as Fig.~\protect\ref{Fig:Ex1:ppol} but for s-polarized light. The values of the reconstructed surface roughness parameters are $\delta_\star = \SI{22.99(0.25)}{\nano\metre}$ and  $a_\star = \SI{454.3(4.3)}{\nano\metre}$ for $\theta_0=\ang{0}$; and $\delta_\star = \SI{22.81(0.23)}{\nano\metre}$ and  $a_\star = \SI{457.1(4.8)}{\nano\metre}$ for $\theta_0=\ang{40}$.
    [Input values: $\delta =\lambda/20=\SI{22.90}{\nano\meter}$ and  $a =\lambda=\SI{457.90}{\nano\meter}$].
  }%
  \label{Fig:Ex1:spol}
\end{figure}

The first scattering system that we consider is a silver substrate bounded by a randomly rough surface to vacuum which is the incident medium. An $\alpha$-polarized  plane incident wave of wavelength $\lambda=\SI{457.90}{nm}$ (in vacuum) illuminates the surface from the vacuum. At the wavelength of the incident light [$\lambda=2\pi/\omega$] the dielectric function of silver is $\varepsilon(\omega)=-7.50+0.24\imu$~\cite{Book:Palik1997}. The rms-roughness of the surface is taken to be $\delta=\lambda/20=\SI{22.90}{nm}$ and the surface-height autocorrelation function $W(|\pvec{x}|)$ is assumed to have the Gaussian form, Eq.~\eqref{eq:28_Gaussian}, and characterized by a transverse correlation length $a = \lambda=\SI{457.90}{nm}$.  For these roughness and geometrical parameters, the scattering problem was solved by a rigorous, non-perturbative and direct numerical solution of the reduced Rayleigh equation by the method of Ref.~\onlinecite{Simonsen2011-05}. In this way, we calculated by computer simulations the in-plane angular dependence of the contribution to the mean DRCs from  $\alpha$-polarized light scattered incoherently by the rough surface. When \num{2500} surface realizations were used in calculating the average, we obtained the results for $\left< \partial R_{\alpha\alpha}/\partial\Omega_s\right>_{\textrm{incoh}}$ presented as open symbol in Figs.~\ref{Fig:Ex1:ppol}(a) and \ref{Fig:Ex1:spol}(a) for p and s polarization, respectively; in both cases the polar angle of incidence was $\theta_0=\ang{0}$ or $\theta_0=\ang{40}$. These data constitute the input functions $\left<\partial R_{\alpha\alpha}(\theta_s)/\partial\Omega_s\right>_{\textrm{incoh,input} }$ for our first sets of reconstruction examples. It should be remarked that computer simulation results for the in-plane, cross-polarized scattering were essentially indistinguishable from zero on the scales of Figs.~\ref{Fig:Ex1:ppol}(a) and \ref{Fig:Ex1:spol}(a) and hence such results were not presented. Since in-plane, cross-polarized scattering is a multiple scattering effect~\cite{McGurn1996}, we find it reasonable to assume that the results presented in Figs.~\ref{Fig:Ex1:ppol}(a) and \ref{Fig:Ex1:spol}(a) do not receive a significant contribution from multiple scattering processes.

The first of these data sets that we reconstructed was for p-polarized light and normal incidence [Fig.~\ref{Fig:Ex1:ppol}(a) open circles]. To this end, it was assumed that the trial function $W(|\pvec{x}|)$ had the Gaussian form~\eqref{eq:29_Gaussian}, and therefore the set of variational parameters we assumed is ${\mathcal P}=\{\delta_\star,a_\star\}$. The minimization procedure was started from the initial values $\delta_\star=\SI{8.00}{nm}$ and $a_\star=\SI{150.00}{nm}$ and the minimization of the cost function $\chi^2({\mathcal P})$ for p-polarized data,  defined in Eq.~\eqref{eq:27}, resulted in the reconstructed parameters $\delta_\star=\SI{22.72(0.20)}{nm}$ and $a_\star=\SI{458.6(3.6)}{nm}$~\footnote{The confidence intervals reported in this work correspond to the 95\% confidence level.}. When these values are compared to the values $\delta=\SI{22.90}{nm}$ and $a = \SI{457.90}{nm}$ used to generate the input data, one finds good agreement. The corresponding reconstructed mean DRC curve, obtained by using Eq.~\eqref{eq:MDRC_PPT_total} and the number values of the reconstructed roughness parameters $\delta_\star$ and $a_\star$, is indicated as a solid line in Fig.~\ref{Fig:Ex1:ppol}(a); the main difference between the input and reconstructed mean DRC curves is found for the normal scattering direction. This discrepancy we speculate is due to an inaccuracy in the input data that is due to the use of a finite number of surface realizations for their calculation. The input and reconstructed surface-height autocorrelation function $W(|\pvec{x}|)$ are shown in Fig.~\ref{Fig:Ex1:ppol}(b) as open symbols and a solid line, respectively. The absolute difference between these two functions is presented as the gray shaded regions in Fig.~\ref{Fig:Ex1:ppol}(b) where ``$\times 5$'' indicates that this difference is multiplied by a factor \num{5} for reasons of clarity.

Next we performed reconstruction of p-polarized scattering data corresponding to the polar angle of incidence $\theta_0=\ang{40}$; this input data set is presented as open squares in Fig.~\ref{Fig:Ex1:ppol}(a).  When the reconstruction was performed in a completely equivalent way to how it was done for the scattering data corresponding to normal incidence, for instance by assuming the same initial values for the variational parameters, we obtained the following values for the reconstructed parameters $\delta_\star=\SI{22.96(0.23)}{nm}$ and $a_\star=\SI{463.5(4.7)}{nm}$. Also these reconstructed parameters are in agreement with the values assumed for the input roughness parameters. Still it is found that the number value of reconstructed transverse correlation length $a_\star$ deviates slightly more from the corresponding input value $a$ when  $\theta_0=\ang{40}$ than what is found when data for normal incidence is used in the reconstruction; this is reflected in a larger absolute difference between the input and reconstructed $W(|\pvec{x}|)$ as presented in Fig.~\ref{Fig:Ex1:ppol}(c).

\begin{figure}[tbp] 
  \centering
  \includegraphics*[width=0.453\columnwidth]{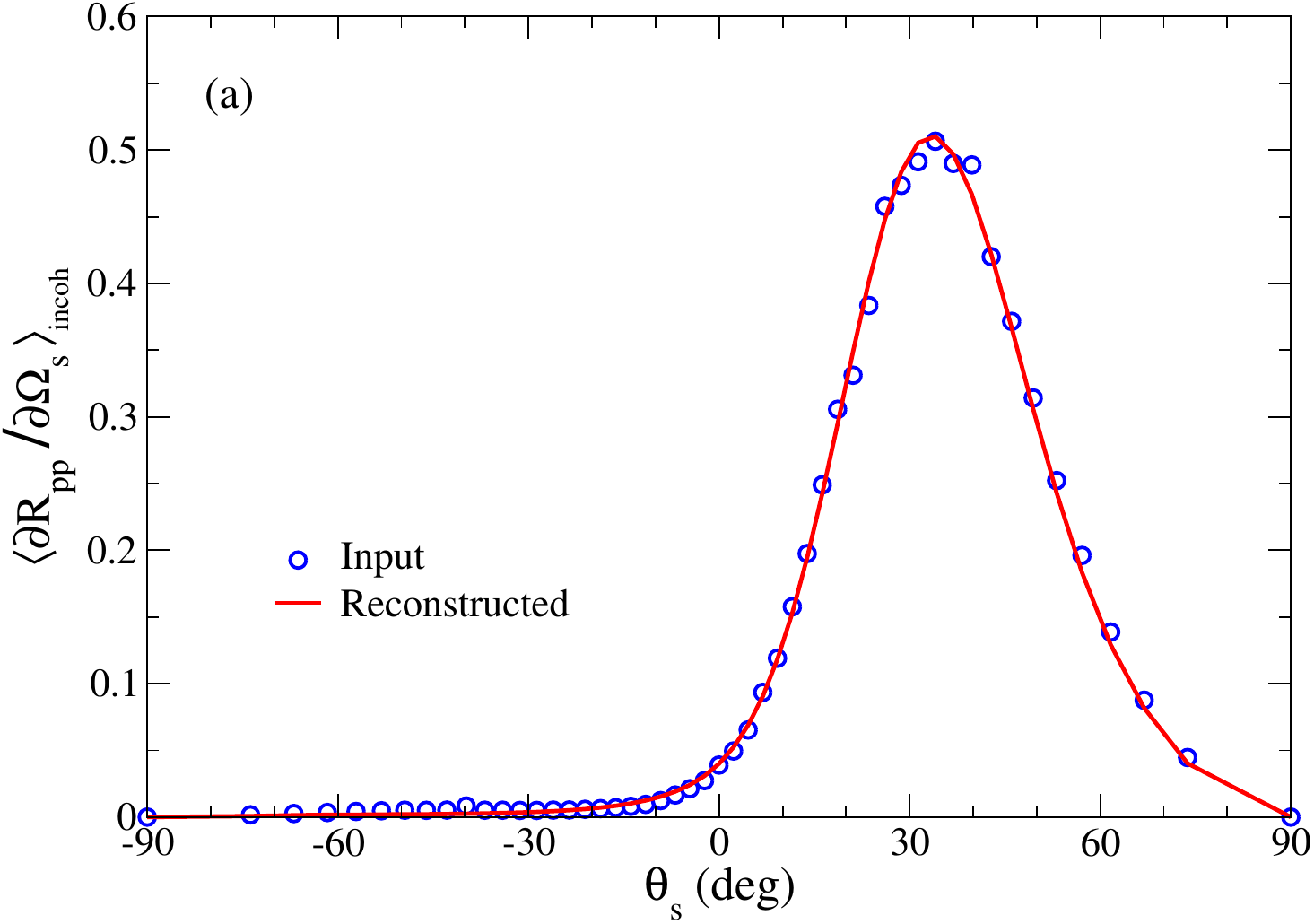}
  \qquad
  \includegraphics*[width=0.45 \columnwidth]{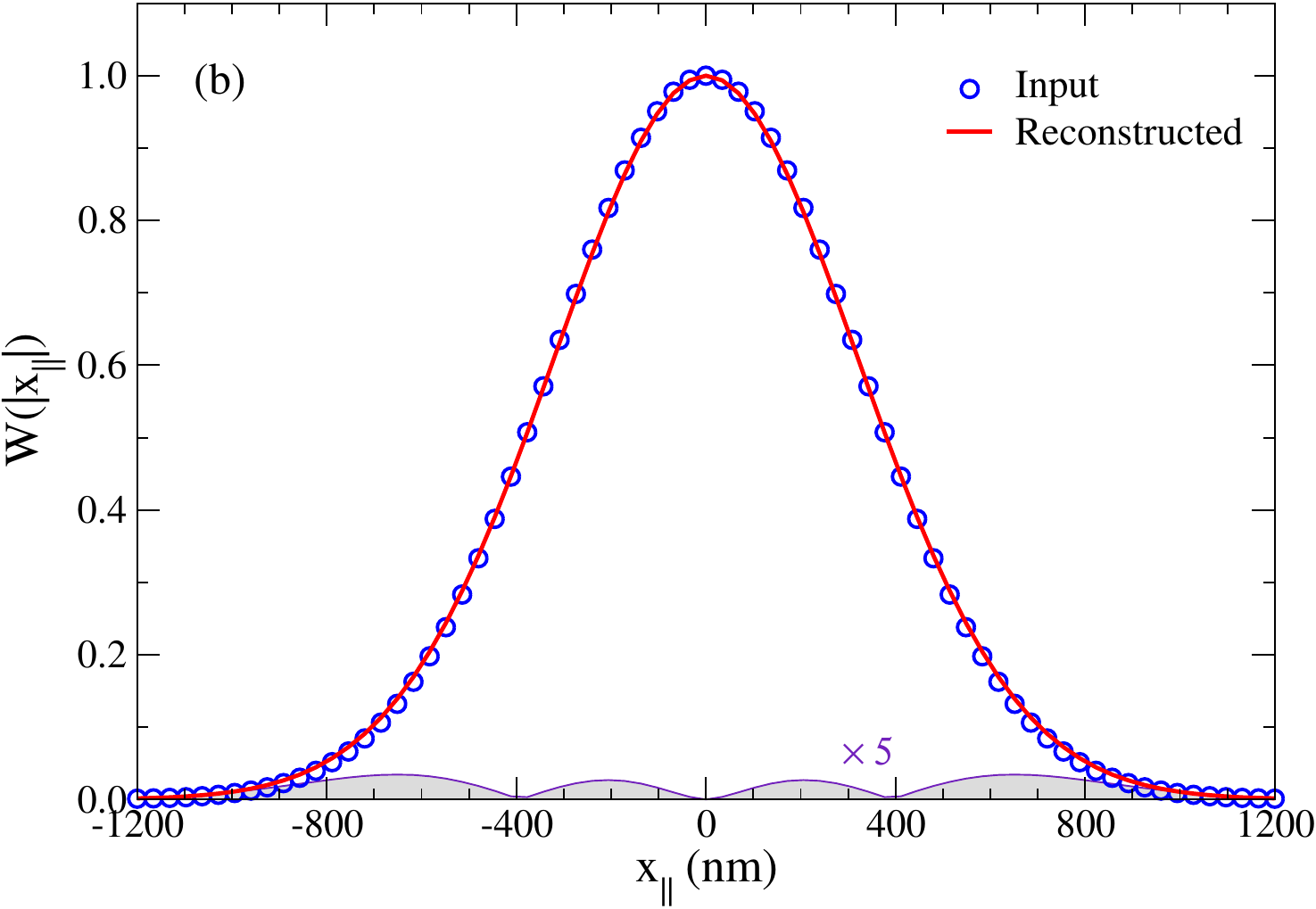}
  \caption{(Scattering system no.~1) Reconstruction of the rms-roughness $\delta_\star$, the transverse correlation length $a_\star$, and the exponent $\gamma_\star$ from in-plane p-to-p scattering data obtained for a Gaussian correlated silver surface. The parameters defining the scattering system, including the surface roughness, are identical to those given in the caption of Fig.~\protect\ref{Fig:Ex1:ppol}.
    %
    (a) The incoherent components of the in-plane, co-polarized (p-to-p) mean differential reflection coefficient $\left< \partial R_{pp}/\partial\Omega_s\right>_{\textrm{incoh}}$ as a function of the polar angle of scattering $\theta_s$ obtained from computer simulations for the polar angles of incidence $\theta_0=\ang{40}$~(open circles) [same data as in Fig.~\protect\ref{Fig:Ex1:ppol}(a)], and from second-order phase perturbation theory with the use of the reconstructed surface roughness parameters (solid lines) for a two-dimensional randomly rough silver surface whose correlation function is assumed to have the form~\eqref{eq:30}.  The values of the reconstructed surface roughness parameters are $\delta_\star = \SI{23.11(0.30)}{\nano\metre}$, $a_\star = \SI{459.7(6.8)}{\nano\metre}$, and $\gamma_\star=\SI{1.95(0.06)}{}$. In obtaining these results, the minimization was started from the values $\delta_\star=\SI{8.00}{nm}$, $a_\star=\SI{150.00}{nm}$ and  $\gamma_\star=1.00$,
    (b) The input~(open circles) and reconstructed~(solid line) surface-height autocorrelation function $W(|\pvec{x}|)$ for the random surface. The shaded gray region represents the absolute difference between the input and reconstructed surface-height autocorrelation functions. The notation ``$\times 5$'' means that this difference has been multiplied by a factor \num{5} for better visibility.
  }%
  \label{Fig:Ex1:ppol_gamma}
\end{figure}

For the same input data set, that is, for p polarization and $\theta_0=\ang{40}$ [Figs.~\ref{Fig:Ex1:ppol}(a) and \ref{Fig:Ex1:ppol_gamma}(a)], we also performed reconstruction based on the assumption that the  surface-height autocorrelation function $W(|\pvec{x}|)$ had the stretched exponential form~\eqref{eq:30} so that the set of variational parameters is ${\mathcal P}=\{\delta_\star, a_\star, \gamma_\star\}$. The motivation for using another trial function in the reconstruction of the same data set is that one often does not know in advance the form of the correlation function of a surface. Therefore, the use of the stretched exponential form~\eqref{eq:30} and the additional variable parameter that defines it, provides additional degree of freedom in the variation calculation over the assumption of using the Gaussian trial function~\eqref{eq:29_Gaussian}. When the minimization was started from the values $\delta_\star=\SI{8.00}{nm}$, $a_\star=\SI{150.00}{nm}$ and  $\gamma_\star=1.00$, where the initial values of the two former parameters are identical to the initial values assumed in producing the results in Fig.~\ref{Fig:Ex1:ppol} and the value of $\gamma_\star$ implies an initial exponential correlation function, the reconstruction approach  produced $\delta_\star=\SI{23.11(0.30)}{nm}$, $a_\star=\SI{459.7(6.8)}{nm}$ and  $\gamma_\star=\SI{1.95(0.06)}{}$ [Fig.~\ref{Fig:Ex1:ppol_gamma}]. These results are fully consistent with both the input values assumed when generating the scattering data and the results obtained when assuming the Gaussian trial function. Recall that the value of $\gamma$ in Eq.~\eqref{eq:30} that corresponds to the Gaussian correlation function~\eqref{eq:28_Gaussian} is  $\gamma=2.00$. It is interesting to observe that when basing the reconstruction on the stretched exponential~\eqref{eq:30}, the result for $\delta_\star$ is only marginally different from the result for the same parameter obtained when assuming the Gaussian form for $W(|\pvec{x}|)$. Furthermore, a similar comparison for $a_\star$ reveals that the use of the stretched exponential trial function produce a number value for this parameter that is closer to the input value, while at the same time, the width of the confidence interval increases only slightly. That we obtain consistent results when reconstructing the same scattering data using different forms of the trial function, testifies to the robustness of the used approach.  A summary of the reconstructed roughness parameters for p-polarized light can be found in the upper half of Table~\ref{tab:Ex01}.

%
\begin{table}
  \caption{\label{tab:Ex01}  (Scattering system no.~1) Summary of the scattering system parameters obtained during the different reconstruction scenarios based on in-plane and co-polarized scattering data corresponding to an \textit{Gaussianly} correlated silver surface; $\delta_\star$, $a_\star$, and $\gamma_\star$.
    The incident light, of wavelength $\lambda=\SI{457.90}{nm}$, was $\alpha$-polarized and the polar angle of incidence was $\theta_0=\ang{0}$ or $\theta_0=\ang{40}$. The dielectric function of silver was $\varepsilon(\omega)=-7.50+0.24\imu$~\cite{Book:Palik1997}. The surface roughness parameters assumed in generating the input data were:  $\delta=\lambda/20=\SI{22.90}{nm}$ and  $a=\lambda=\SI{457.90}{nm}$.    The last column indicates the relevant figure where the results of the reconstruction in question is presented. The symbol ``---'' indicates that the corresponding variable was not reconstructed and instead had the value assumed in the input data (numerical simulations). When $\gamma_\star$ was among the reconstructed parameters, the trial correlation function of the form~\eqref{eq:30} was assumed; in all other cases the Gaussian form~\eqref{eq:29_Gaussian} was assumed for this function. Note that a Gaussian correlation function~\eqref{eq:28_Gaussian} corresponds to the exponent $\gamma_\star=\num{2}$ for the stretched exponential in Eq.~\eqref{eq:30}.
    The initial values for  $\{ \delta_\star, a_\star, \gamma_\star\}$ used in performing the reconstructions were $\{ \SI{8}{\nano\meter}, \SI{150}{\nano\meter}, 1\}$. 
 }
  \begin{ruledtabular}
    \begin{tabular}{lrcccc}
      $\alpha$ & \quad$\theta_0$(deg)  & $\delta_\star \; \si{(nm)}$ & $a_\star \; \si{(nm)}$  & $\gamma_\star$ & Comments \quad \\
      \hline 
      %
      p  &   0  &   \SI{22.72(0.20)}{}     & \SI{458.6(3.6)}{}        & ---                  &  Fig.~\protect\ref{Fig:Ex1:ppol}   \\
      p  &  40  &   \SI{22.96(0.23)}{}     & \SI{463.5(4.7)}{}        & ---                  &  Fig.~\protect\ref{Fig:Ex1:ppol}   \\
      p  &  40  &   \SI{23.11(0.30)}{}     & \SI{459.7(6.8)}{}        & \SI{1.95(0.06)}{}    &  Fig.~\protect\ref{Fig:Ex1:ppol_gamma}  \\      
      s  &   0  &   \SI{22.99(0.25)}{}     & \SI{454.3(4.3)}{}        & ---                  &  Fig.~\protect\ref{Fig:Ex1:spol} \\
      s  &  40  &   \SI{22.81(0.23)}{}     & \SI{457.1(4.8)}{}        & ---                  &  Fig.~\protect\ref{Fig:Ex1:spol} \\
      s  &  40  &   \SI{23.53(0.44)}{}     & \SI{439.6(9.9)}{}        & \SI{1.84(0.08)}{}    &  Fig.~\protect\ref{Fig:Ex1:spol_gamma}  
    \end{tabular}
  \end{ruledtabular}
\end{table}

\begin{figure}[tbp] 
  \centering
  \includegraphics*[width=0.452\columnwidth]{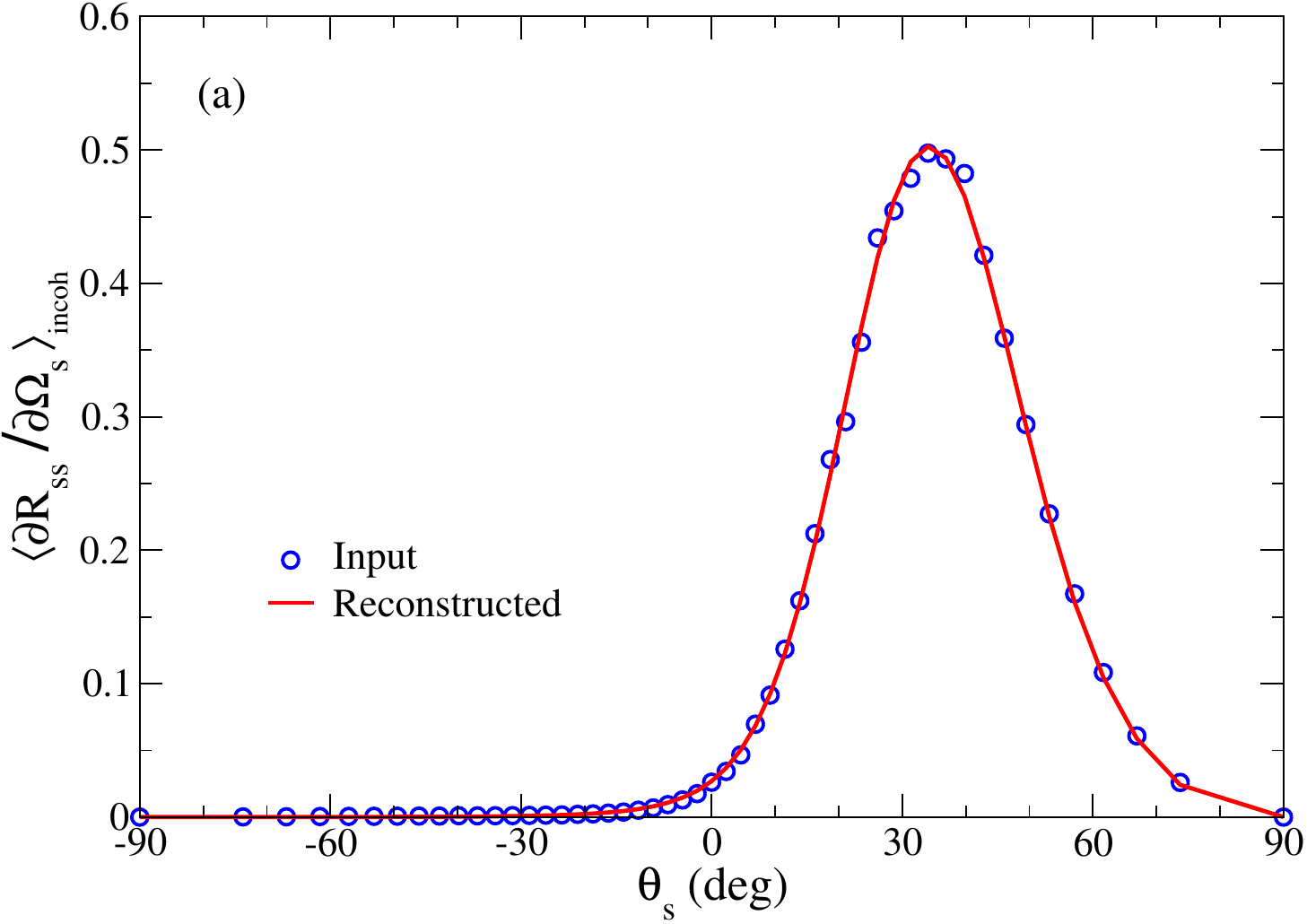}
  \qquad
  \includegraphics*[width=0.45 \columnwidth]{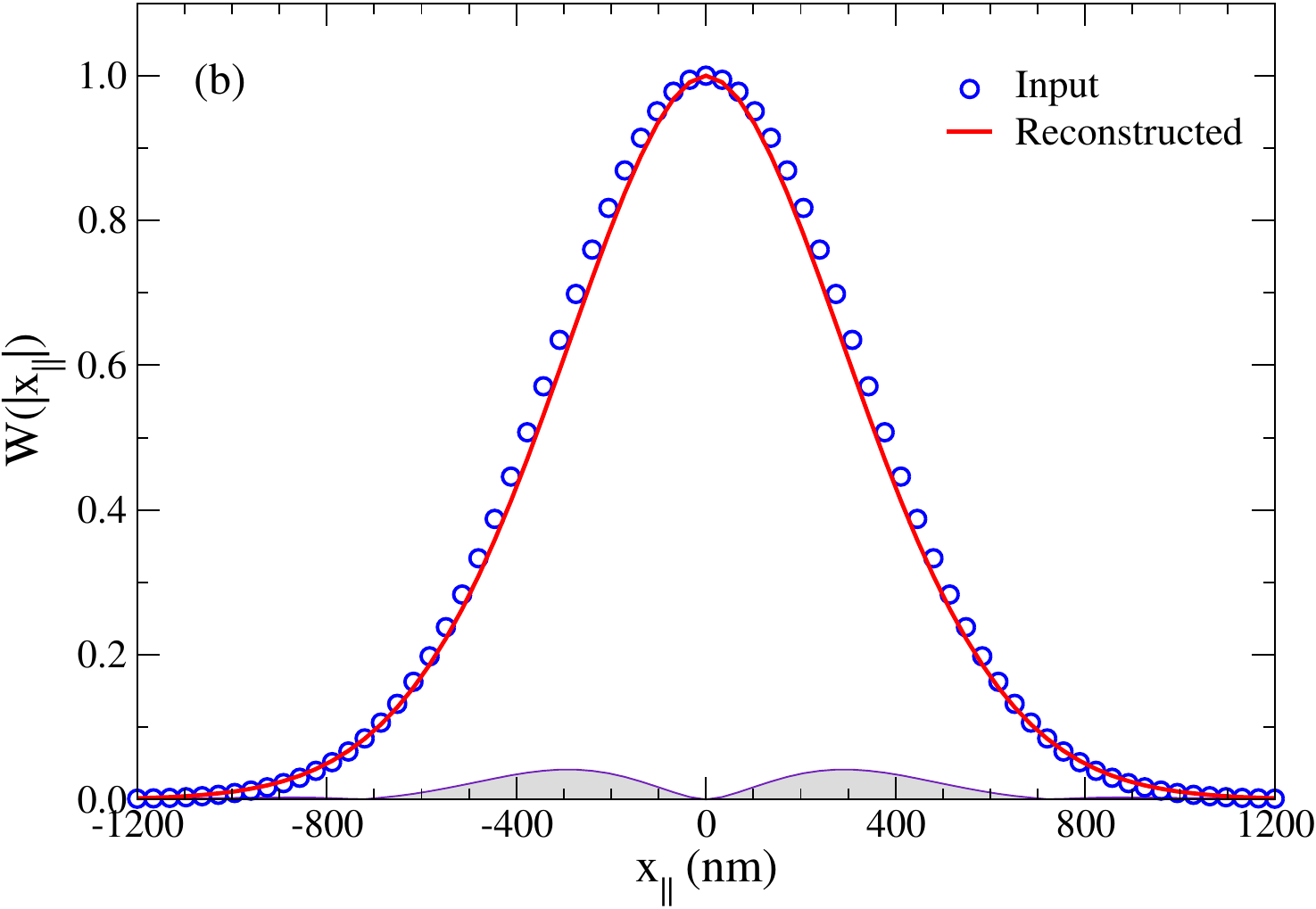}
  \caption{(Scattering system no.~1) Reconstruction of the rms-roughness $\delta_\star$, the transverse correlation length $a_\star$, and the exponent $\gamma_\star$ from in-plane s-to-s scattering data. The same as Fig.~\protect\ref{Fig:Ex1:ppol_gamma} but for s-polarized light. The values of the reconstructed surface roughness parameters are $\delta_\star = \SI{23.53(0.44)}{\nano\metre}$, $a_\star = \SI{439.6(9.9)}{\nano\metre}$, and $\gamma_\star=\num{1.84(0.08)}$. [Input values: $\delta =\SI{22.90}{\nano\meter}$, $a =\SI{457.90}{\nano\meter}$ and $\gamma=2$].
  }%
  \label{Fig:Ex1:spol_gamma}
\end{figure}

%
\medskip
We now turn to the reconstruction based on scattering data obtained when s-polarized light, instead of p-polarized light, is scattered from this rough silver surface. The input data sets for s polarization corresponding to $\theta_0=\ang{0}$ and $\theta_0=\ang{40}$ are presented in Fig.~\ref{Fig:Ex1:spol}(a) as open circles and squares, respectively. One observes that these scattering data are not very different from the corresponding scattering data produced when the incident light is p-polarized [see Fig.~\ref{Fig:Ex1:ppol}(a)]; this  is due to the relatively long correlation length that was assumed in generating them. Performing reconstruction on the basis of these s-polarized scattering data, done in a completely equivalent manner to how it was performed for the corresponding p-polarized data sets, produced the results presented in  Figs.~\ref{Fig:Ex1:spol} and \ref{Fig:Ex1:spol_gamma}. In this way we obtained the reconstructed roughness parameters listed for s-polarization [$\alpha=s$] in the lower half of Table~\ref{tab:Ex01}. By comparing the values presented in this table, or by comparing the results presented in Figs.~\ref{Fig:Ex1:ppol}--\ref{Fig:Ex1:spol_gamma}, will uncover that the results obtained by basing the reconstruction of the surface roughness parameters on p- or s-polarized scattering data produce rather similar results; at least this is the case for the angles of incidence and roughness parameters that were assumed here. All reconstructions that we have presented for this scattering system are internally consistent and do agree with the assumed input parameters except for the reconstruction using s polarization and the stretched exponential form for $W(|\pvec{x}|)$. In this latter case, the values are still reasonable, but the result for $\delta_\star$ is slightly too large and $a_\star$ is too small (and with the largest error bars of the cases we have studied so far).

\medskip
It is remarked that we have also performed reconstruction based on unpolarized incident light, joint inversion using input data for several polar angles of incidence, and joint inversion  based on simultaneously use of input data corresponding to p- and s-polarized incident light (see Ref.~\cite{Simonsen2014-05}). The values obtained in this way for the reconstructed parameters did only improve marginally relative to the results reported in Table~\ref{tab:Ex01}; hence, we do not give additional details here. Furthermore, we have also performed reconstruction of simulated scattering data obtained for some non-Gaussian correlation functions with satisfactory results~(results not shown).

%

%
%
\subsubsection{Scattering system no.~2}

\begin{figure}[tbp] 
  \centering
  \includegraphics*[width=0.456\columnwidth]{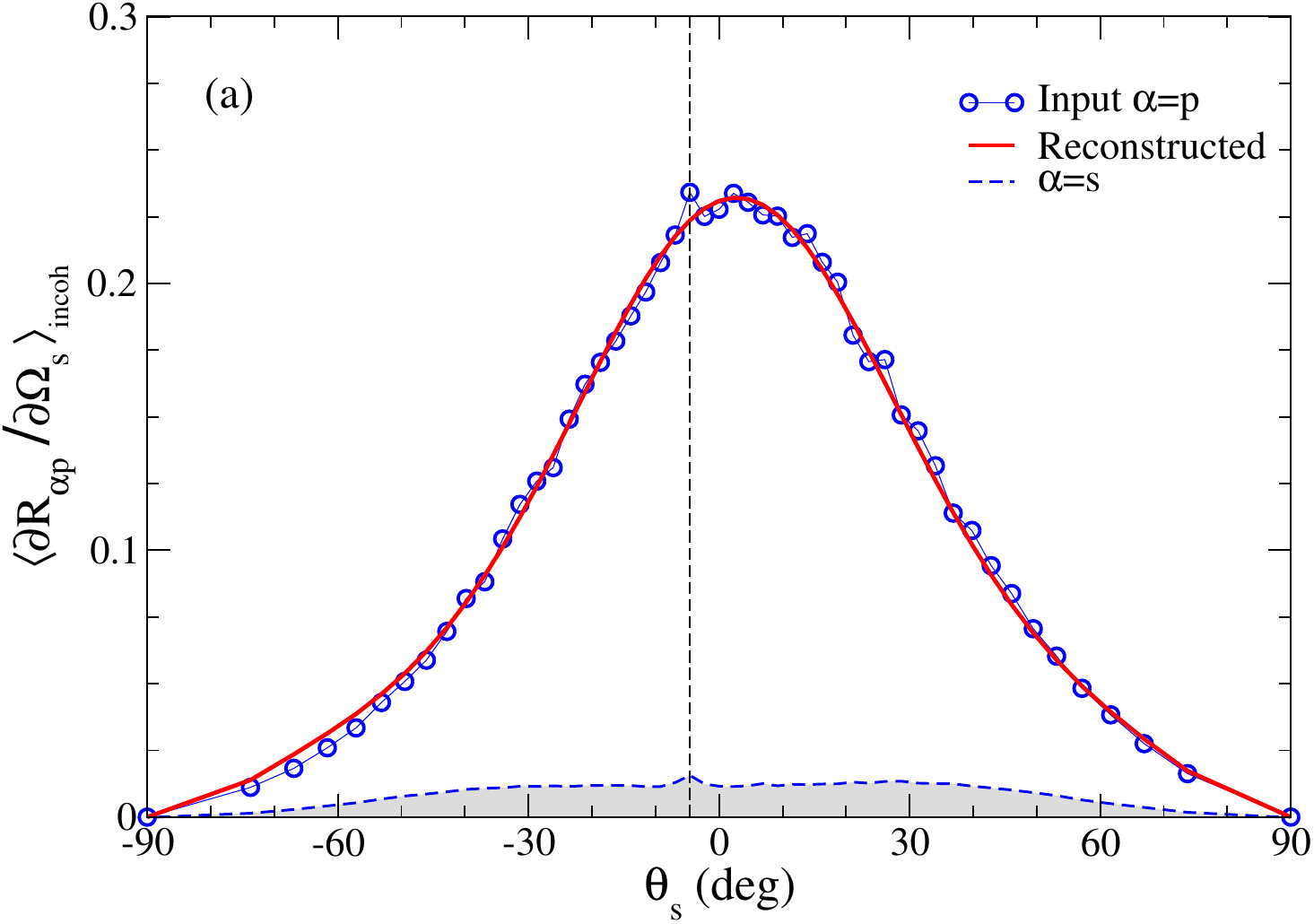}
  \qquad
  \includegraphics*[width=0.45 \columnwidth]{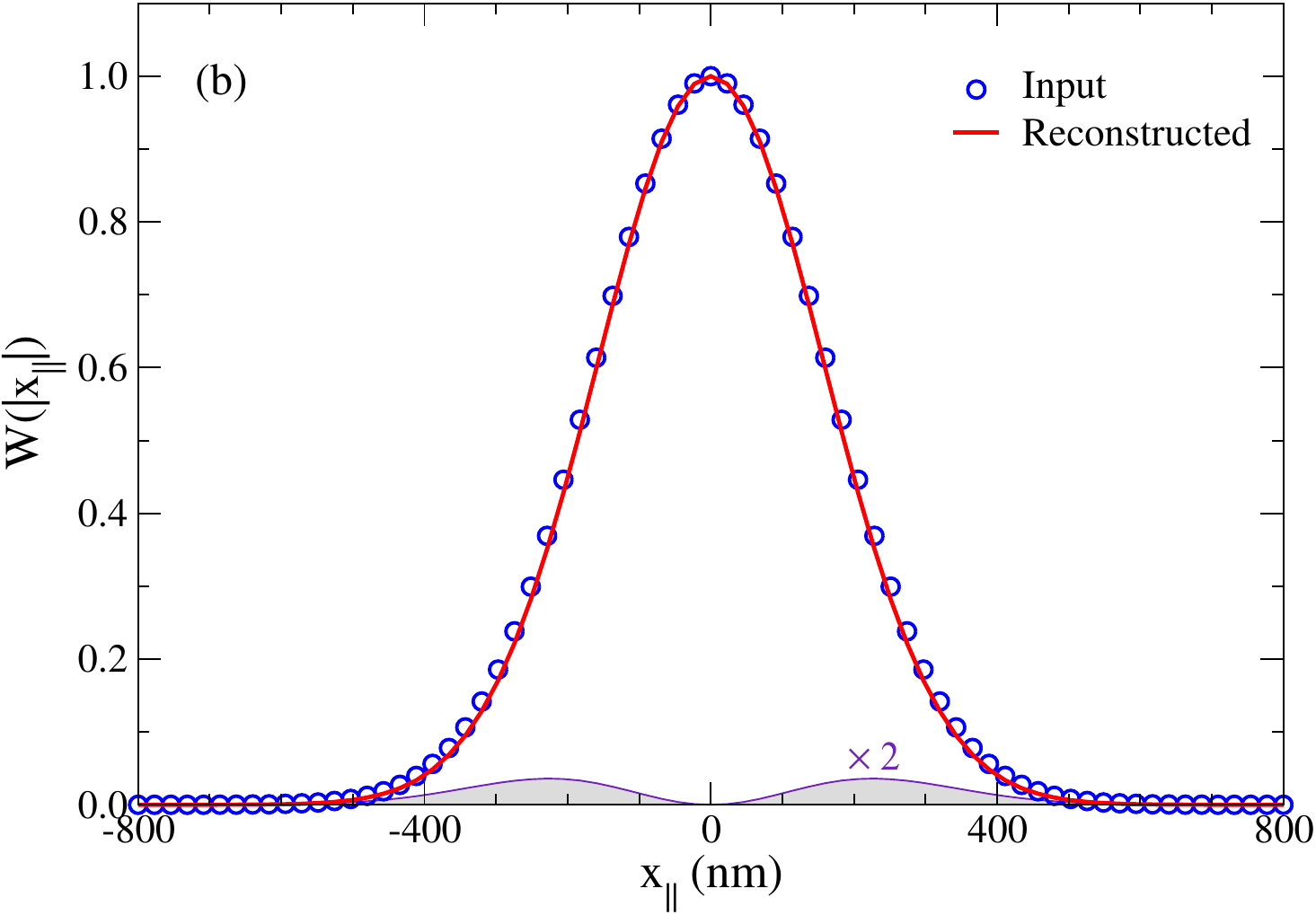}
  \caption{(Scattering system no.~2) Reconstruction of the rms-roughness $\delta_\star$ and the transverse correlation length $a_\star$ from in-plane p-to-p scattering data obtained for a Gaussian correlated silver surface. The wavelength (in vacuum) of the incident light is $\lambda = \SI{457.90}{\nano\meter}$ for which the dielectric function of silver is  $\varepsilon(\omega) = -7.50+0.24\imu$. The surface roughness parameters assumed in the computer simulations have the values $\delta =\lambda/20=\SI{22.90}{\nano\meter}$ and $a = \lambda/2 =\SI{228.95}{\nano\meter}$.
    (a) The incoherent components of the in-plane, co-polarized (p-to-p) mean differential reflection coefficient $\left< \partial R_{pp}/\partial\Omega_s\right>_{\textrm{incoh}}$
    obtained from computer simulations~(open circles) and from second-order phase perturbation theory with the use of the reconstructed surface roughness parameters~(solid line), for the polar angles of incidence $\theta_0=\ang{4.56}$ and as functions of the polar angle of scattering $\theta_s$  for a two-dimensional randomly rough silver surface whose correlation function is defined by Eq.~\protect\eqref{eq:29_Gaussian}. The values of the reconstructed surface roughness parameters are $\delta_\star = \SI{25.69(0.55)}{\nano\metre}$ and $a_\star = \SI{223.4(3.0)}{\nano\metre}$. Computer simulation results for the in-plane angular dependence of $\left< \partial R_{sp}/\partial\Omega_s\right>_{\textrm{incoh}}$ (p-to-s cross-polarization) are shown as a dashed blue line (and gray shaded region); this contribution is a result of multiple scattering. The vertical dashed line represents the backscattering direction $\theta_s=-\theta_0$. The mean DRC results produced by computer simulations were obtained on the basis of \num{2500} surface realizations. 
    (b) The input~(open circles) and reconstructed~(solid line) surface-height autocorrelation function $W(|\pvec{x}|)$ for the random surface. The shaded gray region represents the absolute difference between the input and reconstructed surface-height autocorrelation functions. The notation ``$\times 2$'' means that this difference has been multiplied by a factor \num{2} for better visibility.
  }
  \label{Fig:Ex2:ppol}
\end{figure}

\begin{figure}[tbp] 
  \centering
  \includegraphics*[width=0.452\columnwidth]{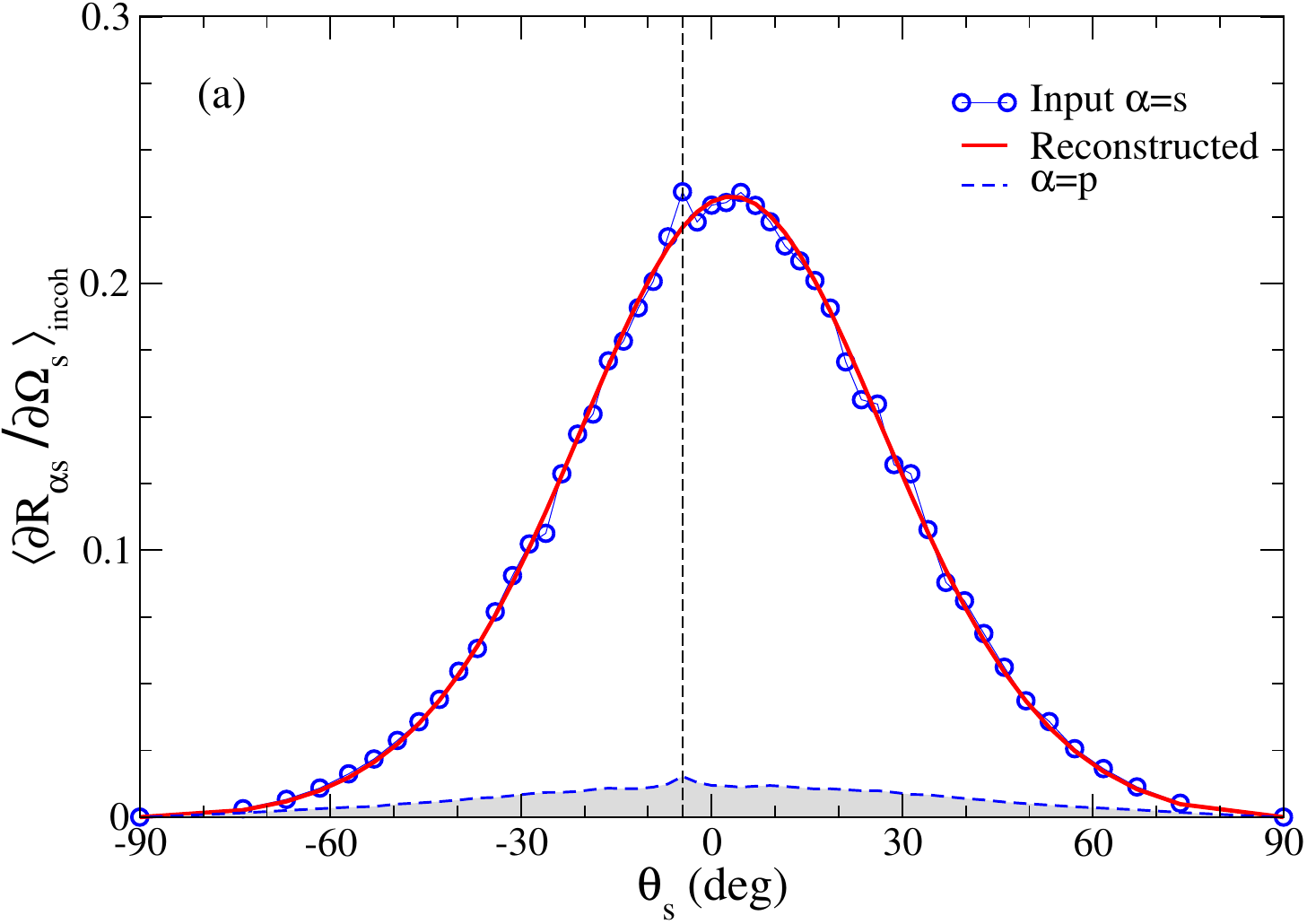}
  \qquad
  \includegraphics*[width=0.45 \columnwidth]{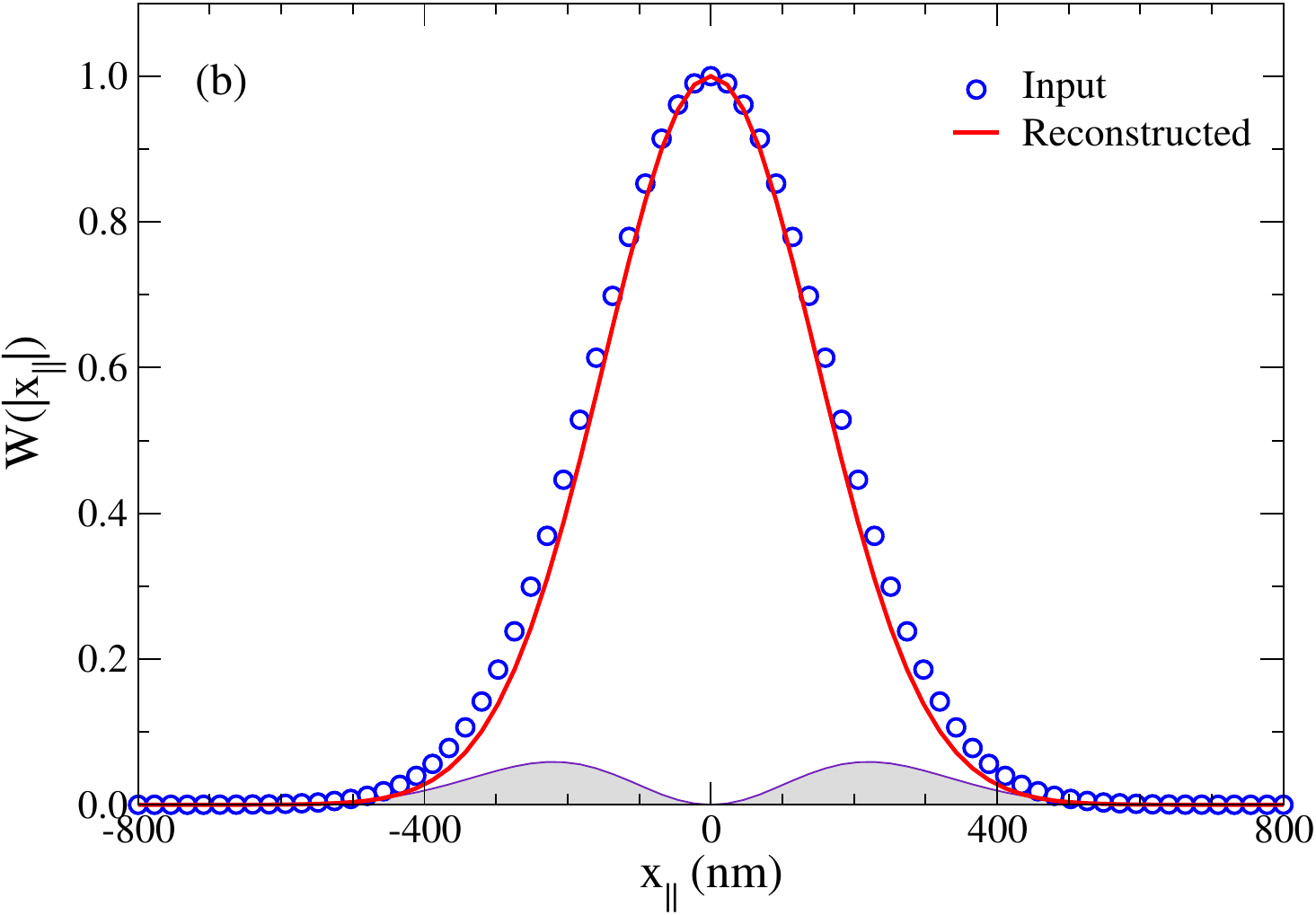}
  \caption{(Scattering system no.~2)
    The same as Fig.~\protect\ref{Fig:Ex2:ppol} but for s-polarized incident light.  
    The values of the reconstructed surface roughness parameters are $\delta_\star = \SI{29.15(1.05)}{\nano\metre}$ and $a_\star = \SI{211.3(3.6)}{\nano\metre}$.
    [Input values: $\delta=\lambda/20=\SI{22.90}{nm}$ and $a = \lambda/2 = \SI{228.95}{nm}$].
  }%
  \label{Fig:Ex2:spol}
\end{figure}

The second scattering system that we  consider is identical to the first except that the correlation length of the rough surface now is half of what it was for the first scattering system; hence, the rms-roughness and correlation length for this scattering system are assumed to be $\delta=\lambda/20=\SI{22.90}{nm}$ and $a = \lambda/2=\SI{228.95}{nm}$, respectively. For these roughness parameters, a computer simulation approach~\cite{Simonsen2011-05} was again used to generate scattering data. By averaging the results obtained on the basis of \num{2500} surface realizations, we calculated the contributions to the mean DRC from the light scattered incoherently by the rough surface. When the polar angle of incidence is $\theta_0=\ang{4.56}$, such results for co-polarized scattering are presented for p- and s-polarized light in Figs.~\ref{Fig:Ex2:ppol}(a) and \ref{Fig:Ex2:spol}(a), respectively. These scattering data constitute the input functions $\left<\partial R_{\alpha\alpha}(\theta_s)/ \partial\Omega_s \right>_{\textrm{incoh,input}}$ for the reconstruction we will perform based on these data. In Figs.~\ref{Fig:Ex2:ppol}(a) and \ref{Fig:Ex2:spol}(a) the vertical dashed lines indicate the backscattering direction and the simulation results show a weak enhanced backscattering peak. To confirm for this scattering system that multiple scattering processes contribute to the in-plane scattering, we have also in Figs.~\ref{Fig:Ex2:ppol}(a) and \ref{Fig:Ex2:spol}(a) included the cross-polarized components of the mean DRCs~(blue dashed lines and shaded regions); for both p- and s-polarized incident light we observe that the in-plane, cross-polarized components of the mean DRCs do not vanish. This, in addition to the presence of a weak  enhanced backscattering peak, is a clear indication that multiple scattering contributes to the in-plane scattered intensity. It is stressed that in the plane of incidence, but not outside of it, the cross-polarized component of the mean DRC is zero within single scattering~\cite{NavarreteAlcala2009}; finding such a component to be non-zero signals multiple scattering processes.

As our first example of reconstruction based on scattering data for this scattering geometry, we assume p-polarized incident light and a trial function $W(|\pvec{x}|)$ of a Gaussian form given by Eq.~\eqref{eq:29_Gaussian}. The set of variational parameters is therefore ${\mathcal P}=\{ \delta_\star, a_\star \}$. The minimization procedure of the cost function $\chi^2({\mathcal P})$, Eq.~\eqref{eq:27},  was started from the values $\delta_\star=\SI{8.00}{nm}$ and $a_\star=\SI{75.00}{nm}$. In this way, it was found that the minimum of $\chi^2({\mathcal P})$ occurred for the values $\delta_\star = \SI{25.69(0.55)}{\nano\metre}$ and $a_\star = \SI{223.4(3.0)}{\nano\metre}$, that should be compared with the values $\delta=\SI{22.90}{nm}$ and $a=\SI{228.95}{nm}$ used to generate the input data. The quality of the reconstructed surface roughness parameters is fair and practically useful; the absolute relative errors in the number values of the reconstructed surface roughness parameters $\delta$ and $a$ are \SI{12.2}{\percent} and \SI{2.4}{\percent}, respectively. However, the obtained results are not consistent with the input roughness parameters and clearly less accurate than what was obtained for the first scattering system where multiple scattering processes contributed less. We find that $\delta_\star$ is slightly too large while, at the same time, $a_\star$ is too small. The corresponding reconstructed correlation function is compared to the input correlation function in Fig.~\ref{Fig:Ex2:ppol}(b). 

By performing a similar reconstruction based on the corresponding input data for s-polarized light~[Fig.~\ref{Fig:Ex2:spol}(a)] we obtained the following values for the reconstructed parameters $\delta_\star = \SI{29.15(1.05)}{\nano\metre}$ and $a_\star = \SI{211.3(3.6)}{\nano\metre}$. Neither these results are consistent with the assumed input roughness parameters and the values are less accurate than what was found when basing the inversion on p-polarized scattering data; this is the same conclusion that we drew when reconstructing scattering data from our first scattering geometry.The corresponding reconstructed correlation function is compared to the input correlation function in Fig.~\ref{Fig:Ex2:spol}(b). The absolute relative errors in the number values of the reconstruction of $\delta$ and $a$ when basing it on s-polarized scattering data is \SI{27.3}{\percent} and \SI{7.7}{\percent}, respectively. Also when the incident light is s-polarized the rms-roughness is found to be reconstructed less accurately than the correlation length. Moreover, for the roughness and geometrical parameters we assume, we find that the reconstruction based on p-polarized scattering data overall produces a more accurate reconstruction than when basing it on s-polarized scattering data.

%
\subsubsection{Scattering system no.~3}

\begin{figure}[tbp] 
  \centering
  \includegraphics*[width=0.468\textwidth]{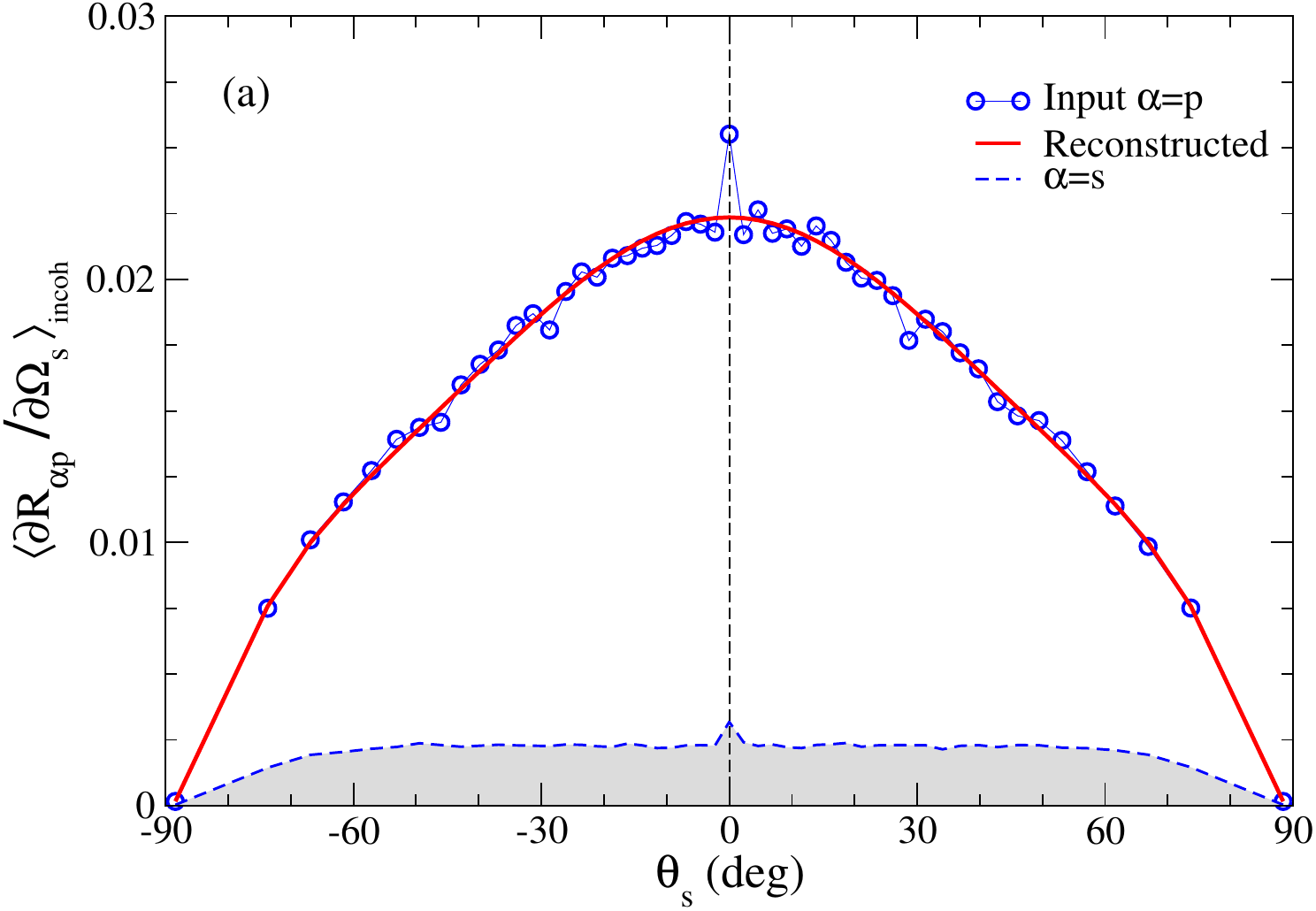}
  \qquad
  \includegraphics*[width=0.45 \textwidth]{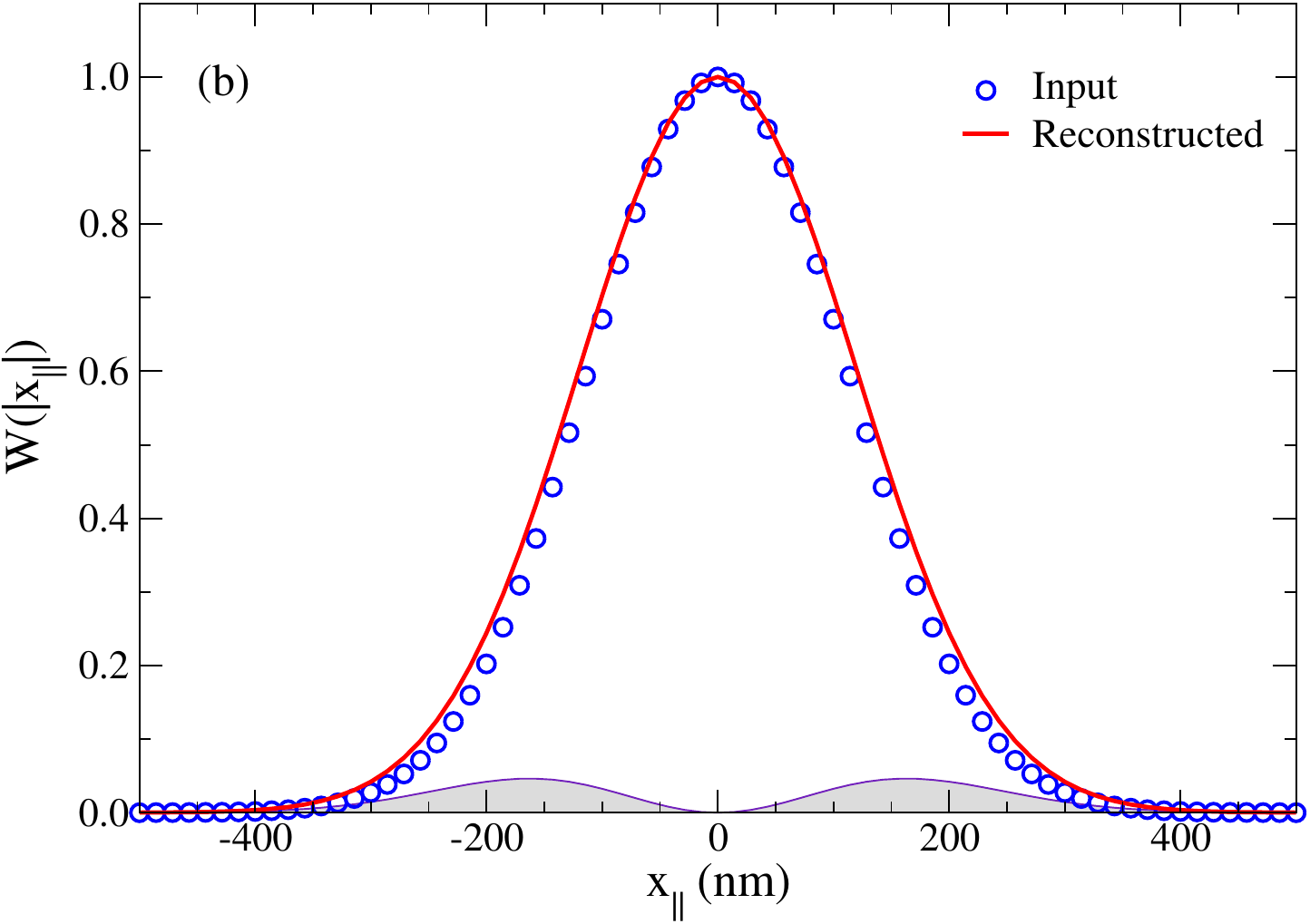}
  \caption{(Scattering system no.~3) Reconstruction of the rms-roughness $\delta_\star$ and the transverse correlation length $a_\star$ from in-plane p-to-p scattering data obtained for a Gaussian correlated silver surface. The wavelength (in vacuum) of the incident light is $\lambda = \SI{632.80}{\nano\meter}$ for which the dielectric function of silver is  $\varepsilon(\omega) = -18.28+0.48\imu$. The surface roughness parameters assumed in the computer simulations have the values $\delta =\lambda/40=\SI{15.82}{\nano\meter}$ and $a = \lambda/4 =\SI{158.20}{\nano\meter}$.
    (a) The incoherent components of the in-plane, co-polarized (p-to-p) mean differential reflection coefficient $\left< \partial R_{pp}/\partial\Omega_s\right>_{\textrm{incoh}}$
    obtained from computer simulations~(open circles) and from second-order phase perturbation theory with the use of the reconstructed surface roughness parameters~(solid line), for the polar angles of incidence $\theta_0=\ang{0}$ and as functions of the polar angle of scattering $\theta_s$  for a two-dimensional randomly rough silver surface whose correlation function is defined by Eq.~\protect\eqref{eq:29_Gaussian}. The values of the reconstructed surface roughness parameters are $\delta_\star = \SI{16.54(0.42)}{\nano\metre}$ and $a_\star = \SI{168.5(4.8)}{\nano\metre}$. Computer simulation results for the in-plane angular dependence of $\left< \partial R_{sp}/\partial\Omega_s\right>_{\textrm{incoh}}$ (p-to-s cross-polarization) are shown as a dashed blue line (and gray shaded region); this contribution is a result of multiple scattering. The vertical dashed line represents the backscattering direction $\theta_s=-\theta_0$. The mean DRC results produced by computer simulations were obtained on the basis of \num{2500} surface realizations. 
    (b) The input~(open circles) and reconstructed~(solid line) surface-height autocorrelation function $W(|\pvec{x}|)$ for the random surface. The shaded gray region represents the absolute difference between the input and reconstructed surface-height autocorrelation functions.}
  \label{Fig:Ex4:ppol}
\end{figure}

\begin{figure}[tbp] 
  \centering
  \includegraphics*[width=0.468\textwidth]{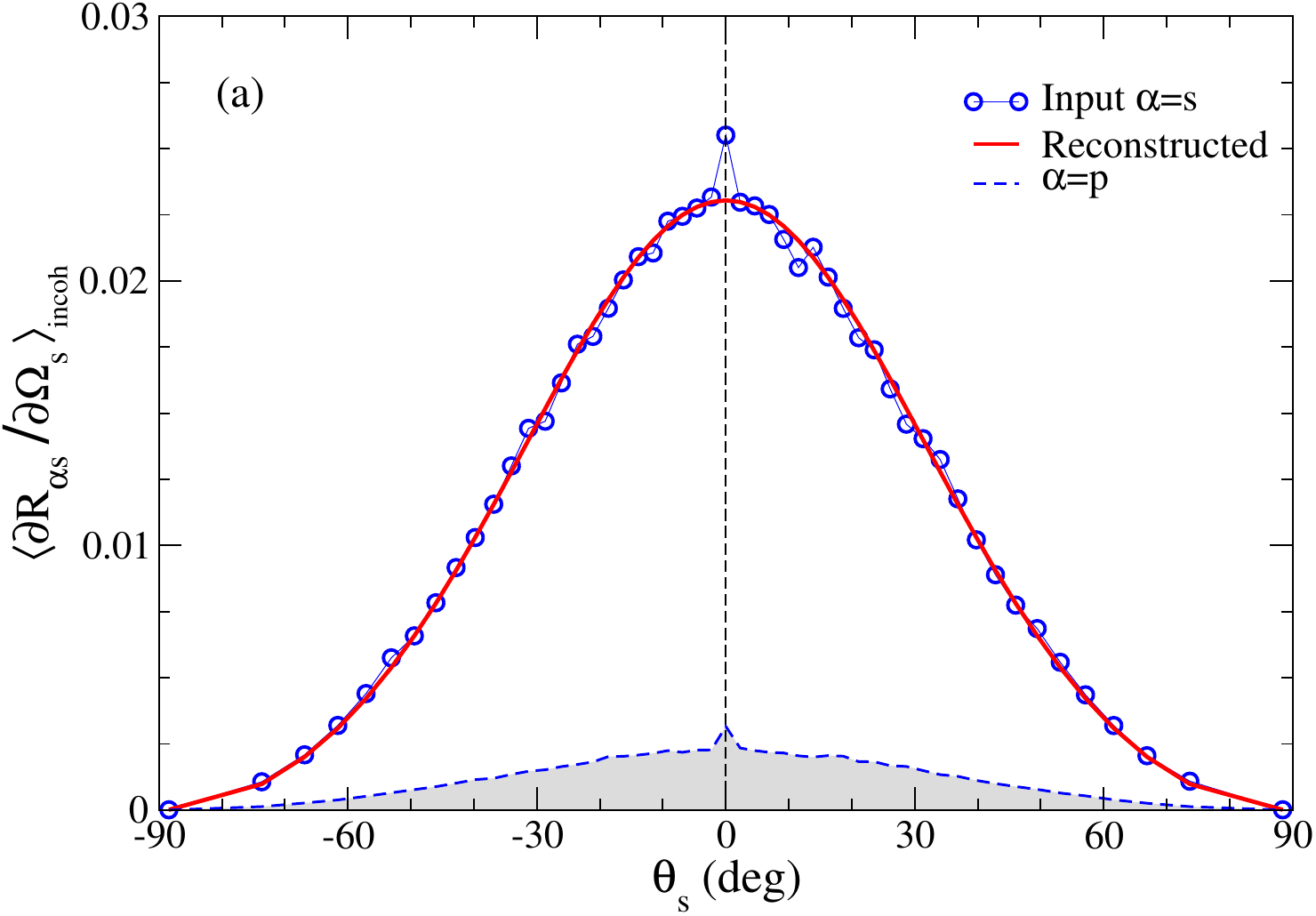}
  \qquad
  \includegraphics*[width=0.45 \textwidth]{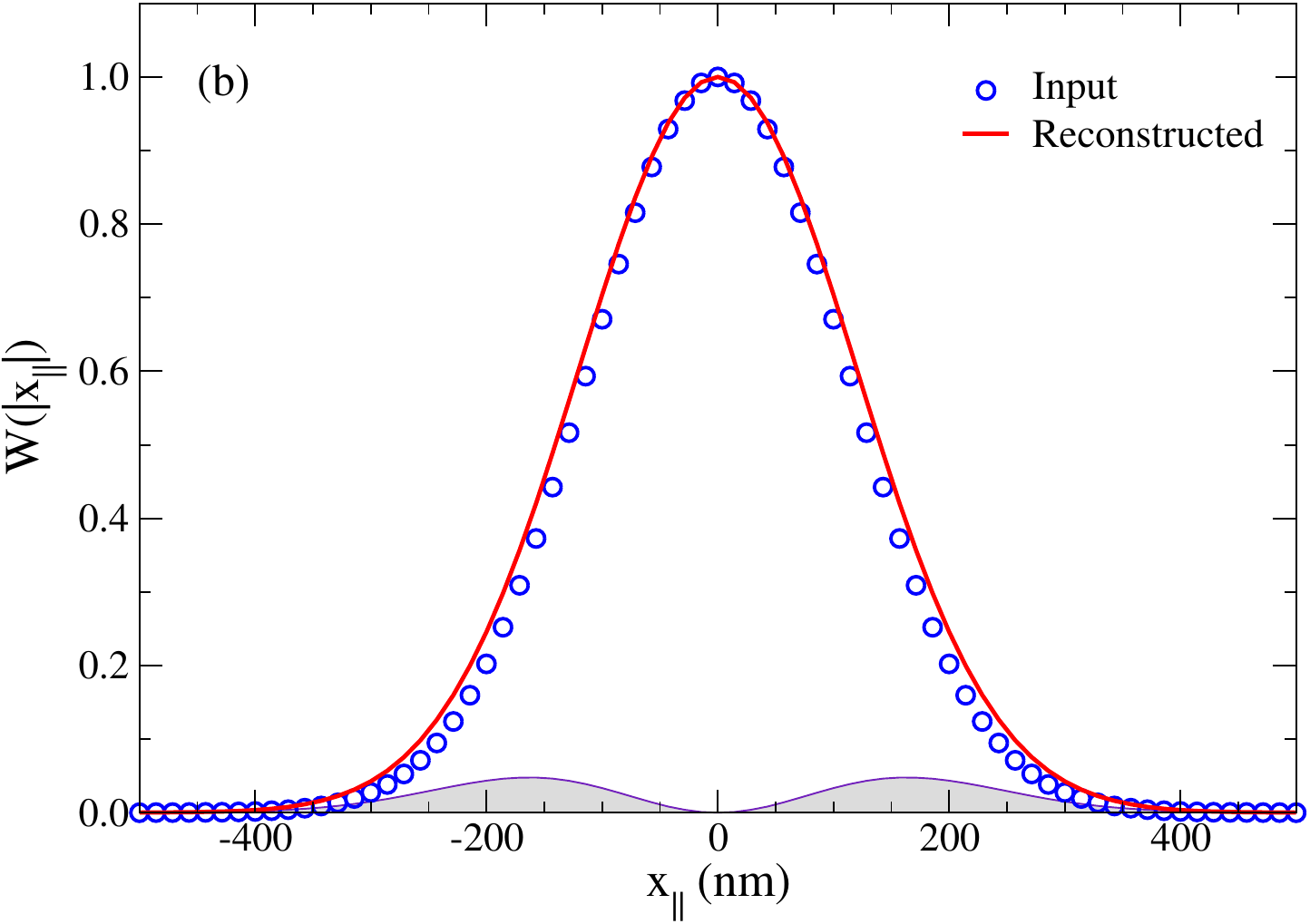}
  \caption{(Scattering system no.~3) The same as Fig.~\protect\ref{Fig:Ex4:ppol} but for s-polarized incident light.  
    The values of the reconstructed surface roughness parameters are $\delta_\star = \SI{16.78(0.67)}{\nano\metre}$ and $a_\star = \SI{168.8(7.2)}{\nano\metre}$.
    [Input values:  $\delta = \lambda/40 = \SI{15.82}{nm}$ and $a=\lambda/4=\SI{158.20}{nm}$].
  }%
  \label{Fig:Ex4:spol}
\end{figure}

The final scattering system that we will use for the purpose of generating scattering data by computer simulation is again a rough silver surface. However, in this case the wavelength of the incident light is $\lambda=\SI{632.80}{nm}$ (in vacuum) and the dielectric function of silver at this wavelength we take to be  $\varepsilon(\omega)= -18.28+0.48\imu$~\cite{Johnson1972}. The surface roughness parameters that we assume are $\delta = \lambda/40=\SI{15.82}{nm}$ and $a=\lambda/4=\SI{158.20}{nm}$; they are identical to those assumed in a recent study of the reconstruction based on scattering data from rough dielectric surfaces~\cite{Simonsen2014-05}. For these parameters, computer simulations, averaged over \num{2500} surface realizations, were performed~\cite{Simonsen2011-05} to generate scattering data for p- and s-polarized incident light for a polar angle of incidence $\theta_0=\ang{0}$. Figures~\ref{Fig:Ex4:ppol}(a) and \ref{Fig:Ex4:spol}(a) present the results of such simulations for the in-plane angular distributions of co-polarized (blue open symbols) and cross-polarized (blue dashed lines and gray shaded regions) scattering; moreover, for better clarity thin sold lines connects the data points of each of the data sets. For both polarizations and for both co- and cross-polarized scattering,  $\left< \partial R_{\alpha\beta}/\partial \Omega_s\right>_{\textrm{incoh}}$ show enhanced backscattering peaks at the polar angle of scattering $\theta_s=\ang{0}$ (dashed vertical lines in the figures). These enhanced backscattering peaks, which are the most pronounced for in-plane, co-polarized scattering, appear due to the multiple scattering of SPPs propagating along the weakly rough silver surface~\cite{Simonsen2004-3}. From the results presented in Figs.~\ref{Fig:Ex4:ppol}(a) and \ref{Fig:Ex4:spol}(a) it is readily observed that cross-polarized scattering is significant for the whole angular interval $-\ang{90}<\theta_s<\ang{90}$. Furthermore, in this case, there is a readily detectable difference between the in-plane angular dependence of the co-polarized scattering data corresponding to p- and s-polarized incident light; this difference is mainly due to the smaller value of the lateral correlation length that characterize the surface roughness [$a=\lambda/4$]. One also observes that the amplitude of the enhanced backscattering peak over its background is higher in p-to-p scattering than what it is in s-to-s scattering.  

In order to reconstruct the parameters of the surface roughness for this scattering system, we started by considering the data corresponding to p-polarized incident light. Since the reconstruction is based on co-polarized scattering, the input data for this reconstruction is the data set marked by blue open symbols in Fig.~\ref{Fig:Ex4:ppol}(a) and, thus,  it constitutes the input function $\left<\partial R_{pp}(\theta_s)/ \partial\Omega_s \right>_{\textrm{incoh,input}}$ that we will use. The trial function was again assumed to have the Gaussian form~\eqref{eq:29_Gaussian} and hence the variational parameter set is ${\mathcal P}=\{\delta_\star, a_\star\}$. With initial values $\delta_\star = \SI{2.00}{\nano\metre}$ and $a_\star = \SI{75.00}{\nano\metre}$, the minimization procedure of the cost function $\chi^2({\mathcal P})$ produced the surface roughness parameters $\delta_\star = \SI{16.54(0.42)}{\nano\metre}$ and $a_\star = \SI{168.5(4.8)}{\nano\metre}$. These values are rather close to the surface roughness parameters used to generate the input scattering data even if the reconstructed parameters technically are not consistent with the input values. For the absolute relative difference between the number value of the reconstructed parameters and the input parameters we find  \SI{4.5}{\percent} for the rms-roughness and \SI{6.5}{\percent} for the lateral correlation length. Given the amount of contribution from multiple scattering that is present in the input scattering data, we find this result quite encouraging. The resulting co-polarized mean DRC and $W(|\pvec{x}|)$ obtained on the basis of the reconstructed surface roughness parameters are presented as solid lines in the two panels of Fig.~\ref{Fig:Ex4:ppol}.

 %
 %

%
\smallskip
When a completely equivalent reconstruction to what was just done for p polarization is performed based on the corresponding s-to-s scattering data shown in Fig.~\ref{Fig:Ex4:spol}(a), one obtains the reconstructed values $\delta_\star = \SI{16.78(0.67)}{\nano\metre}$ and $a_\star = \SI{168.8(7.2)}{\nano\metre}$ for the surface roughness parameters. These results are rather similar to those obtained when reconstructing p-to-p scattering data; the main difference being that the error bars on the reconstructed parameters are larger in the case of s polarization. Also for s polarization  the reconstructed roughness parameters deviate slightly from the input parameters. The solid lines in the two panels of Fig.~\ref{Fig:Ex4:spol} represent the ss component of the mean DRC, calculated on the basis of Eq.~\eqref{eq:MDRC_PPT_total}, and the correlation function $W(|\pvec{x}|)$, calculated on the basis of Eq.~\eqref{eq:29_Gaussian}, when the values of the surface roughness parameters reconstructed on the basis of s-to-s scattering data are used. A fairly good agreement are found between these functions and those corresponding to the input functions.

It should be apparent from the results in Figs.~\ref{Fig:Ex4:ppol}(a) and \ref{Fig:Ex4:spol}(a) that the second-order phase perturbation theory that we base the inversion on is not capable to predict the enhanced backscattering peaks that the input data in these figures show. Except for this multiple scattering feature of the input data, the phase perturbation theoretical results seems to agree well with the input data over the whole range of scattering angles $\theta_s$; in particular, the difference in width of the angular dependencies of the co-polarized scattered intensities  for p and s polarized light is well captured by it.

 %
 %
 %

\smallskip 
We end this subsection with the remark that there is one important message we hope the reader will take away from the different reconstruction examples presented here, and in particular, in Figs.~\ref{Fig:Ex4:ppol} and \ref{Fig:Ex4:spol}. Even if the input scattering data receive significant contributions from multiple scattering, the proposed reconstruction procedure for the surface roughness parameters based on second-order phase perturbation theory is in principle capable of obtaining useful, and often rather accurate, results for the value of these parameters; at least, this was the case for the scattering systems and the roughness and geometrical parameters that we have assumed.

%
%

\subsection{Inversion of experimental scattering data}
So far in this and our recent related works on the reconstruction of surface roughness parameters based on scattering data for dielectric surfaces~\cite{Simonsen2014-05,Simonsen2016-06}, we have exclusively based the inversion on scattering data obtained by rigorous computer simulations. In generating such data, the statistical properties of the used randomly rough surfaces were well controlled both when it comes to the form of the correlation function and the height distribution and the parameters that define these functions. Now we address the more practically relevant case where the inversion is performed on the basis of \emph{experimental} scattering data obtained when polarized light is scattered from a rough metal surface. 

In a series of experiments, Navarrete Alcal\'{a}~\etal\cite{NavarreteAlcala2009} performed measurements of the in-plane and co-polarized angular dependence of the mean~DRC for incident light of wavelength $\lambda=\SI{10.6}{\micro\meter}$ (in vacuum) that was scattered from two-dimensional randomly rough gold surfaces (also see Ref.~\onlinecite{Navarrete2002}). At this wavelength, the value of the dielectric function of gold is $\varepsilon(\omega)=\num{-2489.77}+\num{2817.36}\imu$~\cite{Book:Palik1997}. The rough surfaces used in these experiments were fabricated by the method described in Refs.~\onlinecite{Gray1978,ODonnell1987} and characterized by a Gaussian distribution of heights and an isotropic surface-height autocorrelation function of an approximately Gaussian form. The difference between the samples produced were therefore, in principle, found in the parameters defining these functions, that is, the values of the roughness parameters $\delta$ and $a$ of each sample. After production, each of the rough surfaces was given a morphological characterization and the experimental values for the roughness parameters $\delta$ and $a$ were obtained (see Refs.~\onlinecite{NavarreteAlcala2009,Navarrete2002}).

%
%

\begin{figure}[tbp] 
  \centering
  \includegraphics*[width=0.462\textwidth]{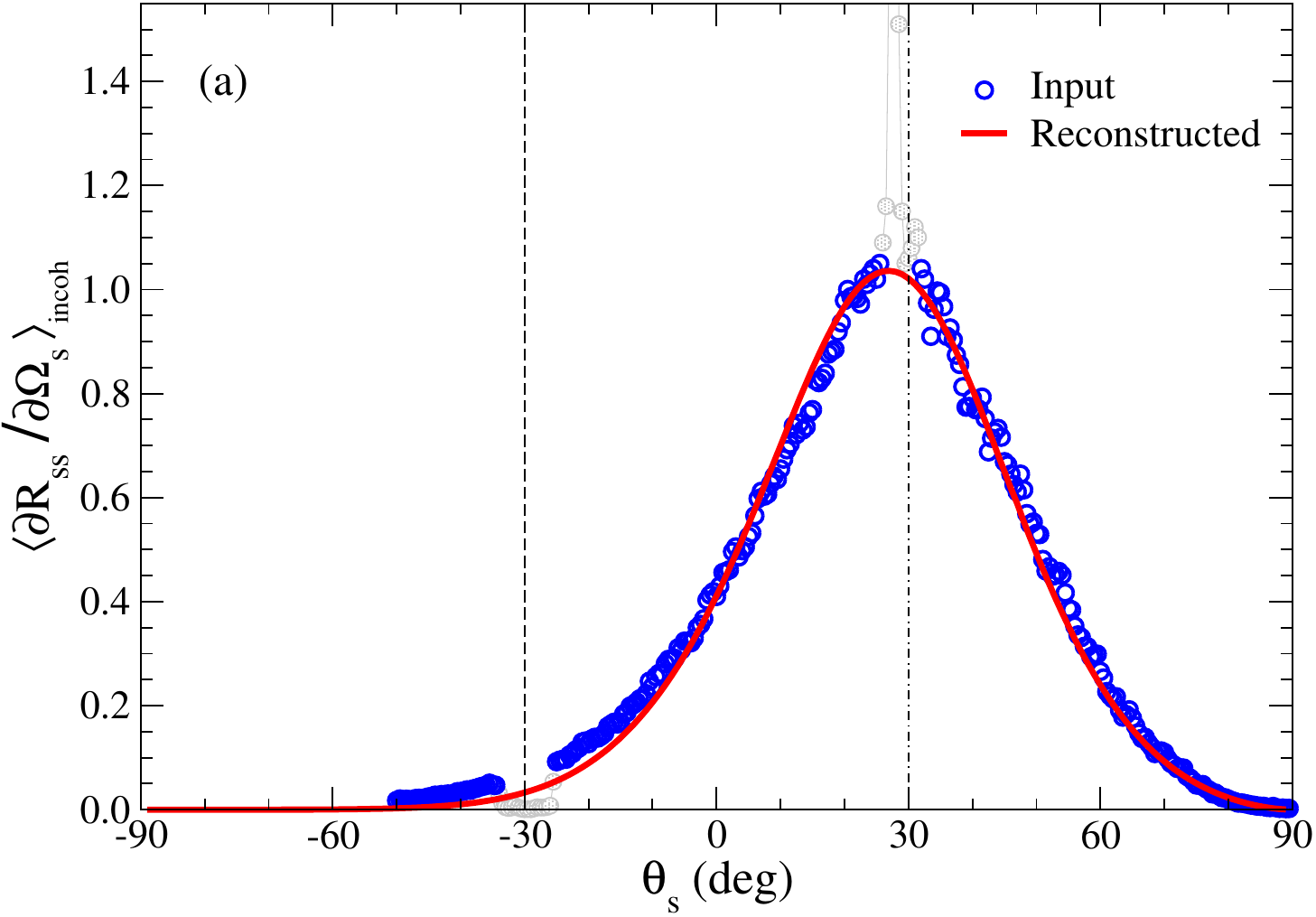}
  \qquad
  \includegraphics*[width=0.45 \textwidth]{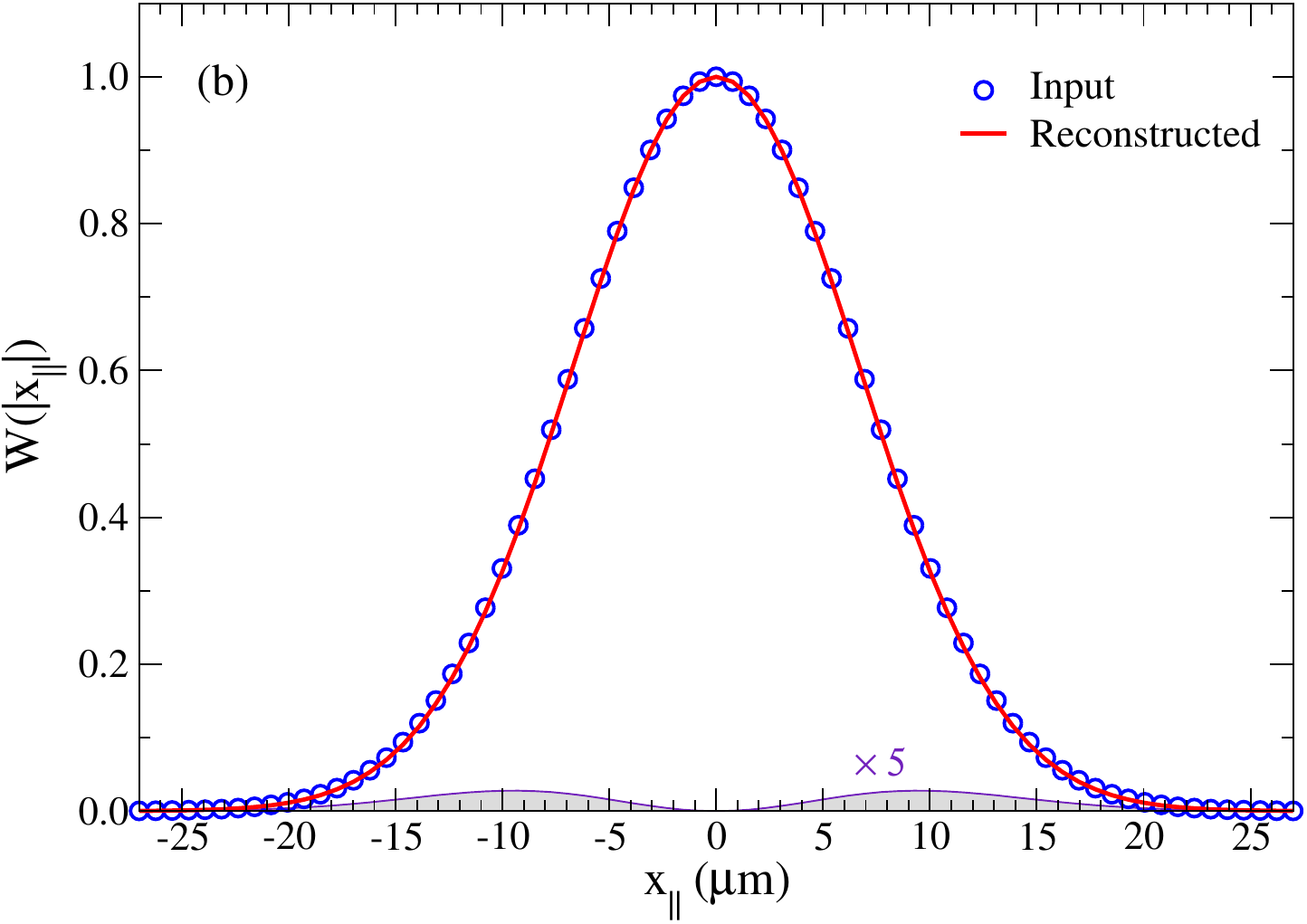}
  \caption{(Sample 7047) Reconstruction of the rms-roughness $\delta_\star$ and transverse correlation length $a_\star$ from measured scattering data in s-polarization obtained for a Gaussian correlated gold surface. The wavelength of the incident light (in vacuum) is $\lambda = \SI{10.6}{\micro\meter}$ for which the dielectric function of gold is assumed to be $\varepsilon(\omega)=\num{-2489.77}+\num{2817.36}\imu$~\cite{Book:Palik1997}. The values of the surface roughness parameters of the sample used in the experiment were determined to be $\delta=\SI{1.60 \pm 0.05}{\micro\meter}$ and $a=\SI{9.50 \pm 1.30}{\micro\meter}$ (or  $\delta\approx \lambda/7$ and $a\approx \lambda$)~\cite{NavarreteAlcala2009}.
    (a) The mean differential reflection coefficient $\left< \partial R_{ss}/\partial\Omega_s\right>_{\textrm{incoh}}$ as a function of the polar angle of scattering $\theta_s$ obtained experimentally for the polar angles of incidence $\theta_0=\ang{30}$~(symbols), and from second-order phase perturbation theory for $\theta_0=\ang{28}$ with the use of the reconstructed surface roughness parameters (solid lines), for a two-dimensional randomly rough gold surface whose correlation function is defined by Eq.~\protect\eqref{eq:29_Gaussian}. The experimental data points (symbols) are taken from Ref.~\onlinecite{NavarreteAlcala2009} for sample~7047. When performing the reconstruction, however,  only the experimental data points marked by blue open symbols were included. The values of the reconstructed surface roughness parameters are  $\delta_\star=\SI{1.62(0.04)}{\micro\meter}$ and $a_\star=\SI{9.46(0.11)}{\micro\meter}$ when the minimization procedure was started from the values $\delta_\star=\SI{0.50}{\micro\meter}$ and $a_\star=\SI{1.00}{\micro\meter}$. The dash-dotted and dashed vertical lines indicate the specular and anti-specular (backscattering)  directions respectively. 
    (b)~The input (open circles) and reconstructed (solid line) surface-height autocorrelation function $W(|\pvec{x}|)$ for the random surface; the former was obtained from Eq.~\eqref{eq:28_Gaussian} using the experimentally determined value for $a$.}
  \label{Fig:Sample7047:spol}
\end{figure}
%
%

\smallskip
The first sample we consider was named sample~7047 by the authors of Ref.~\onlinecite{NavarreteAlcala2009}. Their morphological characterization confirmed that surface roughness of the sample was approximately characterized by a Gaussian height distribution and a Gaussian correlation function. The experimental values for the rms-roughness and the lateral correlation length were determined from the measured surface morphology to be $\delta=\SI{1.60 \pm 0.05}{\micro\meter}$ and $a=\SI{9.50 \pm 1.30}{\micro\meter}$, respectively. In terms of the wavelength of the incident light, this means that $\delta\approx \lambda/7$ and $a\approx \lambda$. The results for the measurements of the in-plane dependence of the s-to-s component of the mean DRC for sample 7047 when the light is incident at a polar angle $\theta_0=\ang{30}$ are presented as symbols in Fig.~\ref{Fig:Sample7047:spol}(a); the dash-dotted and dashed lines represent the specular and backscattering (anti-specular) directions, respectively. For the purpose of the inversion, we are primarily interested in the contribution to the mean DRC from the light that has been scattered incoherently by the surface. Hence, only the experimental data points presented as blue open symbols in Fig.~\ref{Fig:Sample7047:spol}(a) have been used for the inversion; the data points shown as gray filled symbols around the specular and anti-specular directions in this figure have been excluded. The reason that we do not include the data points around the backscattering direction in the inversion is that here reliable measurements cannot be performed~\cite{NavarreteAlcala2009}.

The set of data points shown as blue open symbols in Fig.~\ref{Fig:Sample7047:spol}(a) constitute the input function $\left<\partial R_{ss}(\theta_s)/\partial\Omega_s \right>_{\textrm{incoh,input}}$ for our first reconstruction example based on experimental data. To perform the reconstruction of this data set, we assumed the trial function $W(|\pvec{x}|)$ of the Gaussian form~\eqref{eq:29_Gaussian}, and the set of variational parameters is therefore ${\mathcal P} = \{\delta_\star, a_\star \}$. The reconstruction procedure was started with the values $\delta_\star=\SI{0.50}{\micro\meter}$ and $a_\star=\SI{1.00}{\micro\meter}$ and the values of these parameters that minimize the cost function $\chi^2({\mathcal P})$, Eq.~\eqref{eq:27}, were found to be $\delta_\star=\SI{1.62(0.04)}{\micro\meter}$ and $a_\star=\SI{9.46(0.11)}{\micro\meter}$. These values are in excellent agreement with the corresponding values found from the analysis of the surface topography map. This agreement is reflected in the good correspondence between the functions  $\left<\partial R_{ss}(\theta_s)/\partial\Omega_s \right>_{\textrm{incoh,input}}$ [open symbols in Fig.~\ref{Fig:Sample7047:spol}(a)] and the function  $\left<\partial R_{ss}(\theta_s)/\partial\Omega_s \right>_{\textrm{incoh,calc}}$ calculated on the basis of Eq.~\eqref{eq:MDRC_PPT} (for $\alpha=s$) using the reconstructed values for the surface roughness parameters [solid line in Fig.~\ref{Fig:Sample7047:spol}(a)]. Figure~\ref{Fig:Sample7047:spol}(b) compares the input and reconstructed surface-height autocorrelation function $W(|\pvec{x}|)$. It should be mentioned that to arrive at the results presented in Fig.~\ref{Fig:Sample7047:spol} it was assumed that the polar angle of incidence was $\theta_0=\ang{28}$, not $\theta_0=\ang{30}$ for which the experiments were claimed to have been performed. We speculate that this behavior could be due to an alignment issue of the sample; for instance, it is observed from the experimental data in Fig.~\ref{Fig:Sample7047:spol}(a) that the specular peak is at an angle less than  $\ang{30}$.

%
%

\begin{figure}[tbp] 
  \centering
  \includegraphics*[width=0.469\textwidth]{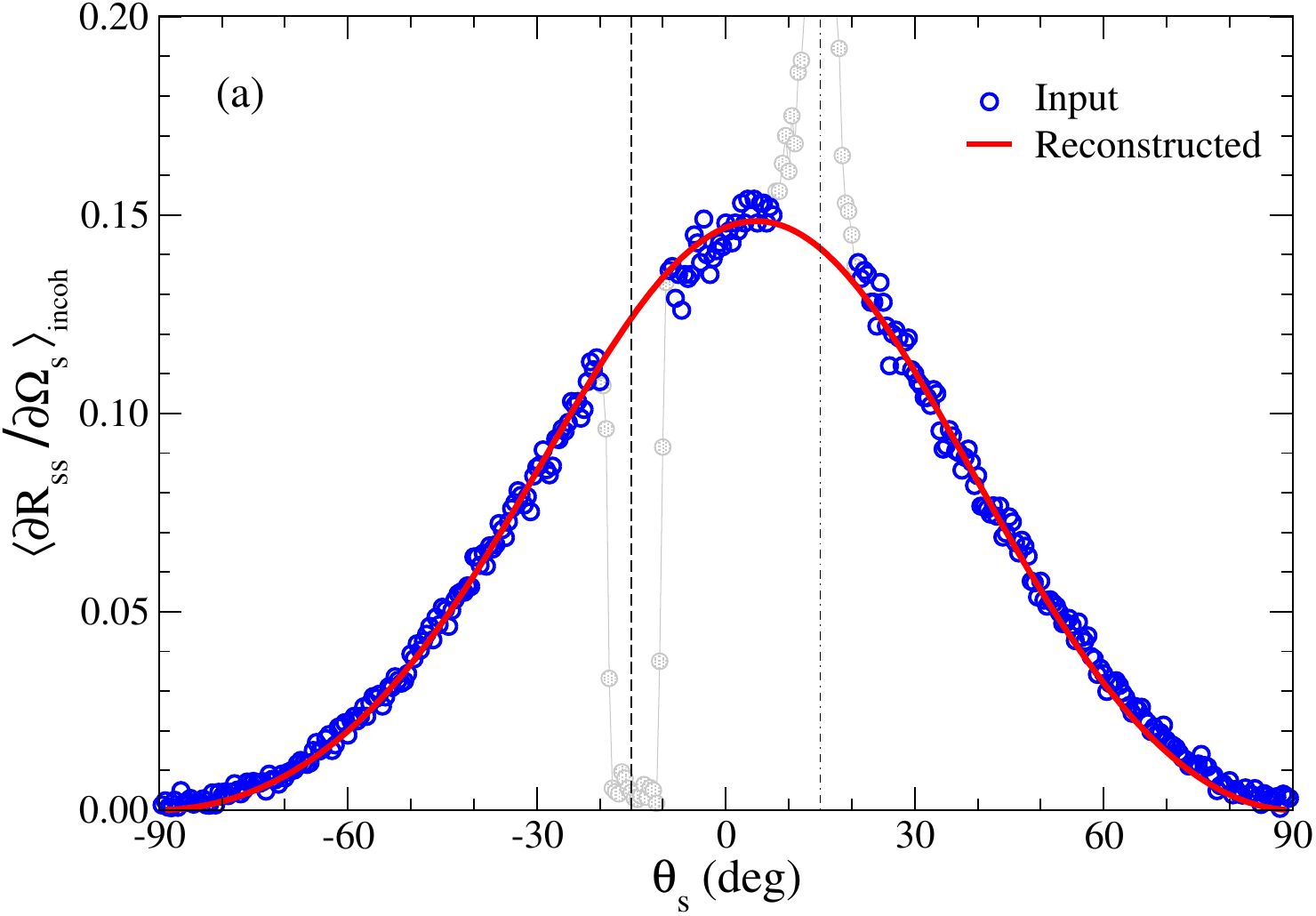}
  \qquad
  \includegraphics*[width=0.45 \textwidth]{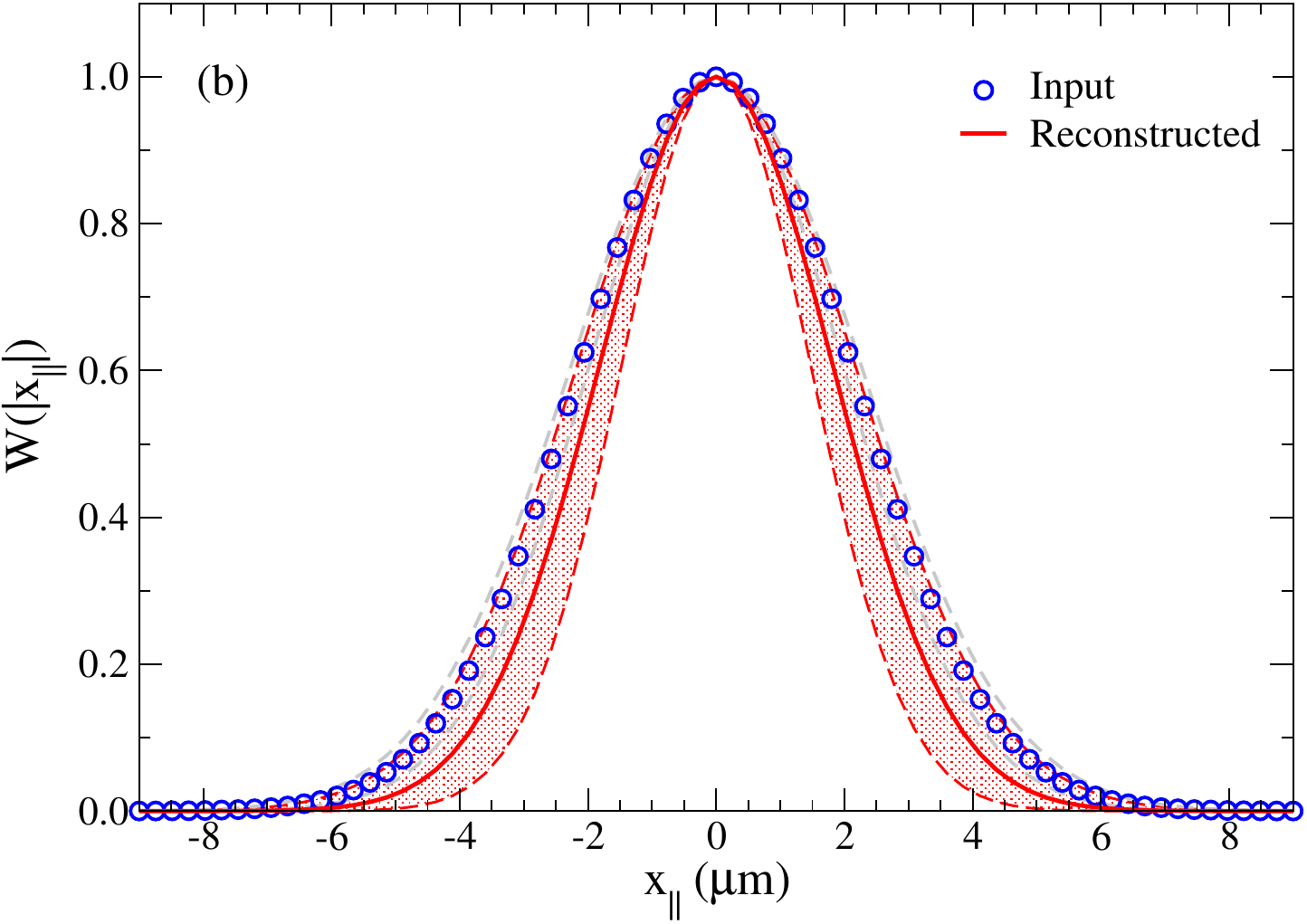}
  \caption{(Sample 8053) The same as Fig.~\ref{Fig:Sample7047:spol} but for sample~8053 from Ref.~\onlinecite{NavarreteAlcala2009} for the polar angle of incidence $\theta_0=\ang{15}$. In this publication, this sample is reported to have approximately Gaussian statistics and  characterized by the rms-roughness $\delta=\SI{0.75 \pm 0.04}{\micro\meter}$ and the lateral correlation length $a=\SI{3.00 \pm 0.20}{\micro\meter}$. The reconstruction based on the experimental data points (open symbols in Fig.~\protect\ref{Fig:Sample8052:spol}(a)) and the form~\protect\eqref{eq:29_Gaussian} for $W(|\pvec{x}|)$, resulted in the values $\delta_\star=\SI{0.92(0.13)}{\micro\meter}$ and $a_\star=\SI{2.58(0.49)}{\micro\meter}$ when the minimization procedure was started from the values $\delta_\star=\SI{0.10}{\micro\meter}$ and $a_\star=\SI{0.50}{\micro\meter}$. The region bounded by the two gray dashed lines represents the variation in $W(|\pvec{x}|)$, when assuming the form~\protect\eqref{eq:29_Gaussian}, that is due to the experimental uncertainty on the value of the lateral correlation length $a$. Similarly, the shaded region bounded by the two red dashed lines represents the bounds on $W(|\pvec{x}|)$ stemming from the upper and lower bounds of the confidence interval on the reconstructed parameter $a_\star$.  }
  \label{Fig:Sample8052:spol} 
\end{figure}
 %

\smallskip
The second experimental sample that we consider is sample~8053 from Ref.~\onlinecite{NavarreteAlcala2009}. This is also a gold sample and the morphological characterization of it revealed the rms-roughness $\delta=\SI{0.75 \pm 0.04}{\micro\meter}$ and the lateral correlation length $a=\SI{3.00 \pm 0.20}{\micro\meter}$~\cite{NavarreteAlcala2009}; in terms of the wavelength of the incident light~[$\lambda=\SI{10.6}{\micro\meter}$] one has $\delta\approx\lambda/14$ and $a\approx2\lambda/7$. Also for this sample both the height distribution and the correlation function were found to be well described by the Gaussian forms.

The symbols in Fig.~\ref{Fig:Sample8052:spol}(a) are the measured data points for the in-plane dependence of the co-polarized component of the mean DRC when s-polarized light is incident (from vacuum) on the surface of sample~8053 at the polar angle of incidence $\theta=\ang{15}$. As was done when basing the reconstruction on data for sample~7047, we neglected some of the measured data points around the specular and anti-specular directions (filled symbols) so that only the experimental data points shown as blue open symbols in Fig.~\ref{Fig:Sample8052:spol}(a) are considered part of the function $\left<\partial R_{ss}(\theta_s)/\partial\Omega_s \right>_{\textrm{incoh,input}}$. Using these input data, the minimization of the cost function~\eqref{eq:27} with respect to the set of variational parameters ${\mathcal P} = \{\delta_\star, a_\star \}$, resulted in the reconstructed surface roughness parameters $\delta_\star=\SI{0.92(0.13)}{\micro\meter}$ and $a_\star=\SI{2.58(49)}{\micro\meter}$. These results were obtained by assuming a Gaussian form~\eqref{eq:29_Gaussian} for $W(|\pvec{x}|)$ and the initial values  $\delta_\star=\SI{0.10}{\micro\meter}$ and $a_\star=\SI{0.50}{\micro\meter}$. Moreover, for this data set, the polar angle of incidence assumed in the reconstruction is the same as was used in performing the experiment~[$\theta_0=\ang{15}$].

The reconstructed roughness parameters obtained for sample~8053 are significantly less precise (larger error bars) than what was found previously for sample~7047. By considering the confidence intervals, one finds that the reconstructed roughness parameters $\delta_\star$ and $a_\star$ are still consistent with the experimentally obtained values reported in Ref.~\onlinecite{NavarreteAlcala2009} for $\delta$ and $a$ even if the confidence intervals of the reconstructed and experimentally obtain roughness parameters barely overlap. The results in Fig.~\ref{Fig:Sample8052:spol}(a) show a good agreement between the input and reconstructed in-plane angular dependence of the mean DRC. The solid line in Fig.~\ref{Fig:Sample8052:spol}(b) shows that the number value of the lateral correlation length is underestimated as compared to what is found by a direct examination of the surface topography. The red dashed lines in this figure represent the Gaussian correlation function for values of the correlation length corresponding to upper and lower bounds on the confidence interval for $a_\star$. It is seen from Fig.~\ref{Fig:Sample8052:spol}(b), as expected, that the input correlation function~(open symbols) is inside the shaded region bounded by the red dashed lines. Furthermore, in the same figure the region between the two dashed gray lines represents the variation in $W(|\pvec{x}|)$, when assuming the form~\protect\eqref{eq:29_Gaussian}, that is due to the experimental uncertainty on the value of the lateral correlation length $a$. The absolute relative errors in the number values of the reconstructed and experimentally obtained surface roughness parameters $\delta$ and $a$ are found to be  \SI{16.5}{\percent} and \SI{7.9}{\percent}, respectively.

\section{Conclusions}
\label{Sec:Conclusions}

A detailed derivation of the expressions for the in-plane, p-to-p incoherent light scattering from a two-dimensional randomly rough metal surface is presented within second-order phase perturbation theory. Similar expressions was recently obtained  in Ref.~\onlinecite{Simonsen2014-05} for the scattering of s-polarized light. On the basis of second-order phase perturbation theory, it is shown that the mean differential reflection coefficient can be calculated accurately for a wide range of two-dimensional randomly rough metal surfaces. In particular, this is also the case for p-to-p scattering close to the ``pseudo-Brewster'' scattering angle for which the reflectivity of the corresponding planar metal surface has a minimum. In contrast, in a recent study focusing on p-to-p incoherent light scattering from \textit{dielectric} surfaces, it was shown that the same theory produces unreliable results for polar scattering angles around the Brewster angle~\cite{Simonsen2016-06}.

An inversion approach based on in-plane, co-polarized scattering data is introduced for the normalized surface-height autocorrelation function $W(|\pvec{x}|)$ and the rms-roughness of a two-dimensional randomly rough metal surface. To this end, a cost function is defined as the in-plane angular integral of the square difference between the input mean DRC and the similar quantity calculated within second-order phase perturbation theory for the same scattering angles. By minimizing this cost-function, the surface rms-roughness and the correlation function $W(|\pvec{x}|)$ could be obtain for an assumed parametrized form of the latter function. The forms for $W(|\pvec{x}|)$ that we considered were the Gaussian and stretched exponential forms.

In this way, several sets of inversions were performed on two classes of scattering data.  The first class of scattering data were produced by rigorous computer simulations based on the non-perturbative numerical solution of the reduced Rayleigh equation~\cite{Simonsen2011-05}. For several sets of roughness parameters, normal and non-normal incidence, and p and s polarization of the incident light, inversion based on such scattering data resulted in quite accurate reconstruction of both the correlation function and the rms-roughness. This was also the case when we deliberately chose the roughness parameters such that the scattering data displayed well-pronounced enhanced scattering peaks which are signatures of multiple scattering. Even if second-order phase perturbation theory is not able to produce such peaks, its use in the inversion of the in-plane scattering data still produced rather useful, and often accurate, results. At least, this was the case for the roughness parameters we assumed in producing the scattering data on which the inversion was performed.

The second class of scattering data were obtained by experimentally measuring the intensity of the light that is diffusely scattered in-plane from two-dimensional randomly rough gold surfaces that were specially prepared to be characterized by Gaussian correlation functions. The reconstruction of the first scattering data set from this class that we considered produced statistical properties of the randomly rough surface that agreed quite well with those assumed in fabricating the rough surface of the sample. The second sample we considered assumed roughness parameters for which second-order phase perturbation theory is expected to be less reliable due to the higher local slopes that the surface has. Hence, it is not surprising that the results that we obtain by the inversion are less accurate than for the first sample~(wider confidence intervals). However, the roughness parameters obtained by inversion were still consistent with the input parameters even if the input values were close to the boundaries of the confidence intervals. 

The inversion approach that we have presented is practical, rather accurate (in many cases), computational efficient and therefore fast to perform. However, the drawback of the present formulation of the approach is that it relies on a  parametrization for the autocorrelation function $W(|\pvec{x}|)$. Further research should, therefore, focus on the non-parametric reconstruction of the autocorrelation of the surface profile function. Moreover, one also needs to determining the wavelength, ranges of roughness and correlation lengths for which the proposed approach can produce reliable results. These issues will be addressed in future work.

\begin{acknowledgments}
  The authors are indebted to Dr.\ A.G.\ Navarrete Alcal\'{a}  and  Dr.\ E.I.\ Chaikina for providing the experimental scattering data from Ref.~\onlinecite{NavarreteAlcala2009} and to allow us to use them in this study. 
  This research was supported in part by NTNU and the Norwegian metacenter for High Performance Computing (NOTUR) by the allocation of computer time.
The research of I.S. was supported in part by The Research Council of Norway (Contract No.~216699) and  the French National Research Agency (ANR-15-CHIN-0003).
\end{acknowledgments}

\appendix*
\section{Derivation of Eq.~\eqref{eq:MDRC_PPT_total}}

The expressions derived in this Appendix are valid for a scattering geometry where the substrate is either metallic or dielectric. To simplify the notation, $\varepsilon$ will here denote the dielectric function of the substrate suppressing any explicit reference to the frequency dependence of this quantity. 

The starting point for our derivation of the pp element of Eq.~\eqref{eq:MDRC_PPT_total} is the pp elements of Eqs.~(12), (15), and (16)--(19) from Ref.~\onlinecite{McGurn1996}, and the definition~\eqref{eq:11} of the scattering matrix $S(\pvec{q} |\pvec{k} )$ in terms of the matrix of reflection amplitudes ${\bf R}(\pvec{q} |\pvec{k} )$.  From these equations we obtain for the pp element of the scattering matrix the expansion
\begin{align}
  S_{pp}(\pvec{q} |\pvec{k} ) 
  &=
  S^{(0)}_{pp}(\pvec{q} |\pvec{k} ) - \imu S^{(1)}_{pp}(\pvec{q} |\pvec{k} ) - \frac{1}{2} S^{(2)}_{pp}(\pvec{q} |\pvec{k} ) + \cdots 
  \label{eq:A.1}
\end{align}
where the superscript denotes the order of the corresponding term in the surface profile function $\zeta(\pvec{x})$.  The coefficient $S^{(0)}_{pp}(\pvec{q} |\pvec{k} )$ is
\begin{align}
  S^{(0)}_{pp}(\pvec{q} |\pvec{k} ) 
  &= 
  (2\pi)^2\delta (\pvec{q} -\pvec{k} ) 
  \frac{\varepsilon \alpha_0( k_\parallel ) - \alpha ( k_\parallel )}{\varepsilon\alpha_0( k_\parallel ) + \alpha ( k_\parallel )}
  \nonumber\\
  &= 
  (2\pi )^2 \delta (\pvec{q} - \pvec{k} ) 
  \frac{
    \left[\varepsilon\alpha_0( q_\parallel ) - \alpha ( q_\parallel )\right]^{\frac{1}{2}}
  }{
    \left[\varepsilon\alpha_0( q_\parallel ) + \alpha ( q_\parallel )\right]^{\frac{1}{2}}}
  \frac{ 
    \left[\varepsilon\alpha_0( k_\parallel ) - \alpha ( k_\parallel )\right]^{\frac{1}{2}}
  }{
    \left[\varepsilon\alpha_0 ( k_\parallel ) + \alpha ( k_\parallel )\right]^{\frac{1}{2}}
  }
  \nonumber\\
  &= 
  (2\pi )^2 \delta (\pvec{q} - \pvec{k} ) (\varepsilon - 1) 
  \frac{ f_{p}^{\frac{1}{2}} ( q_\parallel )}{d_p( q_\parallel )}\frac{f_{p}^{\frac{1}{2}} ( k_\parallel )}{d_p( k_\parallel )}, 
  \label{eq:A.2}
\end{align}
where the $f_{p}( q_\parallel )$, $\alpha_0( q_\parallel )$ and $\alpha ( q_\parallel )$ are defined by Eqs.~\eqref{eq:f_p}, \eqref{eq:def-alpha0} and \eqref{eq:def-alpha}, respectively.
The coefficient $S^{(1)}_{pp}(\pvec{q} | \pvec{k} )$ is obtained as
\begin{align}
  S^{(1)}_{pp}(\pvec{q} |\pvec{k} ) 
  =& 
  -2(\varepsilon - 1)\alpha^{\frac{1}{2}}_{0}( q_\parallel ) \alpha^{\frac{1}{2}}_0 ( k_\parallel ) 
  \frac{ 
    \varepsilon q_\parallel k_\parallel - \alpha ( q_\parallel )
    \pvecUnit{q}\cdot\ \pvecUnit{k}  
    \alpha ( k_\parallel )
  }{
    d_p( q_\parallel ) d_p( k_\parallel )
  } 
  \hat{\zeta}(\pvec{q} - \pvec{k} ),
  \label{eq:A.5}
\end{align}
while the coefficient $S^{(2)}_{pp}(\pvec{q} |\pvec{k} )$ is given by
\begin{align}
  S^{(2)}_{pp}(\pvec{q} |\pvec{k} ) 
  &= 
  -4(\varepsilon - 1) \frac{\alpha^{\frac{1}{2}}_0( q_\parallel )\alpha^{\frac{1}{2}}_0( k_\parallel )}{d_p( q_\parallel )d_p( k_\parallel )}
    \int \dfint[2]{p_\parallel}{(2 \pi )^2}   
    \hat{\zeta}(\pvec{q} - \pvec{p} )\hat{\zeta}(\pvec{p} - \pvec{k} )
  \nonumber \\  & \quad
  \times \bigg\{ \frac{1}{2} \left[\alpha ( q_\parallel )+\alpha ( k_\parallel ) \right]
  \left[ q_\parallel k_\parallel - \alpha ( q_\parallel ) \pvecUnit{q} \cdot \pvecUnit{k} 
    \alpha ( k_\parallel ) \right]
  \nonumber\\ & \quad \qquad
  + \left( \frac{\varepsilon -1}{\varepsilon}\right) 
  \bigg[ 
      \alpha ( q_\parallel )
      \pvecUnit{q} \cdot \pvecUnit{p}
      \alpha ( p_\parallel )
      \pvecUnit{p} \cdot \pvecUnit{k}
      \alpha ( k_\parallel )
      \nonumber\\ & \quad \qquad \qquad \qquad \qquad
      - \frac{
        \left[
          \varepsilon q_\parallel p_\parallel 
          - \alpha ( q_\parallel )
          \pvecUnit{q} \cdot \pvecUnit{p}  
          \alpha ( p_\parallel )
        \right]
        \left[
          \varepsilon  p_\parallel k_\parallel 
          - \alpha ( p_\parallel )
          \pvecUnit{p} \cdot \pvecUnit{k}
          \alpha ( k_\parallel )
        \right]
      }{
        d_p( p_\parallel )
      }
      \nonumber\\ & \quad \qquad \qquad \qquad \qquad
      - \varepsilon 
      \left( \frac{\omega}{c}\right)^2 
      \frac{
        \alpha ( q_\parallel )
        \big[ \pvecUnit{q} \times \pvecUnit{p} \big]_3
        \big[ \pvecUnit{p} \times \pvecUnit{k} \big]_3
        \alpha ( k_\parallel )
      }{
        d_s( p_\parallel )
      } 
      \bigg] \bigg\} . \label{eq:A.6}
\end{align}

On substituting Eqs.~\eqref{eq:A.2}, \eqref{eq:A.5}, and \eqref{eq:A.6} into Eq.~\eqref{eq:A.1} we find that through terms of second order in the surface profile function $S_{pp}(\pvec{q} |\pvec{k} )$ is given by
\begin{align}
  S_{pp}(\pvec{q} |\pvec{k} ) 
  =
    &\sgnQK
    (\varepsilon - 1)\frac{f_{p}^{\frac{1}{2}} ( q_\parallel )f_{p}^{\frac{1}{2}} ( k_\parallel )}{d_p( q_\parallel )d_p( k_\parallel )}
  \bigg\{ (2\pi )^2 \delta (\pvec{q} - \pvec{k} )
  \sgnQK
  \nonumber\\ & \quad
  + 2 \imu \frac{\alpha^{\frac{1}{2}}_0( q_\parallel )\alpha^{\frac{1}{2}}_0( k_\parallel )}{f_p^{\frac{1}{2}}( q_\parallel )f_p^{\frac{1}{2}} ( k_\parallel )}
  \sgnQK                
  \left[\varepsilon q_\parallel k_\parallel - \alpha ( q_\parallel )
    \pvecUnit{q} \cdot \pvecUnit{k}  
    \alpha ( k_\parallel ) 
  \right]
  \hat{\zeta}(\pvec{q} - \pvec{k} ) 
  \nonumber\\& \quad 
  + 2 \frac{\alpha^{\frac{1}{2}}_0( q_\parallel )\alpha^{\frac{1}{2}}_0( k_\parallel )}{f_p^{\frac{1}{2}} ( q_\parallel )f_p^{\frac{1}{2}} ( k_\parallel )} 
   \sgnQK
   \int \dfint[2]{p_\parallel}{(2\pi )^2}     
   \hat{\zeta}(\pvec{q} - \pvec{p} ) \hat{\zeta}(\pvec{p} - \pvec{k} )
  \nonumber\\ & \quad \quad 
  \times \bigg[ \frac{1}{2} \left[ \alpha ( q_\parallel ) + \alpha ( k_\parallel ) \right] 
  \left(
    q_\parallel k_\parallel - \alpha ( q_\parallel )
    \pvecUnit{q} \cdot \pvecUnit{k}  
    \alpha ( k_\parallel )
    \right)
  \nonumber\\& \quad \qquad 
  +
  \left( \frac{\varepsilon - 1}{\varepsilon}\right) 
  \bigg( \alpha ( q_\parallel ) 
  \pvecUnit{q} \cdot \pvecUnit{p}  
  \alpha ( p_\parallel ) 
  \pvecUnit{p} \cdot \pvecUnit{k}  
  \alpha ( k_\parallel )
  \nonumber\\ & \quad \qquad \quad 
  - \frac{
    \left[
      \varepsilon q_\parallel p_\parallel - \alpha ( q_\parallel )
      \pvecUnit{q} \cdot \pvecUnit{p}  
      \alpha ( p_\parallel ) 
    \right]
    \big[
      \varepsilon p_\parallel k_\parallel - \alpha ( p_\parallel) 
      \pvecUnit{p} \cdot \pvecUnit{k}  
      \alpha ( k_\parallel ) \big]
  }{
    d_p( p_\parallel )
  }
  \nonumber\\& \quad \qquad \quad 
  - \varepsilon \left(\frac{\omega}{c}\right)^2 
  \frac{
    \alpha ( q_\parallel )
    \big[ \pvecUnit{q} \times \pvecUnit{p} \big]_3   
    \big[ \pvecUnit{p} \times \pvecUnit{k} \big]_3   
    \alpha ( k_\parallel )
  }{
    d_s( p_\parallel )
  }
   \bigg) \bigg]\bigg\}.
\label{eq:A.7}
\end{align}
This expression for $S_{pp}(\pvec{q} |\pvec{k} )$ is manifestly reciprocal, i.e. it satisfies Eq.~\eqref{eq:12}.
In writing the expression~\eqref{eq:A.7}, we have for reasons of later convenience factored out a phase $\!\sgnQK$, where $\sgn(\cdot)$ denotes the signum (or sign) function defined by $x=\sgn(x)|x|$.

We next express Eq.~\eqref{eq:A.7} in the from of a Fourier integral,
\begin{align}
  S_{pp}(\pvec{q} |\pvec{k} ) 
  &=
  \sgnQK  
  (\varepsilon - 1)
  \frac{ 
    \left[f_{p}( q_\parallel )f_{p}( k_\parallel ) \right]^{\frac{1}{2}}
  }{
    d_p( q_\parallel )d_p( k_\parallel )
    }
    \int \dint[2]{x_\parallel}     
    \, \exp \left[- \imu (\pvec{q} - \pvec{k} ) \cdot \pvec{x} \right] 
  \nonumber\\ & \quad  
  \times \bigg\{ 1 + 2 \imu \bigg[ \frac{\alpha_0( q_\parallel )\alpha_0( k_\parallel )}{f_{p}( q_\parallel )f_{p}( k_\parallel )} \bigg]^{\frac{1}{2}} H_p(\pvec{q} |\pvec{k} )\zeta(\pvec{x}) 
  + 2 \bigg[ \frac{\alpha_0( q_\parallel )\alpha_0( k_\parallel )}{f_{p}( q_\parallel )f_{p}( k_\parallel )}\bigg]^{\frac{1}{2}}
   \nonumber\\& \quad \qquad
   \times                          
    \int  \dfint[2]{p_\parallel}{(2\pi )^2}    
    \red{ F_p(\pvec{q} |\pvec{p} |\pvec{k} ) }           
  \int  \dint[2]{u_\parallel}   
  \exp \left[-\imu (\pvec{p} - \pvec{k} )\cdot \pvec{u} \right]
  \zeta ( \pvec{x} ) \zeta(\pvec{x} + \pvec{u} )               
  \bigg\} , \label{eq:A.8}
\end{align}
where the functions $H_p(\pvec{q} |\pvec{k} ) $ and $F_p(\pvec{q} |\pvec{p} |\pvec{k} )$ are defined by the expressions in Eqs.~\eqref{eq:H_p} and ~\eqref{eq:F_p}, respectively. For polarization $\alpha=p,s$, the functions $F_\alpha(\pvec{q} |\pvec{p} |\pvec{k} )$, defined by Eq.~\eqref{eq:F_defintion},
satisfy the relation $F_\alpha(\pvec{q} |\pvec{p} |\pvec{k} ) = F_\alpha(-\pvec{k} | -\pvec{p} | -\pvec{q} )$. Moreover, the quantities $F_\alpha(\pvec{q} |\pvec{p} |\pvec{k} )$ are continuous functions of their first argument $\pvec{q}$ when the wave vector of the scattered light $\vec{q}=\pvec{q}+\alpha_0(q_\parallel)\vecUnit{x}_3$ varies in the plane of incidence. 

From Eq.~\eqref{eq:A.8} and with the use of Eq.~\eqref{eq:1},  we find
\begin{align}
  \left< S_{pp}(\pvec{q} |\pvec{k} )\right> 
  &=
    \sgnQK
    (\varepsilon -1) \frac{[f_{p}( q_\parallel )f_{p}( k_\parallel )]^{\frac{1}{2}}}{d_p( q_\parallel )d_p( k_\parallel )}
    \int \dint[2]{x_\parallel}      
    \exp \left[- \imu (\pvec{q} - \pvec{k} )\cdot \pvec{x} \right]
  \nonumber\\ & \qquad
  \times 
  \bigg\{ 1 + 2\delta^2\bigg[ \frac{\alpha_0( q_\parallel )\alpha_0( k_\parallel )}{f_{p}( q_\parallel )f_{p}( k_\parallel )} \bigg]^{\frac{1}{2}} 
          \int  \dfint[2]{p_\parallel}{(2\pi )^2}  
          F_p(\pvec{q} |\pvec{p} |\pvec{k} ) \, g(|\pvec{p} - \pvec{k} |) 
  \bigg\}
  \nonumber\\ & 
   \cong
    \sgnQK
   (\varepsilon - 1) 
  \frac{[f_{p}( q_\parallel )f_{p}( k_\parallel )]^{\frac{1}{2}}}{d_p( q_\parallel )d_p( k_\parallel )} 
   \int \dint[2]{x_\parallel}  
   \exp \left[ - \imu (\pvec{q} - \pvec{k} )\cdot \pvec{x} \right]
  \nonumber\\ & \qquad 
  \times \exp \bigg\{ 2\delta^2 \bigg [ \frac{\alpha_0( q_\parallel )\alpha_0( k_\parallel )}{f_{p}( q_\parallel )f_{p}( k_\parallel )}\bigg]^{\frac{1}{2}} 
      \int \dfint[2]{p_\parallel}{(2\pi )^2}  
      F_p(\pvec{q} |\pvec{p} |\pvec{k} ) \, g(|\pvec{p} - \pvec{k} |) \bigg\} . \label{eq:A.11}
\end{align}
It follows that
\begin{align}
  \left|\left< S_{pp}(\pvec{q} |\pvec{k} )\right> \right|^2 
  &= 
  \red{ \left| \varepsilon - 1 \right|^2 }  
  \frac{|f_{p}( q_\parallel )f_{p}( k_\parallel )|}{|d_p( q_\parallel )d_p( k_\parallel )|^2}
  \exp \left[-2M_{\red{p}}(\pvec{q} |\pvec{k} ) \right] 
    \int  \dint[2]{x_\parallel} \! \! \int \dint[2]{x_\parallel'}              
    \exp \left[ - \imu (\pvec{q} - \pvec{k} )\cdot (\pvec{x} - \pvec{x}') \right],
  \label{eq:A.12}
\end{align}
where we have introduced the function $M_{p}(\pvec{q} |\pvec{k} )$ defined by Eq.~\eqref{eq:M_alpha-definition} with $\alpha=p$. 

We next find that 
\begin{align}
  \left< \left| S_{pp}(\pvec{q} |\pvec{k} ) \right|^2\right> 
  &= 
   \red{ \left| \varepsilon - 1 \right|^2 }  
    \frac{|f_{p}( q_\parallel )f_{p}( k_\parallel )|}{|d_p( q_\parallel )d_p( k_\parallel )|^2}
    \int  \dint[2]{x_\parallel} \! \! \int \dint[2]{x_\parallel'}          
    \exp \left[- \imu (\pvec{q} -\pvec{k} )\cdot ( \pvec{x} - \pvec{x}') \right]
   \nonumber\\ & \quad
   \times \bigg\{ 1 + 4 \left| \frac{ \alpha_0( q_\parallel )\alpha_0( k_\parallel )}{f_{p}( q_\parallel )f_{p}( k_\parallel )}\right| 
   \left| H_p(\pvec{q} |\pvec{k} ) \right|^2 
   \delta ^2 W(|\pvec{x} - \pvec{x} '|)
   \nonumber\\ & \quad \qquad
   + 4\delta^2 \Re \left[ \frac{\alpha_0 ( q_\parallel )\alpha_0( k_\parallel )}{f_{p}( q_\parallel )f_{p}( k_\parallel )}\right]^{\frac{1}{2}} 
    \int \dfint[2]{p_\parallel}{(2\pi  )^2}  
        F_p(\pvec{q} |\pvec{p} |\pvec{k} )\,
        g(|\pvec{p} - \pvec{k} |)
  \bigg\}
   \nonumber\\ & 
   \cong
   \red{ \left| \varepsilon - 1 \right|^2 }               
   \frac{|f_{p}( q_\parallel )f_{p}( k_\parallel )|}{|d_p( q_\parallel )d_p( k_\parallel )|^2} 
   \int  \dint[2]{x_\parallel} \! \! \int \dint[2]{x_\parallel'}                        
   \exp \left[ - \imu (\pvec{q} - \pvec{k} ) \cdot (\pvec{x} - \pvec{x}') \right]
   \nonumber\\ &  \quad
   \times \exp \bigg\{ 4\delta^2 \Re 
   \left[ 
     \frac{\alpha_0( q_\parallel )\alpha_0( k_\parallel )}{f_{p}( q_\parallel )f_{p}( k_\parallel )} 
   \right]^{\frac{1}{2}} 
    \int  \dfint[2]{p_\parallel}{(2\pi  )^2} 
                 F_p(\pvec{q} |\pvec{p} |\pvec{k} ) \, g(|\pvec{p} - \pvec{k} |) 
   \nonumber\\ &  \quad \qquad \quad 
   + 4\delta^2 \left| \frac{\alpha_0( q_\parallel )\alpha_0( k_\parallel )}{f_{p}( q_\parallel )f_{p}( k_\parallel )}\right| 
   \left| H_p(\pvec{q} |\pvec{k} ) \right|^2 
   W(|\pvec{x} - \pvec{x}'|) 
   \bigg\} 
   \nonumber\\ & 
   =              
   \red{ \left| \varepsilon - 1 \right|^2 }               
   \frac{|f_{p}( q_\parallel )f_{p}( k_\parallel )|}{|d_p( q_\parallel )d_p( k_\parallel )|^2}
   \exp [-2M_{\red{p}}(\pvec{q} |\pvec{k} ) ]
   \int  \dint[2]{x_\parallel} \! \! \int \dint[2]{x_\parallel'}                        
   \exp \left[ - \imu (\pvec{q} - \pvec{k} ) \cdot (\pvec{x} - \pvec{x}') \right]
   \nonumber\\ &  \quad
    \times \exp \bigg\{
     4\delta^2 \left| \frac{\alpha_0( q_\parallel )\alpha_0( k_\parallel )}{f_{p}( q_\parallel )f_{p}( k_\parallel )}\right| 
   \left| H_p(\pvec{q} |\pvec{k} ) \right|^2 
   W(|\pvec{x} - \pvec{x}'|) 
   \bigg\}. 
   \label{eq:A.14}
\end{align}
On subtracting the expression in Eq.~\eqref{eq:A.12} from the expression in Eq.~\eqref{eq:A.14} we obtain
\begin{align}
  \left< \left| S_{pp}(\pvec{q} |\pvec{k} ) \right|^2\right> 
  &- \left| \left< S_{pp}(\pvec{q} |\pvec{k} ) \right> \right|^2 
  \nonumber\\ 
  &=
 \red{ \left| \varepsilon - 1 \right|^2 }                
  \frac{|f_{p}( q_\parallel )f_{p}( k_\parallel )|}{|d_p( q_\parallel )d_p( k_\parallel )|^2} \exp [-2M_{\red{p}}(\pvec{q} |\pvec{k} ) ] 
  \int  \dint[2]{x_\parallel} \! \! \int \dint[2]{x_\parallel'}                                       
  \exp \left[- \imu (\pvec{q} - \pvec{k} )\cdot (\pvec{x} - \pvec{x}') \right]
  \nonumber\\ & \quad 
  \times 
  \left\{ 
    \exp \left[  4\delta^2 
      \left| 
        \frac{\alpha_0( q_\parallel )\alpha_0( k_\parallel )}{f_{p}( q_\parallel )f_{p}( k_\parallel )}
      \right| 
    \left|H_p(\pvec{q} |\pvec{k} )\right|^2 
    W(|\pvec{x} - \pvec{x}'|)
  \right]
  - 1 
\right\}. 
\label{eq:A.15}
\end{align}
Since the integrand on the right-hand-side of Eq.~\eqref{eq:A.15} depends on $\pvec{x}$ and $\pvec{x}'$ only through the difference $\pvec{x}-\pvec{x}'$, it is convenient to make the change of variable $\pvec{x}'=\pvec{x}-\pvec{u}$ in this equation. The integral over $\pvec{x}$ that the resulting equation contains gives the area ${\mathcal S}$ of the plane $x_3=0$ covered by the randomly rough surface. In this way we obtain 
\begin{align}
  \left< \left| S_{pp}(\pvec{q} |\pvec{k} ) \right|^2\right> 
  &- \left| \left< S_{pp}(\pvec{q} |\pvec{k} ) \right> \right|^2 
  \nonumber\\ 
  &=
  {\mathcal S}  
 \red{ \left| \varepsilon - 1 \right|^2 }                
  \frac{|f_{p}( q_\parallel )f_{p}( k_\parallel )|}{|d_p( q_\parallel )d_p( k_\parallel )|^2} \exp [-2M_{\red{p}}(\pvec{q} |\pvec{k} ) ] 
  \nonumber\\ & \quad 
  \times 
    \int  \dint[2]{u_\parallel} 
    \exp \left[- \imu (\pvec{q} - \pvec{k} )\cdot \pvec{u} \right]
  \left\{ 
    \exp \left[  4\delta^2 
      \left| 
        \frac{\alpha_0( q_\parallel )\alpha_0( k_\parallel )}{f_{p}( q_\parallel )f_{p}( k_\parallel )}
      \right| 
    \left|H_p(\pvec{q} |\pvec{k} )\right|^2 
    W(|\pvec{u}|)
  \right]
  - 1 
\right\}. 
                \label{eq:A.15-rewritten}
\end{align}

The double integral over $\pvec{u}$ that this equation contains can be transformed to a single integral over the length of this vector by noting that the part of the integrand in the curly brackets in Eq.~\eqref{eq:A.15-rewritten} is a function of $u_\parallel = |\pvec{u}|$ only. This enables us to perform the directional integration analytically to produce a result depending on the Bessel function of the first kind and order zero [denoted $J_0(\cdot)$ below]. In this way we find that Eq.~\eqref{eq:A.15-rewritten}  can alternatively be written in the equivalent form
\begin{align}
  \left< \left| S_{pp}(\pvec{q} |\pvec{k} ) \right|^2\right> 
  &- \left| \left< S_{pp}(\pvec{q} |\pvec{k} ) \right> \right|^2 
  \nonumber\\ 
  &=
  2 \pi {\mathcal S}  
 \red{ \left| \varepsilon - 1 \right|^2 }                
  \frac{|f_{p}( q_\parallel )f_{p}( k_\parallel )|}{|d_p( q_\parallel )d_p( k_\parallel )|^2} \exp [-2M_{\red{p}}(\pvec{q} |\pvec{k} ) ] 
  \nonumber\\ & \quad 
  \times 
   \int\limits^{\infty}_{0} \dint{u_\parallel} 
           u_\parallel J_0(|\pvec{q} - \pvec{k} | u_\parallel)
  \left\{ 
    \exp \left[  4\delta^2 
      \left| 
        \frac{\alpha_0( q_\parallel )\alpha_0( k_\parallel )}{f_{p}( q_\parallel )f_{p}( k_\parallel )}
      \right| 
    \left|H_p(\pvec{q} |\pvec{k} )\right|^2 
    W(u_\parallel)
  \right]
  - 1 
\right\}. 
\label{eq:A.15-Bessel}
\end{align}
The substitution of the result from Eq.~\eqref{eq:A.15-Bessel}, or equivalent [Eq.~\eqref{eq:A.15-rewritten}], into Eq.~\eqref{eq:13} for $\alpha=\beta=p$ yields Eq.~\eqref{eq:MDRC_PPT_total} for $\alpha=p$.

%
%
\bigskip
We now turn to the case of s polarization. In the Appendix of Ref.~\onlinecite{Simonsen2014-05}, Eq.~(A5), an expression for $S_{ss}(\pvec{q}|\pvec{k})$ was recently derived within second-order phase perturbation theory. In order to harmonize the notation for the expressions for $S_{\alpha\alpha}(\pvec{q}|\pvec{k})$ for p- and s-polarized light, we here recast Eq.~(A5) of Ref.~\onlinecite{Simonsen2014-05} in the alternative form
\begin{align}
  S_{ss}(\pvec{q} |\pvec{k} ) 
  &=
  - \sgnQK  
  (\varepsilon - 1)
  \frac{ 
    \left[f_s( q_\parallel )f_s( k_\parallel ) \right]^{\frac{1}{2}}
  }{
    d_s( q_\parallel )d_s( k_\parallel )
    }
    \int \dint[2]{x_\parallel}     
    \, \exp \left[- \imu (\pvec{q} - \pvec{k} ) \cdot \pvec{x} \right] 
  \nonumber\\ & \quad  
  \times \bigg\{ 1 + 2 \imu \bigg[ \frac{\alpha_0( q_\parallel )\alpha_0( k_\parallel )}{f_s( q_\parallel )f_s( k_\parallel )} \bigg]^{\frac{1}{2}} H_s(\pvec{q} |\pvec{k} )\zeta(\pvec{x}) 
  + 2 \bigg[ \frac{\alpha_0( q_\parallel )\alpha_0( k_\parallel )}{f_s( q_\parallel )f_s( k_\parallel )}\bigg]^{\frac{1}{2}}
   \nonumber\\& \quad \qquad
   \times                          
    \int  \dfint[2]{p_\parallel}{(2\pi )^2}    
    \red{ F_s(\pvec{q} |\pvec{p} |\pvec{k} ) }           
  \int  \dint[2]{u_\parallel}   
  \exp \left[-\imu (\pvec{p} - \pvec{k} )\cdot \pvec{u} \right]
  \zeta ( \pvec{x} ) \zeta(\pvec{x} + \pvec{u} )               
  \bigg\} , \label{eq:A.8_ss}
\end{align}
with $f_s(q_\parallel)$ and $H_s( \pvec{q} | \pvec{k} )$ defined by Eqs.~\eqref{eq:f_s} and \eqref{eq:H_s}, respectively, and 
\begin{align}
  F_s( \pvec{q} | \pvec{p} | \pvec{k} )
  =& - \left( \frac{ \omega }{ c } \right)^2    F( \pvec{q} | \pvec{p} | \pvec{k} ).
     \label{eq:FF_s}
\end{align}
Here the function $F( \pvec{q} | \pvec{p} | \pvec{k} )$ is defined by Eq.~(A6) of Ref.~\onlinecite{Simonsen2014-05} so that the expression for $F_s( \pvec{q} | \pvec{p} | \pvec{k} )$ is given by Eq.~\eqref{eq:F_s}. The expression for $S_{ss}(\pvec{q}|\pvec{k})$ in Eq.~\eqref{eq:A.8_ss} is, up to a phase factor, \textit{form equivalent} to the expression for $S_{pp}(\pvec{q}|\pvec{k})$ given in Eq.~\eqref{eq:A.8}; we only have to make sure that we use the functions $d_\alpha(q_\parallel)$, $f_\alpha(q_\parallel)$, $H_\alpha( \pvec{q} | \pvec{k} )$, and $F_\alpha( \pvec{q} | \pvec{p} | \pvec{k} )$ when writing the expression for $S_{\alpha\alpha}(\pvec{q} |\pvec{k} )$.  The consequence of this is that the expressions in Eqs.~\eqref{eq:A.11}--\eqref{eq:A.16} derived for p-to-p scattering also should hold for s-to-s scattering given that the polarization indices p are replaced by the indices s, and hence the expression within phase perturbation theory for the $\left< \partial R_{ss}/\partial \Omega_s \right>_{\textrm{incoh}}$ is given by Eq.~\eqref{eq:MDRC_PPT_total} with $\alpha=s$.

%
%
\medskip
Before closing this Appendix, we would like to make a few final remarks. For some forms of the correlation function $W(|\pvec{u}|)$, like the Gaussian form, it can be of advantage to expand in powers of $\delta^2$ the exponential function present in the integrands of Eq.~\eqref{eq:A.15-rewritten} or \eqref{eq:A.15-Bessel} and integrate the resulting series term-by-term. Applying this procedure to the integrand in Eq.~\eqref{eq:A.15-Bessel} (also valid for $\alpha=s$) will, for instance, produce  
\begin{align}
  \left< \left| S_{\alpha\alpha}(\pvec{q} |\pvec{k} ) \right|^2 \right> 
  &- \left| \left< S_{\alpha\alpha}(\pvec{q} | \pvec{k} ) \right>  \right|^2
  \nonumber\\  
  &= 
  2\pi {\mathcal S}
  \red{ \left| \varepsilon - 1 \right|^2 }   
  \frac{ 
    \left|f_{\alpha}( q_\parallel )f_{\alpha}( k_\parallel ) \right|
  }{
    \left|d_{\alpha}( q_\parallel )d_{\alpha}( k_\parallel ) \right|^2
  } 
  \exp \left[-2M_{\alpha}(\pvec{q} |\pvec{k} ) \right]
  \nonumber\\ & \quad 
  \times
  \sum\limits^{\infty}_{n=1} \frac{1}{n!} 
  \left[ 
    4\delta^2 
    \left| 
      \frac{
        \alpha_0( q_\parallel )\alpha_0( k_\parallel )
      }{
        f_{\alpha}( q_\parallel )f_{\alpha}( k_\parallel )
      }
      \right| 
      \left| H_{\alpha}( \pvec{q} | \pvec{k} ) \right|^2 \right]^n 
      \int\limits^{\infty}_{0} \dint{u_\parallel}
         u_\parallel J_0(|\pvec{q} - \pvec{k} | u_\parallel) W^n (u_{\|}).
    \label{eq:A.16}
\end{align}
The expression in Eq.~\eqref{eq:A.16} may represent an advantage over the expression in Eq.~\eqref{eq:A.15-Bessel} if the integrals contained in the former equation can be calculated analytically while the integral in the latter has to be calculated numerically. For instance, if the correlation function $W(|\pvec{u}|)$ has the Gaussian form~\eqref{eq:28_Gaussian}, it is straightforward to show that for $n$ a positive integer
\begin{align}
  \int  \dint[2]{u_\parallel}   
    \exp \left[-i(\pvec{q} - \pvec{k} ) \cdot \pvec{u} \right]     W^n(|\pvec{u}|)
  &=
    \frac{\pi a^2}{n} 
    \exp \left[ - \frac{a^2}{4n} \left( \pvec{q}-\pvec{k}\right)^2  \right],
    \label{eq:A.18}
\end{align}
which is the integral that will result after expanding the exponential function of the integrand in Eq.~\eqref{eq:A.15-rewritten}.

%
\bibliographystyle{aipnum4-1}
\bibliography{$HOME/Archive/Papers/BIBLIOGRAPHY,$HOME/Archive/Papers/BOOKS,$HOME/Adm/CV/PubList/Simonsen-Publications,paper2018-06}

\end{document}